%% file: main.tex
\documentclass[%
 reprint,
 amsmath,amssymb,
 aps,
pra,
]{revtex4-2}
\usepackage{upgreek}
\usepackage{amsmath}
\usepackage{graphicx}
\usepackage{dcolumn}
\usepackage{bm}
\usepackage[utf8]{inputenc}
\usepackage{xcolor}
\usepackage[normalem]{ulem}
\usepackage{cancel}
\usepackage{multibib}
\usepackage{longtable}
\usepackage{hyperref}
\begin{document}

\title{Defect Tolerance in Trigonal Selenium Photovoltaics}

\author{Jiban Kangsabanik${}^{1,2}$}
\email{jika@bio.aau.dk}
\author{Kasper Tolborg${}^{2}$}
\author{Thomas Olsen${}^{1}$}
\author{Kristian S. Thygesen${}^{1}$}
\email{thygesen@fysik.dtu.dk}

\affiliation{${}^{1}$CAMD, Computational Atomic-Scale Materials Design, Department of Physics, Technical University of Denmark, 2800 Kgs. Lyngby, Denmark\\
${}^{2}$Department of Chemistry and Bioscience, Aalborg University, 9220 Aalborg, Denmark}

\date{\today}

\begin{abstract}
Understanding how point defects fundamentally influence photovoltaic performance remains a central question for emerging wide-band gap absorbers. Trigonal selenium (t-Se) has recently re-emerged as a promising photovoltaic material due to its near-optimal band gap for tandem and indoor applications. Here we quantify defect-assisted Shockley–Read–Hall (SRH) recombination in t-Se using first principles calculations across a large and chemically diverse set of point defects. Our results suggest that t-Se is intrinsically defect tolerant. Despite the presence of multiple deep levels in the gap, recombination via nonradiative multi-phonon emission processes is strongly suppressed by large lattice reorganizations and large energy releases of at least $0.5E_\mathrm{G}$ per recombination event, while radiative defect-assisted capture also remains too small to account for the observed device losses.
Consequently, SRH recombination mediated by realistic concentrations of point defects cannot account for the observed efficiency limitations in selenium photovoltaics. We explore trends in both radiative and nonradiative SRH recombination rates across the defect data set, highlighting their complex dependence on defect level position, lattice relaxation, charge state, and doping conditions. These findings establish trigonal selenium as a defect-tolerant wide-band-gap absorber and provide transferable design principles for optimizing next-generation photovoltaic materials for tandem and indoor applications.

\end{abstract}

\maketitle

\section{Introduction}
In 1873 W. Smith reported the observation of a photovoltaic effect in selenium, making it the earliest example of a solid-state solar cell absorber material.\cite{smith1873action} Selenium is found in both amorphous (red amorphous and black amorphous)\cite{Lu2024jacs} and crystalline forms ($\alpha$-, $\beta$-, $\gamma$-monoclinic, and trigonal),\cite{cherin1972refinement, marsh1953crystal, foss1980crystal, keller1977effect, Moustafa2024}, which today find applications in various areas of opto-electronics.\cite{Li2024, Hadar2019} The crystalline trigonal phase (t-Se) has recently attracted renewed attention, due to its potential application as the wide-band gap component of a tandem solar cell and as a wide band gap absorber in a single-junction solar cell, e.g. for indoor photovoltaics where the ideal band gap is close to 1.9 eV.\cite{Youngman2021, Nielsen2024prx, Todorov2017, Yan2022} The t-Se phase has a quasi-one dimensional crystal structure consisting of covalently bonded Se chains held together by weaker dispersive forces.\cite{Todorov2017, Moustafa2024} The t-Se phase is the most stable phase at ambient conditions and has a reported band gap of 1.83-2.00 eV.\cite{Todorov2017, Nielsen2022} The material is highly stable, non-toxic, and easy to synthesize at low temperature (below 220 $^{o}$C), making it attractive for photovoltaic applications.\cite{Almora2024, Todorov2017, Hadar2019, Lu2024} 

Although the device efficiency reported by W. Smith in 1873 was less than 1\%,\cite{smith1873action}  improvements in device fabrication techniques resulted in higher single-junction efficiencies reaching 5\% in 1985\cite{nakada1985polycrystalline} and 6.5\% in 2017\cite{Todorov2017} with an open-circuit voltage close to 1 V. The 1985 (ITO)/TiO$_2$/ Se/Au device structure by Nakada \emph{et al.}\cite{nakada1985polycrystalline} with close to 5\% efficiency remained the state of the art for three decades during which the focus was mainly on improving the Se crystal quality. In 2017, Todorov \emph{et al.} reported a significant performance boost obtained by introduction of MoO$_x$ as a hole transport layer and replacement of TiO$_2$ by Zn$_x$Mg$_{1-x}$O as the transparent electron transport layer. These modifications resulted in an open-circuit voltage of 0.97 eV and a 6.5\% device efficiency under 1 Sun illumination.\cite{Todorov2017} 

More recently, rapid advances in materials processing and device engineering have pushed the performance of trigonal selenium photovoltaics substantially forward. Shen \emph{et al.} reported an oxygen-assisted high-temperature deposition strategy that enables the direct growth of highly crystalline, vertically oriented t-Se films with improved grain-boundary properties and reduced electrical losses, leading to a certified efficiency of 7.55\%.\cite{shen2025oxygen} Subsequently, Wen \emph{et al.} demonstrated an illumination-assisted annealing approach that promotes photo-induced crystallization and suppresses film dewetting, producing large-grained Se films ($\sim$2.7~$\mu$m) with low trap density and long carrier lifetimes, enabling selenium solar cells with a certified efficiency of 10.3\% and an open-circuit voltage exceeding 1~V.\cite{wen2026illumination} In parallel, Xia \emph{et al.} developed a vapor-pressure–mediated annealing strategy that induces vertically oriented t-Se films and improves carrier transport along the Se chains, achieving high indoor photovoltaic performance with efficiencies up to 12.2\% under low-intensity illumination.\cite{xia2026vapor} These results highlight the renewed potential of selenium as a wide-band-gap photovoltaic absorber, particularly for tandem and indoor photovoltaic applications.\cite{wen2026illumination}

Despite this encouraging progress, the performance of selenium photovoltaics remains far below the theoretical limits expected for a semiconductor with a band gap of $\sim$1.9~eV ($\approx$25\% Shockley–Queisser limit for single-junction solar cells and $>$50\% detailed-balance limit for indoor photovoltaics).\cite{shockley1961detailed} The microscopic origin of the remaining voltage and efficiency losses remains unresolved, in particular whether the dominant recombination pathways arise from intrinsic point defects in the selenium lattice or from extrinsic sources such as impurities, interfaces, or extended defects.

Point defects are extremely important for semiconductors and often dictate their application potential and performance, not least for for thin-film photovoltaics.\cite{Freysoldt2014, Park2018} Defects that 
form localised states inside the band gap - the so-called 
deep level defects - can facilitate carrier recombination and thereby hinder optimal device operation. Point defects are generated during synthesis of the semiconductor, and although their types and concentrations can be controlled to some extent via the growth conditions, they cannot be completely avoided. It is therefore of paramount importance to be able to probe the defects and understand and evaluate their effects on the semiconductor properties.  

In the 1950's Shockley, Read, and Hall  developed a semi-empirical theory of defect assisted transitions\cite{shockley1952statistics, hall1952electron}. Since then the SRH theory has been widely used to explain carrier capture and recombination phenomena in Si and III-V semiconductors\cite{herman1973thermal,henry1977nonradiative} and it remains a basis for our understanding of such processes.  Recently, first-principles methods based on density functional theory (DFT) have been used to calculate carrier capture rates\cite{Shi2012, Alkauskas2014, Shi2015, Dreyer2020} and thereby enabled an atomic-scale description of the interactions between crystal defects and charge carriers. Because of the sensitivity of the capture rate to the energy and spatial shape of the defect state as well as its coupling to lattice vibrations, it has become evident that non-local hybrid functionals are required to achieve quantitatively reliable results.\cite{heyd2003hybrid, Kim2019, huang2023metastability} The significant cost of such calculations has so far constrained computational studies to a single or a few defects in the materials of interest\cite{Zhang2021, wang2024upper,Zhang2021cellrep, Zhang2021jpcc, Whalley2021, Zhang2021, Liang2022, Zhang2020prb, Zhang2023,Stoliaroff2021, Moustafa2024, kavanagh2025intrinsic}. While such studies are valuable for elucidating the role of very specific defects, they are insufficient for assessing a material’s overall application potential, given the large and diverse range of defects that can influence performance.

In this paper, we present a comprehensive first-principles study of the point-defect landscape in trigonal selenium based on hybrid DFT calculations. Our computational formalism generalizes the current state of the art to defects with multiple charge transition levels (CTLs) in the band gap, as is essential for wide-band-gap semiconductors such as $t$-Se,\cite{doi:10.1021/jacs.5c12981} and to general electron/hole doping conditions. For 34 intrinsic and extrinsic defects spanning multiple chemical families, we calculate structural and thermodynamic properties, CTLs, and carrier-capture rates, and solve the full multi-level recombination-rate equations to evaluate photovoltaic device performance for experimentally relevant carrier and defect concentrations. 

We find that intrinsic point defects in $t$-Se are electronically benign and that most extrinsic impurities likewise remain harmless at realistic concentrations, despite often exhibiting multiple deep levels in the gap. Only a narrow subset of multi-level impurities, most notably Sn-like centers with four CTLs deep inside the band gap, can become detrimental, and only at very high concentrations. Overall, our results show that SRH recombination mediated by realistic concentrations of bulk point defects is insufficient to explain the observed performance limitations of selenium photovoltaics, pointing instead to interfaces, extended defects, or microstructural disorder as the dominant loss channels. More broadly, the work establishes a transferable framework and quantitative design rules for diagnosing defect tolerance in wide-band-gap photovoltaic materials.


\section{Methodology}\label{methodology}

\subsection*{A. Defect formation energy and equilibrium concentration}\label{methodDF}

The formation energy of a defect $\mathrm{D}^q$ is calculated as
\begin{equation}\label{eq:eform}
\begin{split}
E^{f}\left[\mathrm{D}^q\right] = E_{\mathrm{tot}}\left[\mathrm{D}^q\right] - E_{\mathrm{tot}}^\mathrm{bulk} - \sum_i n_{i}\mu_i + \\
q(E_\mathrm{VBM} + \Delta E_\mathrm{F})+ E_\mathrm{corr},
\end{split}
\end{equation}
where $E_{\mathrm{tot}}\left[\mathrm{D}^q\right]$ is the total energy of the optimized supercell with defect D in charge state $q$. $E_{\mathrm{tot}}^\mathrm{bulk}$ is the total energy of the optimized pristine supercell with same size. $\mu_i$ is the elemental chemical potential of the defect species, where $n_i$ represents the number of such atoms added ($n_i > 0$) or removed ($n_i <0$) from the pristine supercell to create the defect supercell. The next term represents the chemical potential of the electrons added or removed to create the charge state $q$, where $E_\mathrm{VBM}$ is the energy of the valence band maximum (VBM) and $\Delta E_\mathrm{F}$ is the Fermi level with respect to the VBM (which can vary from 0 to $E_{\mathrm{gap}}$). $E_\mathrm{corr}$ is the correction term which includes the electrostatic interaction between the charges associated with the periodic repetitions of the supercell as well as the compensating homogeneous background charge and a potential alignment corrections for charged defects. 

From the formation energies in Eq.(\ref{eq:eform}), for a defect D the CTL between charge state $q$ and $q'$ can be calculated as\cite{Moustafa2024},
\begin{equation}\label{eq:ectl}
\begin{split}
E^{q/q'}= \frac{E^{f}\left[\mathrm{D}^q\right]-E^{f}\left[\mathrm{D}^{q'}\right]}{q'-q}
\end{split}
\end{equation}

For a given Fermi level, the equilibrium concentration of a defect at temperature $T$ can be calculated as,
\begin{equation}\label{eq:defc}
N\left[\mathrm{D}^q\right]= N_{\mathrm{site}}g_q \exp(-\frac{E^{f}\left[\mathrm{D}^q\right]}{k_BT})
\end{equation}
Here $N_{\mathrm{site}}$ and $g_q$ are the number of sites available per unit volume to generate defect D and site degeneracy of defect D at charge state q, including spin degeneracy, respectively. $k_B$ is the Boltzmann's constant. As $N\left[\mathrm{D}^q\right]$ is dependent on the position of the Fermi level, it becomes very important to calculate the position of the Fermi level carefully. If all possible defect types are known, the charge neutrality equation (see below) can be solved self-consistently to find the equilibrium Fermi level and hence the relevant defect concentrations.

The charge neutrality equation can be expressed as, 
\begin{equation}\label{eq:neut}
\sum_{D,q}qN\left[\mathrm{D}^q\right]=n_0-p_0,
\end{equation}
where $n_0$ and $p_0$ are the equilibrium concentrations of free electrons and holes in the system, respectively. The latter can be calculated using the density of states, $D(E)$, of the pristine system as,

\begin{equation}\label{eq:freen}
n_0=\int_{E_{\mathrm{CBM}}}^{\infty}\exp(\frac{E_{\mathrm{F}}-E}{k_BT})D(E)dE
\end{equation}

\begin{equation}\label{eq:freep}
p_0=\int_{-\infty}^{E_{\mathrm{VBM}}}\exp(\frac{E-E_{\mathrm{F}}}{k_BT})D(E)dE
\end{equation}

The Fermi level calculated using the charge-neutrality condition can sometimes be misleading due to the incomplete knowledge of the full spectrum of intrinsic and extrinsic defects present during synthesis. In practice, the defect populations incorporated during growth may deviate significantly from thermodynamic equilibrium. As such, it would be more practical to include experimental information about carrier type and concentration, and from that determine the equilibrium Fermi level and defect concentrations. For example, it is known that t-Selenium is typically observed to be a p-type semiconductor with a majority carrier concentration of approximately 10$^{16}$ cm$^{-3}$.  From Eq. \ref{eq:freep} we can then directly calculate the corresponding $E_{\mathrm{F}}$, and use that value to calculate consistent defect formation energies and concentrations. In evaluating SRH recombination and photovoltaic device parameters, we treat the defect concentration as an independent parameter reflecting possible synthesis conditions, while the equilibrium Fermi level is fixed by the experimentally observed carrier density. This approach allows us to systematically explore how recombination rates and device performance depend on both the operating Fermi level and realistic ranges of defect concentrations.

\subsection*{B. Photovoltaic device parameters}\label{methodPV}

Recently, some of us developed a formalism to evaluate photovoltaic device parameters in the presence of defects with any number of CTLs inside the band gap and for general doping conditions. The formalism evaluates SRH recombination by solving the coupled rate equations describing carrier capture and emission processes via all defect charge states under nonequilibrium operating conditions defined by the quasi-Fermi level splitting. As it turns out, most point defects in t-Se have multiple CTLs inside the band gap. As such it becomes essential to use our full formalism instead of the more approximate single-CTL methods typically employed in the literature. In this section, we briefly discuss the most central concepts and parameters of our formalism. For detailed derivations, we refer the reader to the original work.\cite{doi:10.1021/jacs.5c12981}

The photovoltaic device equation in the presence of non-radiative (e.g. SRH) recombination loss, can be expressed as,

\begin{equation}
	\eta=\frac{max(V[J_\text{sc}-J_{r}^\text{rad}(e^{\frac{eV}{k_\text{B}T}}-1)-eR_\text{SRH}(V)d)])}{\int_{0}^{\infty}\hbar^2\omega I_\text{sun}(\omega) d\omega}
	\label{eqPV5}
	\end{equation}

Here, $\eta$ is the defect limited power conversion efficiency, $V$ is the voltage, $J_\text{sc}$ is the short-circuit current density, $J_{r}^{\text{rad}}$ is the dark radiative recombination current density, $k_\text{B}$, and $T$ is the temperature. Further, $R_\text{SRH}$ is the SRH recombination parameter, $e$ is the electronic charge and $d$ is the film thickness. The maximum output power density, $P_\text{max}$ (numerator of Eq. \ref{eqPV5}) is obtained by maximizing the power density, $P=JV$. The denominator represents the input power density ($P_\text{in}$) under the AM1.5G solar spectrum at 298 K, $I_\text{sun}(\omega)$.

The SRH parameter for a deep level defect with a single CTL in the band gap can be expressed as, 

\begin{equation}
R_\text{SRH}^{\mathrm{(1)}}=\frac{N_DC_nC_p(np-n_i^2)}{[C_n(n+n_1)+C_p(p+p_1)]}
\label{eqSR7}
\end{equation}
where $n_1$ ($p_1$) is the number of free electrons (holes) when the Fermi level equals the charge transition level energy. A detailed derivation can be found in our previous work.\cite{doi:10.1021/jacs.5c12981} We note that, according to the mass action law, we have  $n_i^2=n_0p_0=n_1p_1=N_cN_p\exp{(E_v-E_c/kT)}$, where $N_c$ and $N_p$ are the effective density of states for electrons and holes, respectively. The total carrier densities, $n$ and $p$, under operating conditions are expressed as $n=n_0+\Delta n$ and $p=p_0+\Delta p$, where $\Delta n = \Delta p$ are the photo-excited carrier densities. Finally, we note that the operating voltage $V$ equals the difference between the electron and hole quasi Fermi levels.\cite{kim2020upper} An approximate expression for the dependence of the photo-excited carrier density on the voltage reads,
\begin{equation}
	\Delta n(V)=\frac{1}{2}[-n_0-p_0+\sqrt{(n_0+p_0)^2-4n_0p_0(1-e^{\frac{eV}{k_\text{B}T}})]}
	\label{eqSR8}
\end{equation}

The SRH recombination parameter for a defect with possible transitions between charge states -1 and 0 and 0 and 1, reads

\begin{widetext}
\begin{equation}
R_\text{SRH}=\frac{N_D}{G}[\frac{C_n^0C_p^{-1}(np-n_2p_2)}{C_p^{-1}p+C_n^0n_2} + \frac{C_p^0C_n^1(np-n_1p_1)}{C_n^1n+C_p^0p_1}]
\label{eqSRH22}
\end{equation}
\begin{equation*}
G=[1+\frac{C_p^0p+C_n^1n_1}{C_n^1n+C_p^0p_1}+\frac{C_n^0n+C_p^{-1}p_2}{C_p^{-1}p+C_n^0n_2}]
\end{equation*}
\end{widetext}
Here $n_1$ and $p_1$ ($n_2$, $p_2$) are the free electron and hole concentrations when the Fermi level equals the 1/0 (0/-1) CTL. Expressions for $R_\text{SRH}$ in the case of defects with three and four CTLs in the band gap are given in the supplementary information.

In the expressions for $R_\text{SRH}$, $C_n^q$ and $C_p^q$ denote the electron and hole capture rates for the defect in charge state $q$, respectively. For photovoltaic operation both radiative and non-radiative carrier capture will contribute to the carrier loss. The total capture rate can thus be divided as

\begin{equation}
    C^q=C_{\mathrm{rad}}^q+C_{\mathrm{nonrad}}^q
\end{equation}

The radiative capture coefficient ($C_{\mathrm{rad}}^q$) can be calculated using Fermi's golden rule as,
\begin{equation}
	C_{\mathrm{rad}}^q=\frac{\Omega E_{\mathrm{em}}^{3}n_{\mathrm{ref}}\lvert\mu_{nm}\rvert^{2}}{3\pi\epsilon_0\hbar^4c^3},
	\label{eqrad}
\end{equation}
where $\Omega$ is the supercell volume, $E_{\mathrm{em}}$ is the emission energy, and $n_{\mathrm{ref}}$ is the refractive index of the material. $\epsilon_0$ and $\hbar$ are vacuum permittivity and reduced Planck's constant, respectively. Furthermore, $\mu_{nm}$ is the transition dipole moment between the initial and final electronic states,
\begin{equation}
    \mu_{nm}=\langle\psi_n\lvert\hat{r}\rvert\psi_m\rangle=\frac{i\hbar}{m_0}\frac{\langle\psi_n\lvert\hat{p}\rvert\psi_m\rangle}{\varepsilon_n-\varepsilon_m},
\end{equation}
where $\hat{r}$ and $\hat{p}$ are the dipole and momentum operators, respectively, and $m_0$ is the electron mass. Note that the expression for the radiative rate does not involve an overlap between vibrational states in the initial and final states as is commonly present in the expression for the photoluminescence spectrum. That is because we integrate over all emission frequencies leaving only the electronic part of the transition matrix element.

Using Fermi's golden rule from second order perturbation theory, the non-radiative capture coefficient ($C_{\mathrm{nonrad}}^q$) can be calculated as,\cite{alkauskas2014first} 

\begin{widetext}
\begin{equation}
	C=f\frac{2\pi}{\hbar}g\Omega W_{if}^2\sum_{m}w_m\sum_{n}\lvert\langle\chi_{im}\lvert\hat{Q}-Q_0\rvert\chi_{fn}\rangle\rvert^2 \delta(\Delta E_{if}+m\hbar\Omega_{im}-n\hbar\Omega_{fn})
	\label{eqCC2}
	\end{equation}
\end{widetext}
Here, $f$ is the Sommerfeld factor (discussed in supplementray information) and $g$ is the degeneracy factor of the defect. $W_{if}$ is the electron-phonon coupling matrix elements between the initial and final electronic state while $\sum_{m}w_m\sum_{n}\lvert\langle\chi_{im}\lvert\hat{Q}-Q_0\rvert\chi_{fn}\rangle\rvert^2$ accounts for the overlap between the initial and final vibronic states defined by the 1D potential energy diagrams. For the latter, we solve the 1D-Schrodinger equation to
 account for the anharmonicity of the defect potential energy surfaces, which is reported to have significant effect on the calculated carrier capture coefficients. The occupation factors, $w_m$, of the initial state vibronic levels are taken to follow a Boltzmann distribution. The $\delta$-functions (replaced by Gaussians of width 0.8$\hbar\Omega_f$) expresses the energy conservation: The energy difference in electronic energy between the initial and the final states in their equilibrium geometry, $\Delta E_{if}$, must be matched by the difference in vibrational energy.

From Eqs. (\ref{eqSR7}) and (\ref{eqSR8}), we can see that $R_{\mathrm{SRH}}$ is dependent on the operating voltage $V$. As such, Eq. (\ref{eqPV5}) must be solved self-consistently, which will involve calculating $\Delta n$ from Eq. (\ref{eqSR8}) and then evaluating the corresponding $R_{\mathrm{SRH}}$ value from Eq. (\ref{eqSR7}) or (\ref{eqSRH22}), depending on the number of CTLs of the defect. Next, in order to understand how $R_{\mathrm{SRH}}$ varies with $V$ for a particular defect, we define an effective capture coefficient, $C_{\mathrm{eff}}$, which shows how the capture rate varies with the voltage ($V$) and doping concentrations ($n_0$ and $p_0$) and also which of the individual capture coefficients becomes the rate limiting step under different conditions. $R_{\mathrm{SRH}}$ can be expressed in terms of $C_{\mathrm{eff}}$ as,
\begin{equation}
    R_{\mathrm{SRH}}=N_DC_{\mathrm{eff}}\Delta n
\end{equation}

The formalism presented above has been applied to all the 43 defects in t-Se considered in this work spanning defects with one to four CTLs in the band gap. 

\begin{figure*}[t!]
	\centering
	\includegraphics[scale=0.74]{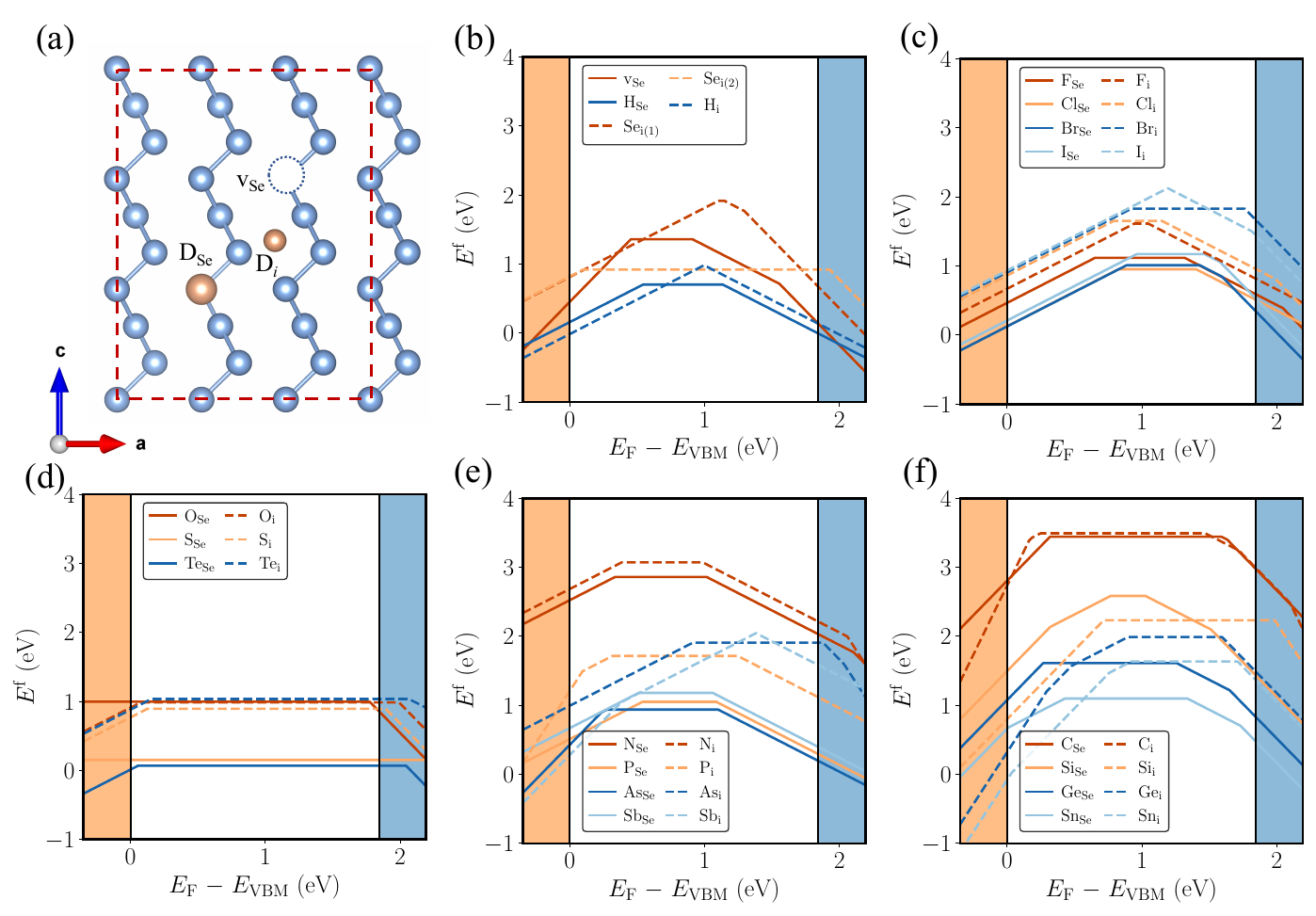}
	\caption{(a) Schematic diagram of possible defects (substitution (D$_\mathrm{Se}$), interstitial (D$_\mathrm{i}$) and vacancies (v$_\mathrm{Se}$) inside Selenium (Se) supercell. (b)-(f) Defect formation energies ($E^f$) as a function of Fermi level ($E_\mathrm{F}$) at dopant-poor, Se-rich growth condition for intrinsic defects, Hydrogen (H), Halogens, Chalcogens, Pnictogens and Carbon Group (tetragens), respectively. For each dopant (D),  substitution (D$_\mathrm{Se}$) and interstial (D$_\mathrm{i}$) are denoted by the same coloured solid and dashed lines respectively.  The charge state $q$ of the defect is shown by the slope of the curve. The $E_\mathrm{F}$ at the turning points gives the charge transition level (CTL) energies. The regions to the left of $E_\mathrm{VBM}$ (x=0 eV) and right of $E_\mathrm{CBM}$ (x=1.843 eV) are valence band below VBM and conduction band above CBM respectively. }
	\label{Fig1}
\end{figure*}

\begin{figure*}[t!]
	\centering
	\includegraphics[scale=0.86]{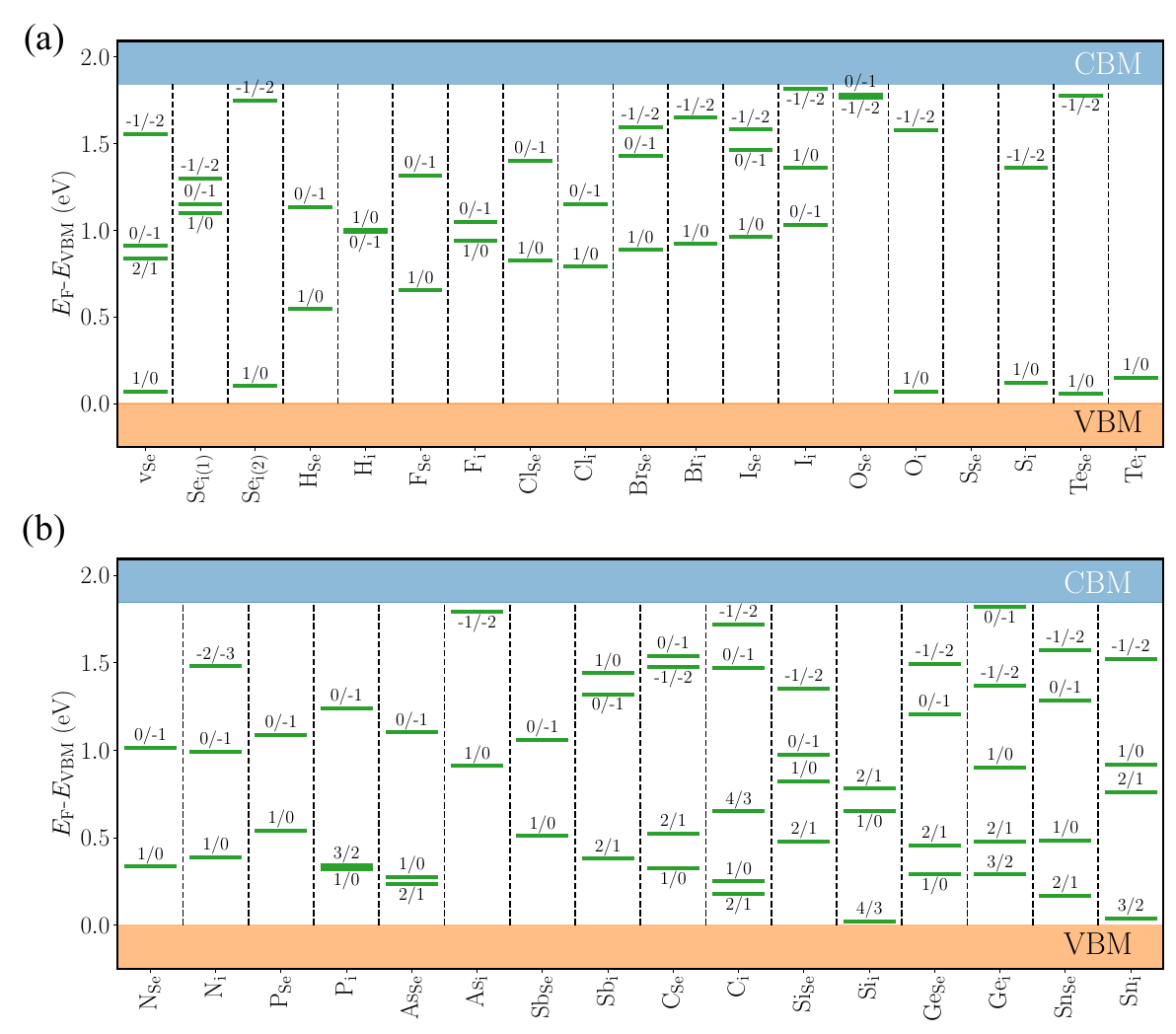}
	\caption{(a), (b) Charge transition levels for all the intrinsic and extrinsic considered for in t-Se. Here VBM and CBM stands for valence band maximum and conduction band minimum respectively. }
	\label{Fig2}
\end{figure*}

\begin{table*}[t!]
\caption{\label{tab:table_1}
Key parameters for five representative defects explored in this work. These include: The charge transition levels, configuration displacement $\Delta Q$, carrier capture coefficients for electron and hole capture in different charge states ($C_{n/p}^q$), semiclassical capture energy barrier, $\Delta E^\ddagger$, and the carrier capture coefficients. The total capture coefficient was obtained by adding the radiative and non-radiative contributions. Efficiency, $\eta$, and open-circuit voltage, $V_{\text{oc}}$, for low (10$^{14}$ cm$^{-3}$) and high (10$^{18}$ cm$^{-3}$) defect concentrations and represented as $\eta^\mathrm{1}$, $\eta^\mathrm{2}$ ($V_\mathrm{oc}^\mathrm{1}$, $V_\mathrm{oc}^\mathrm{2}$) respectively at experimentally observed p-type majority carrier concentration (1.37x10$^{16}$ cm$^{-3}$) (initial equilibrium Fermi level at 0.21 eV above VBM) [300 K and 500 nm film thickness.]}
\begin{ruledtabular}
\begin{tabular}{cccccccccccc}
Defect (D) & CTL & ${\Delta Q}$ & & & Capture coefficients & & & $\eta^\mathrm{1}$ & $\eta^\mathrm{2}$ & $V_\mathrm{oc}^\mathrm{1}$ & $V_\mathrm{oc}^\mathrm{2}$  \\[4.6pt] \cline{4-8}
& & & $C_{n/p}^q$ & $\Delta E^\ddagger$ & Non-rad. & Radiative & Total & (\%) & (\%) & (V) & (V) \\[4.6pt]
 & (eV) & (amu$^{1/2}$\AA) &  & (eV) & (cm$^3$s$^{-1}$) & (cm$^3$s$^{-1}$) & (cm$^3$s$^{-1}$) & & & & \\[4.6pt]
\hline\hline
\\[-8pt]
\multicolumn{12}{c}{\textbf{One CTL}}\\[2pt]
\hline\hline
\\[-8pt]
 As$_\mathrm{i}$ & (1/0)=0.91 & 22.38 & $C_n^1$ & $>$5 & 6.82x10$^{-12}$ & 5.86x10$^{-14}$ & 6.88x10$^{-12}$ & 23.25 & 20.92 & 1.59 & 1.55 \\[4.6pt]
 & & & $C_p^0$ & $>$5 & 6.69x10$^{-31}$ & 4.88x10$^{-15}$ & 4.88x10$^{-15}$ & & & &\\[4.6pt]
\hline\hline
\\[-8pt]
\multicolumn{12}{c}{\textbf{Two CTLs}}\\[2pt]
\hline\hline
\\[-8pt]
F$_\mathrm{Se}$ &(1/0)=0.65 & 10.23 & $C_n^1$ & 1.58 & 1.98x10$^{-12}$ & 4.93x10$^{-14}$ & 2.03x10$^{-14}$ & 23.14 & 19.06 & 1.59 & 1.32 \\[4.6pt]
& & & $C_p^0$ & $>$5 & 1.02x10$^{-8}$ & 5.06x10$^{-16}$ & 1.02x10$^{-8}$ & & & &  \\[4.6pt]
&(0/-1)=1.32 & 8.47 & $C_n^0$ & 0.26 & 2.18x10$^{-10}$ & 5.27x10$^{-16}$ & 2.18x10$^{-10}$ & & & &   \\[4.6pt]
& & & $C_p^{-1}$ & $>$5 & 5.43x10$^{-38}$ & 4.12x10$^{-14}$ & 4.12x10$^{-14}$ & & & & \\[4.6pt]
\hline\hline
\\[-8pt]
\multicolumn{12}{c}{\textbf{Three CTLs}}\\[2pt]
\hline\hline
\\[-8pt]
Br$_\mathrm{Se}$ &(1/0)=0.89 & 12.04 & $C_n^1$ & 4.29 & 1.96x10$^{-10}$ & 2.93x10$^{-14}$ & 1.96x10$^{-10}$ & 23.24 & 17.32 & 1.59 & 1.38 \\[4.6pt]
& & & $C_p^0$ & 2.25 & 3.42x10$^{-15}$ & 2.39x10$^{-15}$ & 5.81x10$^{-15}$ & & & &  \\[4.6pt]
&(0/-1)=1.43 & 19.77 & $C_n^0$ & $>$5 & 2.39x10$^{-7}$ & 2.94x10$^{-16}$ & 2.39x10$^{-7}$ & & & &   \\[4.6pt]
& & & $C_p^{-1}$ & $>$5 & 4.78x10$^{-71}$ & 9.95x10$^{-14}$ & 9.95x10$^{-14}$ & & & &  \\[4.6pt]
&(-1/-2)=1.59 & 20.86 & $C_n^{-1}$ & 0.50 & 7.15x10$^{-16}$ & 1.03x10$^{-17}$ & 7.25x10$^{-16}$ & & & &  \\[4.6pt]
& & & $C_p^{-2}$ & $>$5 & 6.61x10$^{-163}$ & 3.98x10$^{-13}$ & 3.98x10$^{-13}$ & & & &  \\[4.6pt]
\hline\hline
\\[-8pt]
\multicolumn{12}{c}{\textbf{Four CTLs}}\\[2pt]
\hline\hline
\\[-8pt]
v$_\mathrm{Se}$ & (2/1)=0.84 & 21.86 & $C_n^2$ & 0.19 & 2.26x10$^{-12}$ & 3.25x10$^{-13}$ & 2.58x10$^{-12}$ & 23.19 & 20.90 & 1.59 & 1.41  \\[4.6pt]
&  & & $C_p^1$ & 0.14 & 4.18x10$^{-19}$ & 9.56x10$^{-17}$ & 9.60x10$^{-17}$ & & & &  \\[4.6pt]
& (1/0)=0.07 & 6.54 & $C_n^1$ & $>$5 & 8.13x10$^{-127}$ & 8.86x10$^{-13}$ & 8.86x10$^{-13}$ & & & &  \\[4.6pt]
&  & & $C_p^0$ & 0.002 & 1.75x10$^{-6}$ & 5.62x10$^{-18}$ & 1.75x10$^{-6}$ & & & &  \\[4.6pt]
& (0/-1)=0.91 & 9.30 & $C_n^0$ & $>$5 & 1.29x10$^{-19}$ & 1.72x10$^{-15}$ & 1.72x10$^{-15}$ & & & & \\[4.6pt]
&  & & $C_p^{-1}$ & $>$5 & 2.41x10$^{-33}$ & 1.98x10$^{-13}$ & 1.98x10$^{-13}$ & & & & \\[4.6pt]
& (-1/-2)=1.56 & 8.18 & $C_n^{-1}$ & 0.003 & 2.81x10$^{-9}$ & 1.12x10$^{-18}$ & 2.81x10$^{-9}$ & & & & \\[4.6pt]
&  & & $C_p^{-2}$ & $>$5 & 2.77x10$^{-276}$ & 1.98x10$^{-12}$ & 1.98x10$^{-12}$ & & & &  \\[4.6pt]
Si$_\mathrm{Se}$ &(2/1)=0.48 & 8.90 & $C_n^2$ & $>$5 & 4.21x10$^{-15}$ & 4.75x10$^{-12}$ & 4.75x10$^{-12}$ & 23.24 & 18.72 & 1.59 & 1.30 \\[4.6pt]
& & & $C_p^1$ & 0.00001 & 1.13x10$^{-11}$ & 6.83x10$^{-18}$ & 1.13x10$^{-11}$ & & & &   \\[4.6pt]
&(1/0)=0.82 & 15.86 & $C_n^1$ & $>$5 & 2.69x10$^{-15}$ & 1.00x10$^{-12}$ & 1.00x10$^{-12}$ & & & &  \\[4.6pt]
& & & $C_p^0$ & $>$5 & 2.77x10$^{-15}$ & 3.28x10$^{-15}$ & 6.05x10$^{-15}$ & & & &  \\[4.6pt]
&(0/-1)=0.97 & 10.46 & $C_n^0$ & 4.81 & 1.09x10$^{-14}$ & 2.17x10$^{-15}$ & 1.31x10$^{-14}$ & & & &  \\[4.6pt]
& & & $C_p^{-1}$ & $>$5 & 1.80x10$^{-15}$ & 8.48x10$^{-13}$ & 8.50x10$^{-13}$ & & & &   \\[4.6pt]
&(-1/-2)=1.35 & 4.33 & $C_n^{-1}$ & 1.62 & 9.14x10$^{-22}$ & 8.59x10$^{-18}$ & 8.59x10$^{-18}$ & & & &  \\[4.6pt]
& & & $C_p^{-2}$ & $>$5 & 5.46x10$^{-59}$ & 4.56x10$^{-12}$ & 4.56x10$^{-12}$ & & & &   \\[4.6pt]
\end{tabular}
\end{ruledtabular}
\end{table*}

\begin{figure*}[t!]
	\centering
	\includegraphics[scale=1.0]{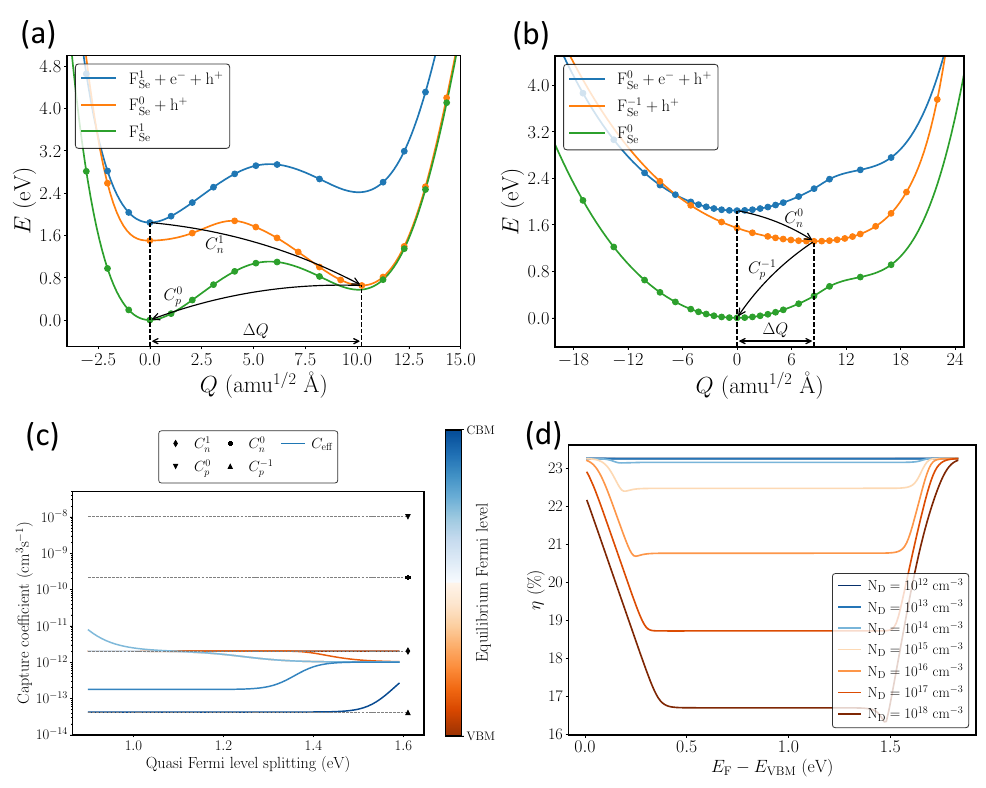}
	\caption{(a)-(b) One-dimensional configuration coordinate diagrams for the $1/0$, $0/-1$ charge state transitions of the F$_{\mathrm{Se}}$ defect in t-Se, respectively. Solid circles denote the data points calculated using HSE06 functional and solid lines are obtained by fitting them with quadratic spline functions accounting for anharmonicity. (c) Carrier capture coefficients  (including both radiative and nonradiative capture processes) at different charge states and associated effective capture coefficient ($C_{\mathrm{eff}}$) with respect to quasi-Fermi level splitting at different majority carrier types (equilibrium Fermi levels denoted by the color scale) at 300 K. (d) Photovoltaic device efficiency ($\mathrm{\eta}$) with respect to the equilibrium Fermi level position at different defect concentrations at 300 K and 500 nm device thickness.}
	\label{Fig5}
\end{figure*}

\begin{figure*}[t!]
	\centering
	\includegraphics[scale=0.89]{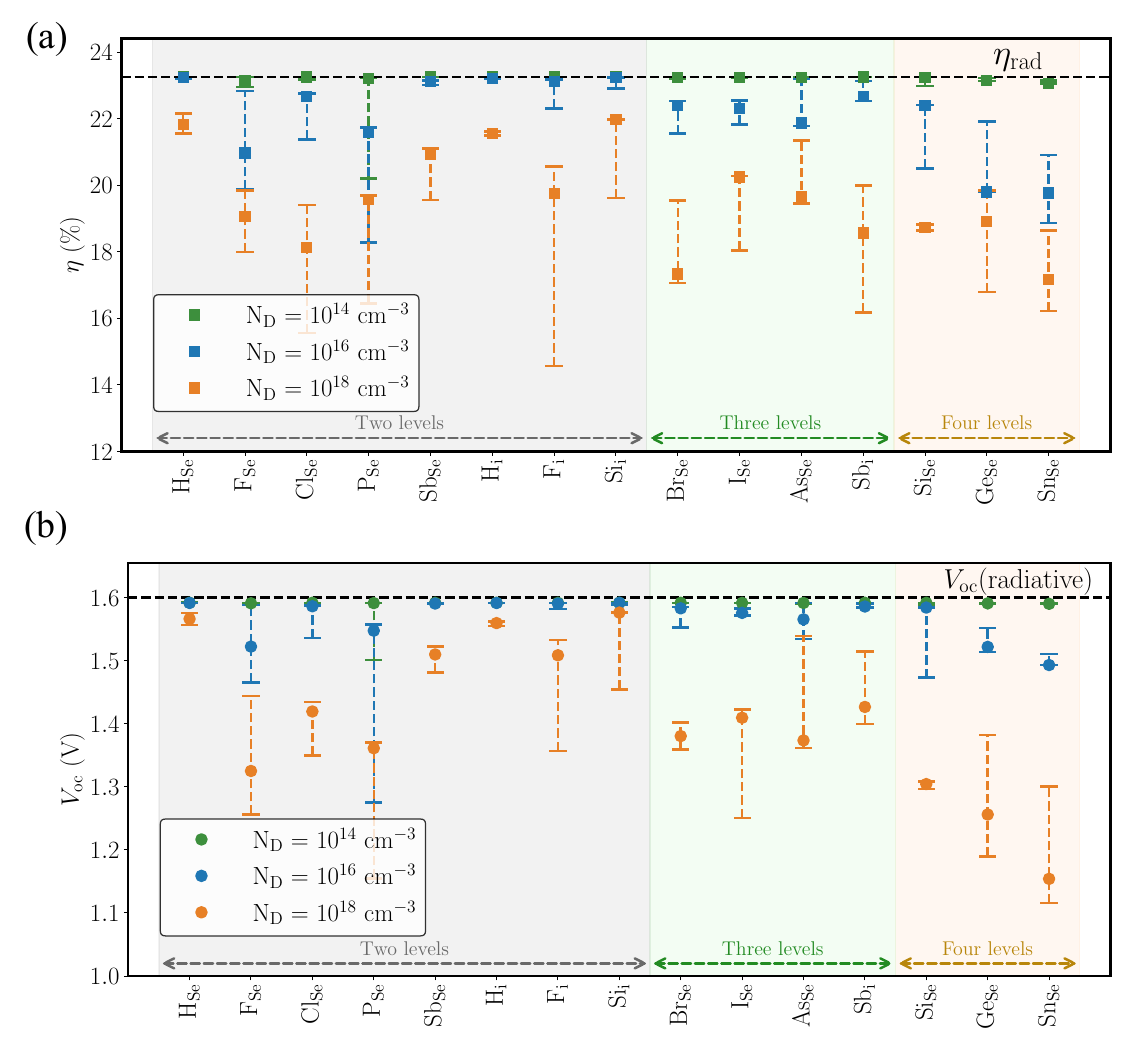}
	\caption{(a) Photovoltaic device efficiency $\eta$ for Se, accounting for the SRH recombination at considered defects, at different defect concentrations (10$^{14}$, 10$^{16}$, and 10$^{18}$ cm$^{-3}$) and at p-type majority carrier concentration (1.37x10$^{16}$ cm$^{-3}$) (initial equilibrium Fermi level at 0.21 eV above VBM). (b) Photovoltaic open-circuit voltage (V$_{\mathrm{oc}}$) for the same as above. Defects which can have defect concentration 10$^{14}$-10$^{18}$ cm$^{-3}$, based on defect formation energies and equilibrium Fermi level, at possible different chemical growth environments, are considered and categorized based on number of charge transition levels (CTLs) inside the band gap. PV device parameters are plotted at 300 K and 500 nm film thickness. Here we have considered a tolerance of 0.2 eV in the defect CTLs. The corresponding carrier capture coefficients are calculated by rigid shifting the CC curves at the considered levels. Associated device parameters are shown as error bars.}
	\label{Fig8}
    
\end{figure*}

\begin{figure*}[t!]
	\centering
	\includegraphics[scale=1.0]{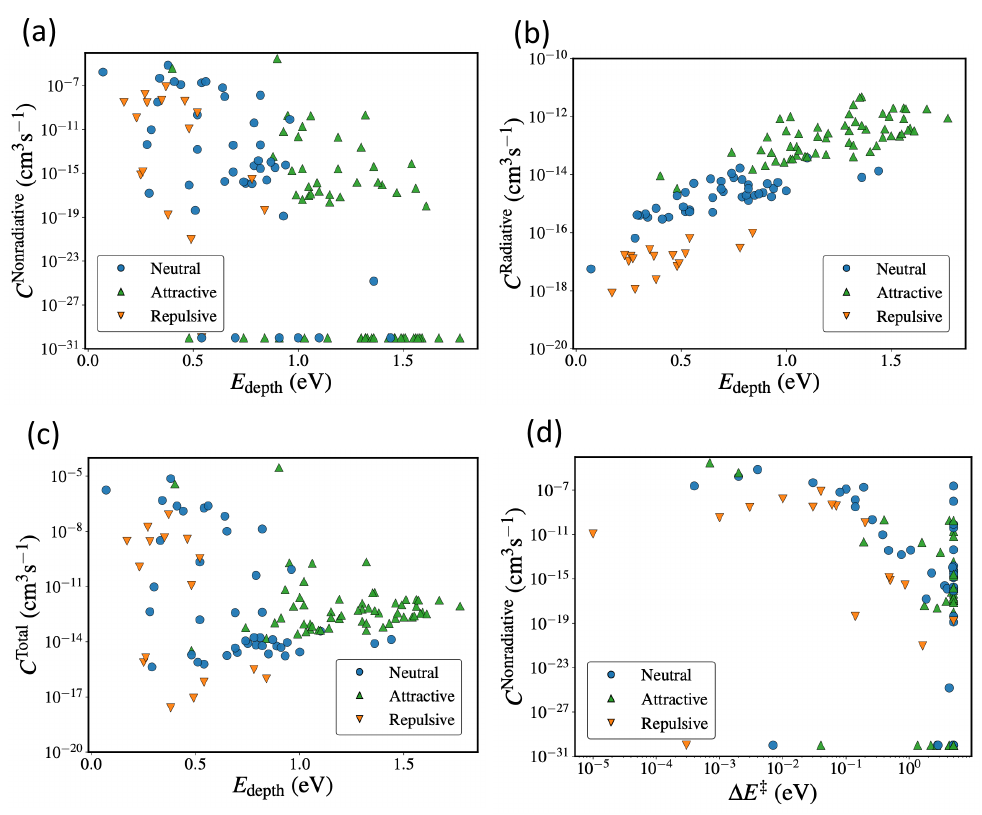}
	\caption{(a) Nonradiative capture coefficients for all the defects are plotted with respect to distance from the band edges (for electron (e$^-$) capture it is from CBM, and for hole (h$^+$) capture, it is from VBM). Data is distributed as as neutral, attractive and repulsive centres, e.g. D$^{-1}$ will be repulsive for e$^-$, but attractive for h$^+$. All the value below 10$^{-30}$ is assumed as negligible and plotted as 10$^{-30}$. (b) Same as (a) with radiative capture coefficients along y-axis. (c) Same as (a) with total capture coefficients along y-axis. (d) Nonradiative capture coefficients vs. semi-classical energy barrier ($\Delta E^\ddagger$) derived from the 1D-configuration coordinate diagrams of the considered defects. }
	\label{Fig9}
    
\end{figure*}

\section*{III. Results and Discussion}

To quantitatively link atomic-scale defect physics with photovoltaic device performance in trigonal selenium, we combine hybrid-DFT defect energetics, anharmonic configuration-coordinate carrier-capture calculations, and a fully self-consistent multi-level Shockley–Read–Hall (SRH) recombination framework. We construct a comprehensive defect library spanning intrinsic vacancies and interstitials together with chemically diverse extrinsic impurities from the halogen, chalcogen, pnictogen, and carbon groups, and determine their charge-state formation energies and thermodynamic CTLs across the full band gap and relevant chemical-potential conditions. These choices reflect both chemical feasibility in chalcogenide growth environments and experimentally relevant doping pathways.\cite{Moustafa2024, Nielsen2022, kavanagh2025intrinsic} Possible defect complexes or multi-defect clusters were not considered, allowing us to focus on the dominant isolated point defects while keeping the configurational space computationally tractable. All defect geometries and configuration-coordinate pathways were relaxed using the HSE06\cite{heyd2003hybrid} functional. Radiative and nonradiative carrier-capture coefficients are then evaluated from first principles along relaxed one-dimensional configuration-coordinate pathways and incorporated into the self-consistent SRH formalism to obtain the effective capture coefficient under operating quasi-Fermi-level splitting. This integrated approach enables a direct, device-relevant assessment of how individual point defects influence carrier lifetimes, open-circuit voltage $V_\mathrm{oc}$, and efficiency $\eta$.

\subsection*{A. Defect Formation Energies and Charge Transition Levels}

We first establish the thermodynamic defect landscape of trigonal selenium under representative growth conditions. Figure~\ref{Fig1} summarizes the formation energies of intrinsic point defects together with a chemically diverse set of extrinsic impurities as a function of the Fermi level $E_\mathrm{F}$, referenced to the valence-band maximum (VBM) under Se-rich, dopant-poor conditions. The corresponding chemical potential limits and competing secondary phases are provided in SI Table~\ref{tab:tables1}. 

The formation energies in Figure~\ref{Fig1}(b–f) reveal three distinct classes of behavior that organize the defect landscape in $t$-Se. First, intrinsic point defects occupy a deep, multicharge-state regime. Both the selenium vacancy ($v_\mathrm{Se}$) and the selenium interstitial (Se$_{i(1)}$) introduce localized states with multiple CTLs inside the band gap (Tables~\ref{tab:table_1} and \ref{tab:tables2}). Among these, $v_\mathrm{Se}$ is thermodynamically favored over most of the Fermi-level range and remains the dominant intrinsic defect under experimentally relevant p-type conditions, where the equilibrium Fermi level lies $\sim$0.2~eV above the VBM. Only within a narrow Fermi-level window near $E_\mathrm{F}\approx0.3$–0.4~eV do $v_\mathrm{Se}$ and Se$_i$ exhibit comparable formation energies.

Second, clear electronic trends emerge among extrinsic substitutions. Isovalent chalcogen impurities (O$_\mathrm{Se}$, S$_\mathrm{Se}$, Te$_\mathrm{Se}$) remain largely neutral across the gap and generate only shallow CTLs near the band edges, consistent with their chemical similarity to Se and suggesting minimal recombination activity. In contrast, most non-isovalent impurities—including hydrogen, halogens, carbon-group elements, and several pnictogens—introduce deep CTLs and frequently exhibit amphoteric behavior, allowing donor- and acceptor-like charge transitions depending on the Fermi-level position. Carbon-group substitutions and selected pnictogens further become energetically favored as $E_\mathrm{F}$ approaches the band edges, reflecting donor-like tendencies under p-type conditions and acceptor-like tendencies under n-type conditions.

Third, dopant size systematically governs site stability. Small impurities (e.g., H, F, O, N, C) show comparable formation energies for interstitial and substitutional configurations, whereas larger atoms preferentially substitute on the Se site, often lowering the substitutional formation energy by more than 1~eV compared to the interstitial formation energy. This trend reflects the limited interstitial free volume imposed by the quasi-one-dimensional chain structure of $t$-Se. Exceptions such as H$_i$ and Se$_i$ can stabilize through split-bond or bridge geometries that reduce local strain. Consistently, substitutional formation energies within a given chemical family decrease as the atomic radius approaches that of Se. For example, Te$_\mathrm{Se}$ and S$_\mathrm{Se}$ are nearly 1~eV more stable than O$_\mathrm{Se}$—highlighting lattice-strain matching as a key determinant of dopant incorporation.

The equilibrium concentrations derived from the calculated formation energies are summarized in Figures~\ref{FigS1} and \ref{FigS2} for representative p-type, intrinsic, and n-type Fermi-level positions and across relevant chemical-potential limits. The concentrations shown correspond to 300~K for reference; however, in practice defect populations are typically established at the growth temperature ($\approx$ 500 K for t-Se) and subsequently freeze in during cooling. Variations in growth conditions primarily shift defect formation energies by nearly constant offsets, leaving the relative concentration of different charge states of a given defect largely unchanged while substantially modifying the absolute concentrations.

Across these limits, many substitutional impurities can reach concentrations spanning $\sim10^{14}$–$10^{18}$~cm$^{-3}$ under experimentally accessible synthesis conditions, with halogens generally favored in p-type environments and pnictogens becoming more prominent toward n-type conditions. In contrast, only a subset of interstitial impurities (notably H-, F-, O-, Sb-, and group-14 interstitials) achieve comparable concentrations, typically under p-type growth. 

These results indicate that a chemically diverse population of impurities may be incorporated at technologically relevant densities if not carefully controlled, motivating a systematic evaluation of their potential impact on carrier recombination and photovoltaic performance.

Having established realistic concentration ranges, we next examine the electronic impact of defects through the positions of their CTLs within the band gap. Each CTL represents a thermodynamic transition between defect charge states and can therefore act as a carrier trapping site. In this work, defects are classified as deep when a CTL lies more than 0.15~eV from a band edge, where SRH recombination becomes possible, and as shallow when the CTL resides within 0.15~eV of a band edge, in which case the defect primarily exchanges carriers with a single band and behaves as a dopant. The 0.15 eV threshold is chosen based on the radiative open-circuit voltage loss, such that levels closer than this to a band edge are not expected to contribute significantly to non-radiative recombination under relevant photovoltaic operating conditions. From Figure~\ref{Fig1}(b–f), it can be seen that several defects, such as C$_{\mathrm{Se}}$, Ge$_{\mathrm{Se}}$, and V$_{\mathrm{Se}}$, exhibit a $2/0$ CTL. However, these and similar multi-charge transitions are not relevant for nonradiative recombination, as they would require the capture of two or more electrons or holes at once, which is not possible.\cite{bagraev1988mechanism} Therefore, only CTLs corresponding to single-carrier capture processes are considered, and the relevant extracted CTLs for all defects are summarized in Figure~\ref{Fig2}. 

Across the defect library, impurities introduce between zero and four deep levels inside the band gap. We therefore denote defects with $n$ deep CTLs as CTL-$n$, providing a compact classification that links chemical identity to recombination potential. Isovalent chalcogen substitutions (O$_\mathrm{Se}$, S$_\mathrm{Se}$, Te$_\mathrm{Se}$) remain shallow (CTL-0) and are expected to be electronically benign. In contrast, many non-isovalent substitutions exhibit amphoteric behavior, meaning that donor-like and acceptor-like charge transitions can both occur depending on the Fermi-level position; such defects typically fall into the CTL-2 category. At the opposite extreme, carbon-group substitutions (C, Si, Ge, Sn) support up to five charge states and form CTL-4 centers with multiple deep levels distributed across the gap.

Importantly, the mere presence of several CTLs does not guarantee efficient recombination. As shown below, capture kinetics depend sensitively on both the energetic depth of the levels and the lattice relaxation accompanying charge transfer, which can strongly suppress individual transitions. Nevertheless, multi-level defects possess a greater intrinsic capacity for carrier exchange with the bands and therefore represent the most likely candidates for detrimental SRH recombination. In the following sections, representative defects from each CTL class (CTL-1 through CTL-4) are analyzed to elucidate their distinct recombination mechanisms. The calculated defect energetics and carrier-capture coefficients discussed below are summarized in Table~\ref{tab:table_1} and Table~\ref{tab:tables2}.

\subsection*{B. Intrinsic Defects}\label{intrinsic}

We first examine the intrinsic point defects—the selenium vacancy ($v_\mathrm{Se}$) and selenium interstitial (Se$_i$), which establish the baseline recombination physics of $t$-Se. Under p-type growth conditions these defects exhibit comparable formation energies and multiple CTLs, with $v_\mathrm{Se}$ supporting four CTLs inside the band gap (Table~\ref{tab:table_1}). Carrier-capture analysis shows that the CTL closest to the VBM enables fast hole capture $C_p^0$ ($1.7\times10^{-6}$~cm$^{3}$s$^{-1}$), while the CTL nearest the conduction band minimum (CBM) supports comparatively slower electron capture $C_n^{-1}$ ($2.81\times10^{-9}$~cm$^{3}$s$^{-1}$). Despite similar semiclassical barriers, larger distance from the band edge and the repulsive character of electron capture reduce the corresponding rate by several orders of magnitude relative to the hole capture at neutral charge state.  Capture at the remaining CTLs is strongly suppressed, leading to an effective capture coefficient $C_\mathrm{eff}$ in the range $10^{-12}$–$10^{-15}$~cm$^{3}$s$^{-1}$ across synthesis and operating conditions (Figure~\ref{Fig3}). Even for defect concentrations as high as $10^{15}$–$10^{16}$~cm$^{-3}$, the resulting SRH lifetimes remain in the millisecond regime, comparable to radiative Shockley–Queisser limits, while thermodynamic formation energies further constrain realistic $v_\mathrm{Se}$ concentrations to $\lesssim10^{14}$~cm$^{-3}$. Thus, although some of the transitions in $v_\mathrm{Se}$ are microscopically active, the overall macroscopic impact of $v_{\mathrm{Se}}$ on device performance is negligible.

The Se$_\mathrm{i}$ exhibits an even more benign recombination character. All three CTLs lie near midgap and are accompanied by large structural relaxations and strongly anharmonic configuration-coordinate potentials (Figure~\ref{FigS3}), producing high capture barriers and uniformly low non-radiative capture coefficients. Radiative capture therefore dominates, and the resulting effective capture rates remain limited by slow minority-carrier processes under both p-type and n-type conditions. Consequently, Se$_\mathrm{i}$ does not introduce significant SRH recombination within experimentally relevant concentration ranges.

Taken together, these results demonstrate that intrinsic point defects in trigonal selenium cannot account for the experimentally observed open-circuit-voltage deficit or efficiency bottleneck. This result establishes an essential baseline for the material and motivates a systematic investigation of extrinsic impurities and general recombination trends, which we address next.

\subsection*{C. Extrinsic Defects}\label{extrinsic}

Having established that intrinsic defects are unlikely to limit device performance, we next examine extrinsic impurities, which introduce a broader spectrum of CTLs and therefore represent the most plausible source of defect-assisted recombination in $t$-Se. To elucidate the governing physics, we analyze representative defects from each CTL class (CTL-1 through CTL-4), focusing on how the number, energetic placement, and lattice relaxation of defect levels determine carrier-capture kinetics. 

A central feature of this analysis is the explicit treatment of multi-level SRH recombination. Unlike the conventional single-level SRH picture—where recombination is controlled by a single midgap “killer” state—defects in wide-gap semiconductors frequently possess multiple CTLs, and the rate-limiting transition can depend sensitively on doping, quasi-Fermi-level splitting, and competing capture pathways. By solving the full multi-level SRH rate equations self-consistently, our framework captures this interplay and enables a direct connection between microscopic defect properties and macroscopic photovoltaic performance. 

In the following subsections, archetypal defects from each CTL category are used to extract general recombination trends and to determine whether any realistic extrinsic impurity can account for the experimentally observed performance limitations of trigonal selenium.

\subsubsection*{I. CTL-1 defects}

As$_\mathrm{i}$ is the only CTL-1 deep defect identified in this study and therefore provides a minimal case for defect-assisted recombination. Its $1/0$ transition lies near midgap (0.91 eV above the VBM), and the corresponding configuration-coordinate potential is strongly anharmonic and accompanied by a large semiclassical capture barrier (Figure~\ref{Fig4}). Consequently, both electron and hole capture coefficients remain very small ($<10^{-12}$~cm$^{3}$s$^{-1}$; Table~\ref{tab:table_1}). 

Despite its midgap energetic position, the slow capture kinetics render As$_\mathrm{i}$ electronically benign: even at unrealistically high defect concentrations of $10^{16}$–$10^{18}$~cm$^{-3}$, the resulting carrier lifetimes remain in the microsecond regime and produce negligible changes in photovoltaic performance. This example highlights a key principle for wide-gap absorbers—midgap position alone does not determine recombination activity; lattice relaxation and capture barriers are equally decisive.

\subsubsection*{II. CTL-2 defects}

The majority of extrinsic impurities in $t$-Se fall into the amphoteric CTL-2 class, characterized by a donor-like $1/0$ transition near the VBM and an acceptor-like $0/-1$ transition near the CBM. We use the halogen substitution F$_\mathrm{Se}$ as a representative example. Its CTLs lie 0.65~eV above the VBM and 0.52~eV below the CBM, respectively (Table~\ref{tab:table_1}). Configuration-coordinate analysis reveals strongly asymmetric capture kinetics: hole capture at the $1/0$ level is fast ($\sim10^{-8}$~cm$^{3}$s$^{-1}$), whereas the corresponding electron capture is slow ($\sim10^{-12}$~cm$^{3}$s$^{-1}$); conversely, the $0/-1$ transition supports moderately fast electron capture but negligible nonradiative hole capture due to a large barrier, leaving radiative processes dominant. 

Despite the presence of individual fast capture channels, the effective recombination rate $C_\mathrm{eff}$ remains limited by the slowest transition and depends sensitively on the operating Fermi-level conditions. Under p-type synthesis, minority-electron capture at the $1/0$ level controls recombination, while under n-type conditions the defect becomes trapped in the $-1$ charge state and hole capture at the $0/-1$ transition becomes rate limiting (Figure~\ref{Fig5}). Consequently, CTL-2 defects such as F$_\mathrm{Se}$ produce negligible efficiency loss across realistic concentrations ($\lesssim10^{16}$~cm$^{-3}$), and only unrealistically high densities approaching $10^{18}$~cm$^{-3}$ lead to modest lifetime reduction. Other CTL-2 impurities exhibit qualitatively similar behavior (Figures~\ref{FigS4}–\ref{FigS13}), indicating that amphoteric two-level defects are generally benign in trigonal selenium.

\subsubsection*{III. CTL-3 defects}

Defects with three deep CTLs represent an intermediate regime between largely benign two-level centers and the more complex multi-level defects discussed below. As a representative example, the halogen substitution Br$_\mathrm{Se}$ exhibits a midgap $1/0$ transition together with $0/-1$ and $-1/-2$ levels located 0.41 and 0.25 eV below the CBM, respectively. As expected, enhanced lattice relaxation ($\Delta Q$) relative to F$_\mathrm{Se}$ is observed. The larger structural reorganization shifts the crossing point of the configuration–coordinate curves closer to the equilibrium geometry, thereby reducing the semiclassical barrier $\Delta E^\ddagger$. Together with the relatively closer defect levels, this results in comparatively large electron-capture coefficients for the near-CBM transitions ($\sim10^{-7}$–$10^{-10}$~cm$^{3}$s$^{-1}$) compared to F$_\mathrm{Se}$, while hole capture remains weak due to the larger energetic separation from the VBM. Under p-type operating conditions, recombination is limited by minority-electron capture at the $1/0$ level, yielding effective capture coefficients in the $10^{-9}$–$10^{-10}$~cm$^{3}$s$^{-1}$ range (see Figure \ref{Fig6}(d)) and carrier lifetimes of order 10–100 ns at very high defect densities ($10^{17}$–$10^{18}$~cm$^{-3}$). 

Although such CTL-3 defects can begin to influence device performance at extreme concentrations, no experimental evidence currently indicates incorporation at these levels. Other CTL-3 centers (e.g., I$_\mathrm{Se}$, As$_\mathrm{Se}$, Sb$_i$) show qualitatively similar behavior, reinforcing the broader conclusion that even moderately complex multi-level defects remain largely benign within realistic concentration ranges. This establishes CTL-4 defects as the primary candidates for potentially detrimental recombination in $t$-Se.

\subsubsection*{IV. CTL-4 defects}

Defects with four deep CTLs constitute the most complex recombination centers in $t$-Se and therefore represent the primary candidates for performance-limiting behavior. Among the group-14 substitutions (C, Si, Ge, Sn)$_\mathrm{Se}$, C$_\mathrm{Se}$ exhibits prohibitively high formation energy and is unlikely to occur in appreciable concentrations. We therefore focus on Si$_\mathrm{Se}$ as an experimentally relevant prototype, while using Ge$_\mathrm{Se}$ and Sn$_\mathrm{Se}$ to establish general trends.

For Si$_\mathrm{Se}$, the outermost transitions ($2/1$ near the VBM and $-1/-2$ near the CBM) lie roughly 0.5 eV from the band edges, whereas the remaining CTLs reside close to midgap. Because of the large band gap of $t$-Se ($\sim$1.85 eV), these midgap levels are nearly 0.9 eV away from both band edges, leading to large transition energies and correspondingly high semiclassical barriers. As a result, these CTLs exhibit negligible capture activity. Although hole capture ($C_p^1$) at the $2/1$ transition shows the largest individual capture coefficient ($\sim10^{-11}$~cm$^{3}$s$^{-1}$), weak electron capture ($C_n^2$) at the complementary transition strongly suppresses the overall recombination rate. Consequently, the effective capture coefficient remains low across operating conditions (see \ref{Fig7}(e)), implying that unrealistically high defect concentrations would be required for Si$_\mathrm{Se}$ to measurably impact device efficiency. Ge$_\mathrm{Se}$ follows a similar pattern: despite several moderately fast individual capture channels, the rate-limiting transitions governing $C_\mathrm{eff}$ remain slow, yielding negligible recombination within realistic concentration ranges.

In contrast, Sn$_\mathrm{Se}$ exhibits a qualitatively different behavior. Multiple near-CBM transition levels combine relatively large non-radiative electron capture coefficients ($\sim10^{-7}$–$10^{-8}$~cm$^{3}$s$^{-1}$) with accessible charge-state populations under p-type operating conditions, allowing $C_\mathrm{eff}$ to approach the fastest capture channels at high quasi-Fermi-level splitting. As a result, Sn$_\mathrm{Se}$ can begin to reduce photovoltaic efficiency at defect concentrations approaching $\sim10^{17}$~cm$^{-3}$, identifying Sn-like multi-level impurities as the worst-case recombination centers in $t$-Se. Importantly, even this limiting scenario requires concentrations near the upper bound of experimental plausibility, reinforcing the broader conclusion that bulk point defects alone are unlikely to account for the observed efficiency deficit in selenium photovoltaics.

\subsection*{D. Device-Level Impact of Point Defects in Trigonal Selenium}

To connect microscopic defect physics with macroscopic photovoltaic performance, we incorporated the calculated defect energetics and carrier-capture coefficients into the self-consistent SRH device model of Eq.~\ref{eqPV5} to evaluate the defect-limited efficiency $\eta$ and open-circuit voltage $V_{\mathrm{oc}}$. Figure~\ref{Fig8} summarizes the impact of representative defects spanning CTL-1 to CTL-4 across concentrations from $10^{14}$ to $10^{18}$~cm$^{-3}$ under p-type synthesis conditions. 

Across the experimentally relevant concentration range ($10^{14}$–$10^{16}$~cm$^{-3}$), all defects produce negligible changes in $\eta$ and $V_{\mathrm{oc}}$, and this conclusion remains robust even when allowing for $\pm0.2$~eV uncertainty in calculated CTL positions. Noticeable performance degradation emerges only at extreme concentrations approaching $10^{18}$~cm$^{-3}$, where multi-level CTL-4 defects—most prominently Sn$_{\mathrm{Se}}$—can reduce $V_{\mathrm{oc}}$ toward $\sim$1.1~V, comparable to the best experimental values. Importantly, such defect densities have not been reported in trigonal selenium, while the most commonly observed dopants (e.g., halogens) remain largely benign even at high concentrations. Calculations for intrinsic and n-type conditions (Figures~S20–S21) lead to the same qualitative conclusion.

These results establish a central finding of this work: bulk SRH recombination mediated by realistic concentrations of point defects is insufficient to explain the long-standing efficiency bottleneck and $V_{\mathrm{oc}}$ deficit in selenium photovoltaics. Instead, the dominant recombination losses are likely governed by extrinsic factors such as extended defects, interfaces, or microstructural disorder, consistent with emerging experimental evidence.\cite{Chen2024, shen2025oxygen, wen2026illumination}

An additional insight emerging from the device-level analysis is that the presence of multiple CTLs does not automatically imply strong recombination. Although multi-level defects provide several possible carrier-capture pathways, the overall SRH rate remains governed by the slowest transition under the relevant quasi-Fermi-level splitting. In many cases, charge trapping in one state can instead suppress recombination by limiting access to faster channels, consistent with the low effective capture coefficients obtained for most CTL-2 and CTL-3 defects.

Beyond ground-state multiphonon capture, alternative recombination pathways may arise from defect excited states or metastable geometries. For wide-band-gap semiconductors, excited-state–mediated trapping has been shown to enhance carrier capture by orders of magnitude in specific cases (e.g., GaN),\cite{alkauskas2016role} suggesting a possible additional recombination channel not captured by ground-state configuration-coordinate analysis. Whether similar mechanisms operate in trigonal selenium remains an open question and represents an important direction for future theoretical and experimental investigation.

\subsection*{E. Trends in Defect-Assisted Recombination}

To extract general principles governing defect-assisted recombination, we analyze the calculated radiative and nonradiative capture coefficients for all defects as a function of defect-level depth from the relevant band edge ($E_\mathrm{depth}$) and the associated lattice relaxation. The resulting dataset—comprising 116 carrier-capture processes across 22 distinct defects evaluated consistently at the hybrid-DFT level with anharmonic configuration-coordinate treatment—provides, to our knowledge, the first statistically meaningful mapping between microscopic defect energetics and quantitative recombination kinetics within a single photovoltaic absorber. In Figure \ref{Fig9}(a-c) we plot the capture coefficients ($C_{\mathrm{nonrad}}^q$, $C_{\mathrm{rad}}^q$, $C_{\mathrm{total}}^q$) of all the trap centers with respect to $E_\mathrm{depth}$, defined as the distance from the CBM for electron ($e^-$) capture and from the VBM for hole ($h^+$) capture, in order to explore possible correlations. They are categorized in three types based on their interaction with the carrier, i.e. attractive, neutral and repulsive. For example a D$^{-1}$ center will be repulsive for $e^-$, but attractive for $h^+$ and all the neutral centers are charge 0 (D$^{0}$).

Several robust correlations emerge. First, non-radiative capture decreases rapidly as defect levels move away from the band edges: for $E_\mathrm{depth} \gtrsim 0.7$–0.8~eV, $C_{\mathrm{nonrad}}^q$ falls below $\sim10^{-10}$~cm$^{3}$s$^{-1}$, rendering such transitions largely irrelevant for photovoltaic recombination. Second, radiative capture exhibits the opposite trend, increasing systematically with $E_\mathrm{depth}$ and showing order-of-magnitude enhancement for Coulomb-attractive centers due to the Sommerfeld factor. As a consequence, recombination via deep levels in $t$-Se is often governed by the radiative rather than the non-radiative mechanism. This behavior is consistent with general defect thermodynamics: donor-like (positively charged) states tend to lie closer to the VBM while acceptor-like (negatively charged) states lie closer to the CBM, meaning that repulsive centers often occur nearer the band edges while Coulomb-attractive centers are typically located deeper in the gap. When both contributions are combined, the total capture coefficient $C_{\mathrm{total}}^q$ does not show a clear monotonic dependence on $E_\mathrm{depth}$ (Figure~\ref{Fig9}c). Instead, the regime with $E_\mathrm{depth} \lesssim 0.7$~eV is dominated by nonradiative capture, whereas for deeper levels ($E_\mathrm{depth} \gtrsim 0.7$~eV) the radiative mechanism becomes the primary contribution. Precisely because nonradiative capture dominates for shallow levels, the variations in $C_{\mathrm{total}}^q$ are largest in this regime.

In addition to defect energetic position and charge states (repulsive vs attractive), the lattice reorganization upon carrier capture is another important parameter. In order to see how the defect local geometry relaxation affects the capture rates we plot $C_{\mathrm{nonrad}}^q$ vs $E_\mathrm{depth}$ with the (mass weighted) configuration-coordinate displacement, $\Delta Q$, as color bar in Figure \ref{FigS22}. Notice that most defect centers with large $C_{\mathrm{nonrad}}^q$ exhibit relatively small structural relaxations ($\Delta Q$), which result in smaller semiclassical barriers $\Delta E^\ddagger$ and therefore higher capture rates. Interestingly, interstitial defects in $t$-Se tend to exhibit larger structural relaxations than substitutional defects, resulting in larger $\Delta Q$ values and consequently lower nonradiative capture coefficients despite often introducing deep states in the gap.

Overall, defects with small $E_\mathrm{depth}$ and small $\Delta Q$ tend to exhibit the largest nonradiative capture coefficients and therefore represent potentially detrimental recombination centers. We emphasize, however, that band-to-band SRH recombination requires two sequential capture events (electron and hole), so a high capture rate for one carrier does not necessarily imply a large overall recombination rate if the complementary capture process is slow. 
 In Figure \ref{Fig9}d we plot the $C_{\mathrm{nonrad}}^q$ with respect to semi-classical energy barrier $\Delta E^\ddagger$, which shows a cut-off close to 1 eV, after which $C_{\mathrm{nonrad}}^q$ starts to reduce rapidly. Although a limited number of outliers are present, which is an expected consequence of the complex interplay between defect energetics, lattice relaxation, and Coulomb interactions, the overall trends remain robust and statistically well defined across the full dataset.

Taken together, these results establish quantitative descriptors for defect tolerance in wide-band-gap absorbers that could not be inferred from defect energetics alone. Detrimental recombination is most likely for defects combining $<$ 0.7 eV energetic depth with minimal lattice relaxation, whereas deeper or more strongly relaxing defects are intrinsically self-limiting. While explicit first-principles capture calculations remain essential for quantitative accuracy, the correlations revealed here provide simpler and transferable principles for identifying defect-tolerant photovoltaic materials in general. The establishment of such principles was made possible by the comprehensive, large and heterogeneous, high-fidelity defect-physics datasets developed in this work.

\section{Conclusions}

In this work, we have developed a comprehensive first-principles framework to assess defect tolerance in trigonal selenium by combining hybrid-DFT defect energetics, anharmonic configuration-coordinate calculations of carrier capture, and a full multi-level SRH formalism directly linked to photovoltaic device performance. This approach establishes a quantitative connection between atomistic defect physics and macroscopic recombination losses, going beyond conventional energetics-based or single-level descriptions.

Our results yield a coherent and quantitative picture of defect physics in trigonal selenium. Intrinsic point defects are electronically benign, and even chemically diverse extrinsic impurities, including multi-level deep centers, produce negligible SRH recombination at experimentally realistic concentrations. More fundamentally, the effective recombination is limited by large lattice relaxation during carrier capture, Coulomb interactions, and rate-limiting sequential carrier capture. Only a narrow subset of multi-level impurities, most notably Sn-like centers, induces appreciable recombination losses, and only at concentrations approaching the upper bounds of experimental plausibility. Consequently, SRH recombination mediated by realistic concentrations of bulk point defects cannot account for the observed open-circuit-voltage deficit and efficiency limitations in selenium photovoltaics.

These results resolve a long-standing question in selenium photovoltaics: bulk point defects alone cannot account for the observed $V_{\mathrm{oc}}$ deficit or efficiency bottleneck in state-of-the-art devices. Instead, it redirects attention toward extrinsic and microstructural recombination pathways, such as interfaces, grain boundaries, and extended defects, as the dominant sources of nonradiative loss, an interpretation consistent with the growing experimental focus on extended defects and interfaces, and subsequent emerging experimental evidence of surface- and boundary-mediated recombination.\cite{Chen2024, shen2025oxygen, wen2026illumination} In this context, stringent control of impurity incorporation and developing targeted passivation strategies for interfaces and structural inhomogeneities will be critical for further improving device performance of t-Se.

The implications of our work carry both immediate technological relevance for t-Se photovoltaics as well as broader conceptual significance. In particular, it provides a transferable and quantitatively predictive framework for diagnosing defect tolerance in wide-band-gap photovoltaic materials. By demonstrating that recombination activity is governed by an intricate interplay of defect level depth, lattice relaxation during charge-state transitions, Coulomb interactions, equilibrium carrier density at synthesis and operating conditions, rather than defect depth alone, we take an important step towards the establishment of design principles for intrinsically defect-tolerant semiconductors. A key observation emerging from the work is that across a large chemically diverse defect set in t-Se, we find that at least one step in the sequential electron–hole capture process is strongly rate limited, most often because nonradiative capture becomes inefficient when the corresponding transition involves a large energy separation from the nearest band edge. Given the vast amount and chemical diversity of defects considered in this work and the observation that no defect is able to efficiently capture both electrons and holes through its CTLs to make a detrimental impact, it is likely that this trend will be transferable to other wide-band-gap semiconductors. As a result, multi-phonon capture via point defects in their electronic ground state is unlikely to be the dominant source of efficiency loss in wide-band-gap photovoltaic absorbers. Instead, recombination is more likely governed by extrinsic pathways such as interfaces, extended defects, or via excited-state-mediated processes. This work establishes a foundation for the rational discovery and optimization of next-generation defect-tolerant photovoltaic absorbers.

\section{Computational Details}
First principles calculations were performed using Density Functional Theory (DFT)\cite{kohn1965self} within the projector augmented wave formalism (PAW) as implemented in GPAW,\cite{mortensen2024gpaw} in combination with Atomic Simulation Environment (ASE) for pre and post processing.\cite{larsen2017atomic} 3 atom t-Se unit cell with P3$_1$12 (\#152) space group was used to calculate the electronic structure and optical properties. The experimental lattice parameters were used for all the calculation and the atomic positions were optimized until the forces become less that 1 meV/Å. A plane-wave cutoff energy of 800 eV and a k- mesh density of 8 Å (12 Å) were used for geometry optimization (ground state calculations). Heyd-Scuseria-Ernzerhoff (HSE06)\cite{heyd2003hybrid} functional was used to accurately estimate the band gap. The accurate exchange mixing parameter was modified ($\alpha$=0.26) to reproduce the experimental indirect band gap (1.843 eV). Optical absorption spectra were calculated using Random Phase approximation with Perdew-Burke-Ernzerhoff (PBE)\cite{perdew1996generalized} exchange correlation functional and was shifted to HSE06 direct band gap (1.95 eV). Phonon assisted absorption at the indirect band gap was computed using the methodology proposed in our previous work \cite{kangsabanik2022indirect}, which involves calculation of electron-phonon coupling, momentum matrix elements, phonon joint density of states, etc. PBE exchange correlation functional along with 16x16x16 (8 Å) k-points (k-mesh density) were used to calculate the electron-phonon coupling matrix elements (phonon spectra). Fermi-Dirac smearing with 0.05 eV was used throughout for electronic structure and optical properties calculations.

For point defect calculations a 3x3x3 (81 atom) supercell of the 3-atom primitive cell was used. Supercell convergence was tested in our previous study. We considered  Se vacancy (${\rm V}_{\rm Se}$), and interstitial (${\rm Se}_{\rm i}$) as the intrinsic defects. For extrinsic defects we have considered hydrogen (H), halogens (F, Cl, Br, I), Chalcogens (O, S, Te), Pnictogens (N, P, As, Sb) and Group-IV elements (C, Si, Ge, Sn) which can be present during the growth of Selenium PV devices. For each extrinsic dopant D we have considered substitutions (${\rm D}_{\rm Se}$) and interstitials (${\rm D}_{\rm i}$) as possible defects, thus a total of 34 defects including the intrinsic ones. For the vacancy and substitutional defects we considered q=2, 1, 0, -1, -2 charge states in the units of $|e|$. For each interstitial defect we considered $q=3|e|$ to $q=-3|e|$ for the pnictogens, $q=+4|e|$ to $q=-4|e|$ for the group-IV elements and  $q=+2|e|$ to $q=-2|e|$ for the rest. We used ShakeNBreak package\cite{mosquera2022shakenbreak} to search for lowest energy defect geometries for all the defects at the considered charge states. This calculation to search for the most stable defect calculation was done using PBE exchange correlation functional. After getting the most stable defect geometries,  we used HSE06 functional ($\alpha$=0.26) for final geometry optimization of the defect supercells with all the charge states (34 defects and 194 charge states). A plane wave cutoff of 800 eV and Gamma-point was used for the all the defect calculations. All the supercells were relaxed until the force convergence of 0.01 eV/Å was reached.  Due to long-range nature of the Coulomb interaction, it is important to include correction for the interaction between the image charges in the finite supercell method. We use the Freysoldt-Neugebauer-van de Walle (FNV) correction scheme to correct for electrostatic interaction and potential alignment term.\cite{freysoldt2009fully} The computational defect and PV workflows were created using Atomic Simulation Recipes \cite{gjerding2021atomic} and executed using the Myqueue task scheduler frontend.\cite{mortensen2020myqueue}

1D Configuration coordinate diagrams for $q/q'$ transition were calculated using HSE06 functional at different geometry points mapped from initial($q$) to final($q'$) relaxed defect geometries. The data was then fitted using harmonic and spline interpolation and 1D schrodinger equation was solved to find the overlap between vibronic wave functions, accounting for anharmonicity when present. For radiative carrier capture the dipole transition matrix elements and for nonradiative capture the electron-phonon couplings were calculated using the formalism as implemented in GPAW. The nonradiative capture calculations were based on the formalisms described in Refs. \cite{turiansky2021nonrad, kim2020carriercapture}, with key components implemented using the NONRAD and carriercapture.jl codes. All the photovoltaic device parameters, our formalism of SRH multi-level defect statistics and relevant parameters were independently implemented in GPAW and have been used throughout this study.

\section{Data Availability}
The data supporting this article have been included as part of the Supplementary Information. The codes used in this work are openly available as referred in the manuscript.

\section{Conflict of interest}
The authors declare no conflict of interest.

\section{Acknowledgements}
J.K and K.S.T acknowledge funding from the Novo Nordisk Foundation Data Science Research Infrastructure 2022 Grant:  A high-performance computing infrastructure for data-driven research on sustainable energy materials (Grant no. NNF22OC0078009) and the Novo Nordisk Foundation Challenge Programme 2021: Smart nanomaterials for applications in life-science, BIOMAG (Grant No. NNF21OC0066526). K.S.T. is a Villum Investigator supported by VILLUM FONDEN (Grant No. 37789). J.K. and K.T. acknowledge funding from the Novo Nordisk Foundation (NNF23OC0087524). The authors thank Prof. Karsten W. Jacobsen for fruitful discussions.




\bibliography{main}
\clearpage
\onecolumngrid
\input{supp.tex}

\end{document}

%% file: supp.tex
\setcounter{page}{1}
\setcounter{figure}{0}
\setcounter{table}{0}
\setcounter{section}{0}
\setcounter{equation}{0}
\renewcommand{\thesection}{S\arabic{section}}
\renewcommand{\thetable}{S\arabic{table}}
\renewcommand{\thefigure}{S\arabic{figure}}
\renewcommand{\theequation}{S\arabic{equation}}
\renewcommand{\arraystretch}{1.2}
\begingroup
\centering
\textbf{Supplementary Material:\\
Defect Tolerance in Trigonal Selenium Photovoltaics}\\[1em]
Jiban Kangsabanik$^*$${}^{1,2}$,
Kasper Tolborg${}^{2}$,
Thomas Olsen${}^{1}$,
Kristian S. Thygesen$^\dagger$${}^{1}$ \\[0.5em]

\textit{${}^{1}$CAMD, Computational Atomic-Scale Materials Design, Department of Physics, Technical University of Denmark, 2800 Kgs. Lyngby, Denmark\\
${}^{2}$Department of Chemistry and Bioscience, Aalborg University, 9220 Aalborg, Denmark}
\par
\vspace{1em}
\endgroup
\vspace{-1.0em}
\onecolumngrid
\begin{center}
\textcolor{blue}{(Email: $^*$jika@bio.aau.dk, $^\dagger$thygesen@fysik.dtu.dk)}
\end{center}

\section*{S1. Methodology}
\subsection*{A. Shockley--Read--Hall (SRH) recombination parameter $R_{\mathrm{SRH}}$ for defects with multiple charge transition levels}\label{methodSRH}

We extend the SRH formalism to defects with multiple charge transition levels (CTLs) inside the band gap. Throughout, $N_D$ denotes the total defect concentration, while $N_q$ is the concentration of the defect in charge state $q$, such that $N_D=\sum_q N_q$. The quantities $C_n^q$ and $C_p^{q-1}$ denote the electron and hole capture coefficients for the transition $q \leftrightarrow q-1$, respectively. Further, $n$ and $p$ are the free electron and hole concentrations under operating conditions, and $n^q$ and $p^q$ are the corresponding carrier concentrations when the Fermi level lies at the CTL $(q/q-1)$. By definition,
\begin{equation}
n^q p^q = n_i^2.
\end{equation}

\subsubsection*{1. Single CTL}
For a defect with one CTL, $(1/0)$, the SRH recombination rate is
\begin{equation}
R_\text{SRH}
=
\frac{N_D C_n^1 C_p^0 (np-n^1p^1)}
{C_n^1(n+n^1)+C_p^0(p+p^1)}
=
\frac{N_D C_n^1 C_p^0 (np-n_i^2)}
{C_n^1(n+n^1)+C_p^0(p+p^1)}.
\label{eqSSRH11}
\end{equation}

\subsubsection*{2. Two CTLs}
For a defect with two CTLs, $(1/0 \mid 0/-1)$, the SRH recombination rate becomes
\begin{equation}
R_\text{SRH}
=
\frac{N_D}{G}
\left[
\frac{C_n^0C_p^{-1}(np-n^0p^0)}{C_p^{-1}p+C_n^0n^0}
+
\frac{C_p^0C_n^1(np-n^1p^1)}{C_n^1n+C_p^0p^1}
\right],
\label{eqSSRH22}
\end{equation}
with
\begin{equation}
G=
1+
\frac{C_p^0p+C_n^1n^1}{C_n^1n+C_p^0p^1}
+
\frac{C_n^0n+C_p^{-1}p^0}{C_p^{-1}p+C_n^0n^0}.
\label{eqSG22}
\end{equation}

Similarly, for a defect with CTLs $(2/1 \mid 1/0)$,
\begin{equation}
R_\text{SRH}
=
\frac{N_D}{G}
\left[
\frac{C_n^1C_p^0(np-n^1p^1)}{C_p^0p+C_n^1n^1}
+
\frac{C_p^1C_n^2(np-n^2p^2)}{C_n^2n+C_p^1p^2}
\right],
\end{equation}
with
\begin{equation}
G=
1+
\frac{C_p^1p+C_n^2n^2}{C_n^2n+C_p^1p^2}
+
\frac{C_n^1n+C_p^0p^1}{C_p^0p+C_n^1n^1}.
\end{equation}

\subsubsection*{3. Three CTLs}
For a defect with three CTLs, $(1/0 \mid 0/-1 \mid -1/-2)$, the SRH recombination rate reads
\begin{equation}
\begin{split}
R_\text{SRH}
=
\frac{N_D}{G}
\Bigg[
&
\frac{
C_p^{-1}C_p^{0}C_n^1(npp-n^1p^1p)
+
C_n^0C_p^{0}C_n^1(npn^0-n^1p^1n^0)
}
{(C_n^{0}n+C_p^{-1}p^0)(C_n^{1}n+C_p^0p^1)}
\\
&+
\frac{C_n^0C_p^{-1}(np-n^0p^0)}{C_n^{0}n+C_p^{-1}p^0}
+
\frac{C_n^{-1}C_p^{-2}(np-n^{-1}p^{-1})}{C_p^{-2}p+C_n^{-1}n^{-1}}
\Bigg],
\end{split}
\label{eqSSRH33a}
\end{equation}
with
\begin{equation}
G=
1+
\frac{C_p^{-1}p+C_n^0n^0}{C_n^0n+C_p^{-1}p^0}
+
\frac{C_n^{-1}n+C_p^{-2}p^{-1}}{C_p^{-2}p+C_n^{-1}n^{-1}}
+
\frac{(C_p^{-1}p+C_n^0n^0)(C_p^0p+C_n^1n^1)}
{(C_n^0n+C_p^{-1}p^0)(C_n^1n+C_p^0p^1)}.
\label{eqSG33a}
\end{equation}

Similarly, for a defect with CTLs $(2/1 \mid 1/0 \mid 0/-1)$,
\begin{equation}
\begin{split}
R_\text{SRH}
=
\frac{N_D}{G}
\Bigg[
&
\frac{
C_p^0C_p^1C_n^2(npp-n^2p^2p)
+
C_n^1C_p^1C_n^2(npn^1-n^2p^2n^1)
}
{(C_n^1n+C_p^0p^1)(C_n^2n+C_p^1p^2)}
\\
&+
\frac{C_n^1C_p^0(np-n^1p^1)}{C_n^1n+C_p^0p^1}
+
\frac{C_n^0C_p^{-1}(np-n^0p^0)}{C_p^{-1}p+C_n^0n^0}
\Bigg],
\end{split}
\end{equation}
with
\begin{equation}
G=
1+
\frac{C_n^0n+C_p^{-1}p^0}{C_p^{-1}p+C_n^0n^0}
+
\frac{C_p^0p+C_n^1n^1}{C_n^1n+C_p^0p^1}
+
\frac{(C_p^0p+C_n^1n^1)(C_p^1p+C_n^2n^2)}
{(C_n^1n+C_p^0p^1)(C_n^2n+C_p^1p^2)}.
\end{equation}

\subsubsection*{4. Four CTLs}
For a defect with four CTLs, $(2/1 \mid 1/0 \mid 0/-1 \mid -1/-2)$, the SRH recombination rate is
\begin{equation}
\begin{split}
R_\text{SRH}
=
\frac{N_D}{G}
\Bigg[
&
\frac{
C_p^0C_p^1C_n^2(npp-n^2p^2p)
+
C_n^1C_p^1C_n^2(npn^1-n^2p^2n^1)
}
{(C_n^1n+C_p^0p^1)(C_n^2n+C_p^1p^2)}
+
\frac{C_n^1C_p^0(np-n^1p^1)}{C_n^1n+C_p^0p^1}
\\
&+
\frac{C_n^0C_p^{-1}(np-n^0p^0)}{C_p^{-1}p+C_n^0n^0}
+
\frac{
C_n^0C_n^{-1}C_p^{-2}(npn-n^{-1}p^{-1}n)
+
C_p^{-1}C_n^{-1}C_p^{-2}(npp^0-n^{-1}p^{-1}p^0)
}
{(C_p^{-1}p+C_n^0n^0)(C_p^{-2}p+C_n^{-1}n^{-1})}
\Bigg],
\end{split}
\label{eqSSRH44}
\end{equation}
with
\begin{equation}
\begin{split}
G=
1
&+
\frac{C_p^0p+C_n^1n^1}{C_n^1n+C_p^0p^1}
+
\frac{C_n^0n+C_p^{-1}p^0}{C_p^{-1}p+C_n^0n^0}
+
\frac{(C_p^0p+C_n^1n^1)(C_p^1p+C_n^2n^2)}
{(C_n^1n+C_p^0p^1)(C_n^2n+C_p^1p^2)}
\\
&+
\frac{(C_n^0n+C_p^{-1}p^0)(C_n^{-1}n+C_p^{-2}p^{-1})}
{(C_p^{-1}p+C_n^0n^0)(C_p^{-2}p+C_n^{-1}n^{-1})}.
\end{split}
\label{eqSG44}
\end{equation}

The above expressions provide the working formulas used in the present study for defects with multiple CTLs. A more detailed derivation is provided in our previous work.\cite{S1}

\clearpage

\section*{S2. Results}\label{results}

\subsection*{A. Defect Concentrations}\label{ctls}

\begin{figure*}[h!]
	\centering
	\includegraphics[scale=0.88]{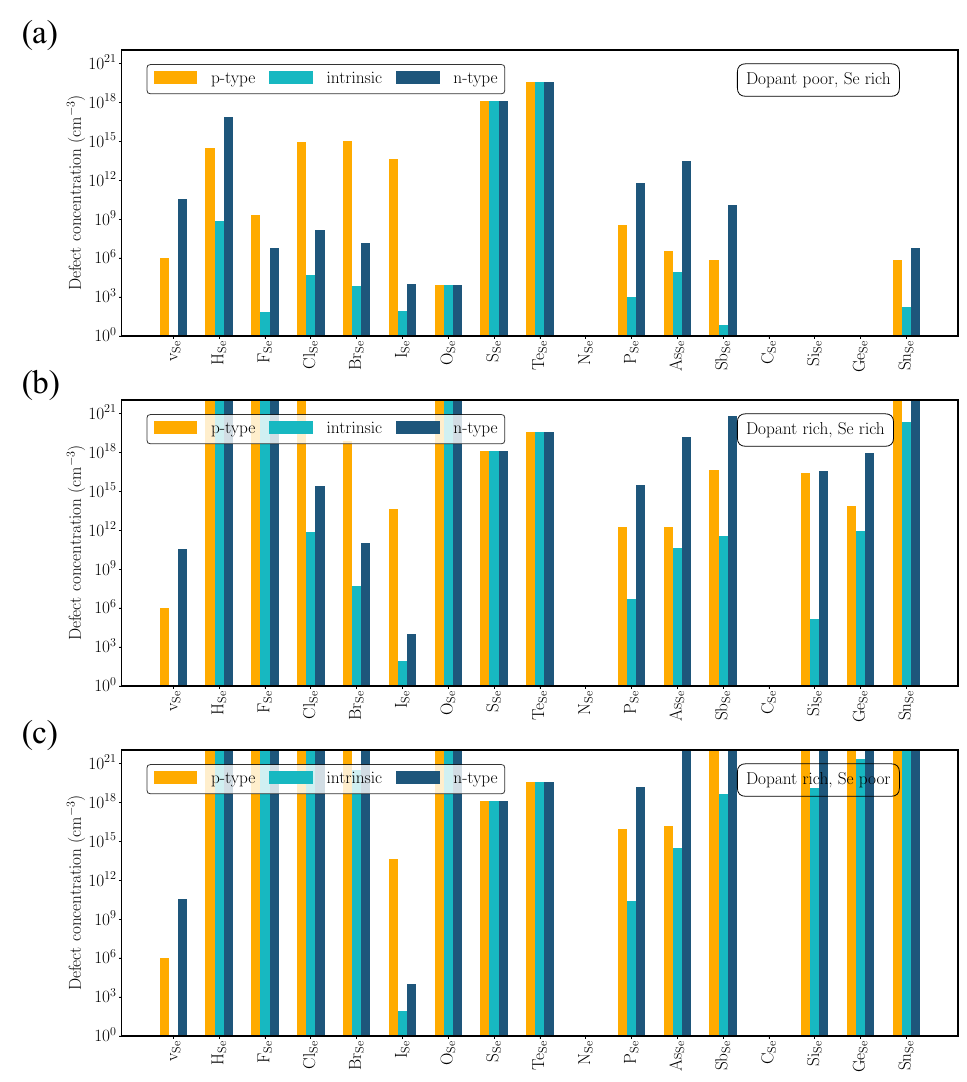}
	\caption{Equilibrium concentrations of substitutional defects at (a) Dopant-poor, Se-rich, (b) Dopant-rich, Se-rich, (c) Dopant-rich, Se-poor conditions. For a particular defect three different histograms represent the different equilibrium Fermi level (E$_F$) positions: p-type (E$_F$ at at 0.21 eV above VBM), intrinsic (E$_F$ at 0.91 eV above VBM), n-type (E$_F$ at 0.23 eV below CBM).}
	\label{FigS1}
\end{figure*}

\begin{figure*}[t!]
	\centering
	\includegraphics[scale=0.88]{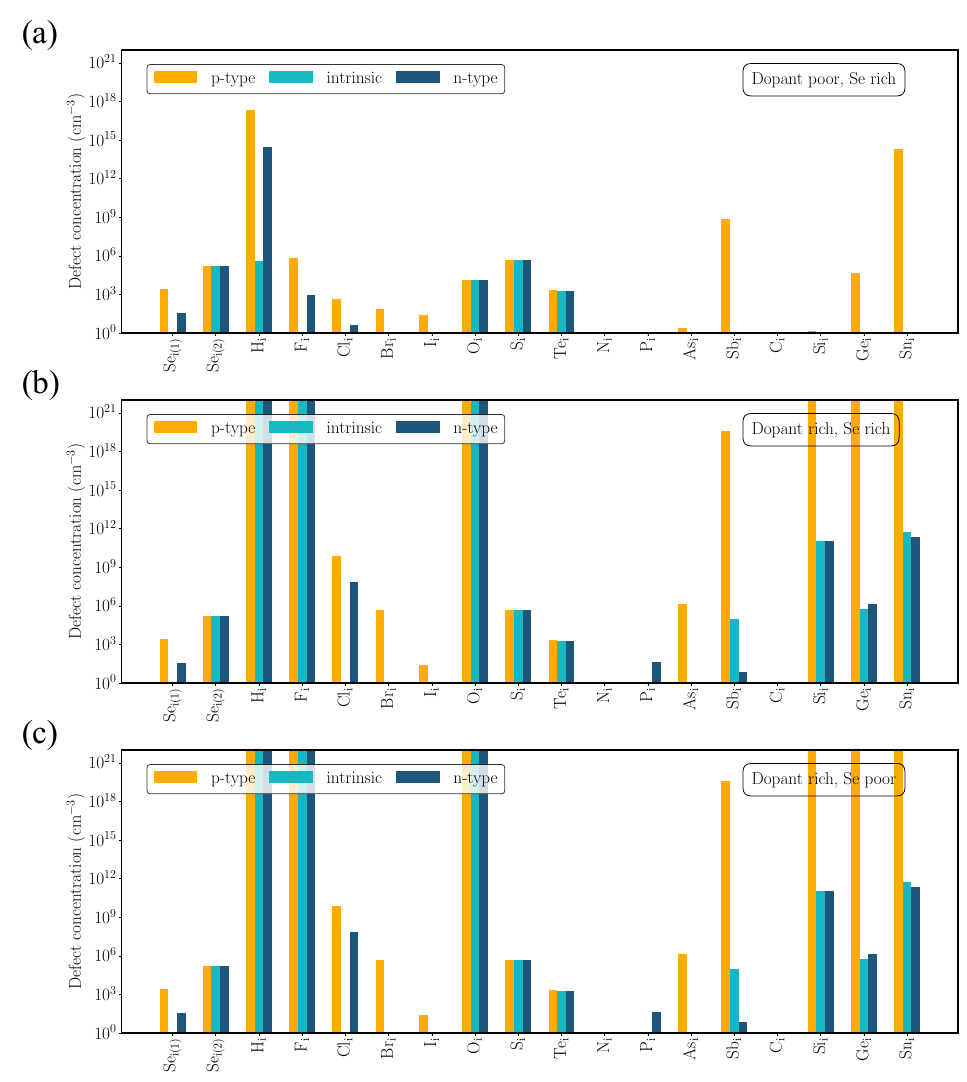}
	\caption{Equilibrium concentrations of interstitial defects at (a) Dopant-poor, Se-rich, (b) Dopant-rich, Se-rich, (c) Dopant-rich, Se-poor conditions. For a particular defect three different histograms represent the different equilibrium Fermi level (E$_F$) positions: p-type (E$_F$ at at 0.21 eV above VBM), intrinsic (E$_F$ at 0.91 eV above VBM), n-type (E$_F$ at 0.23 eV below CBM).}
	\label{FigS2}
\end{figure*}

\clearpage

\subsection*{B. CC diagrams, carrier capture coefficients, photovoltaic device parameters}\label{ccpv}

\begin{figure*}[h!]
	\centering
	\includegraphics[scale=0.95]{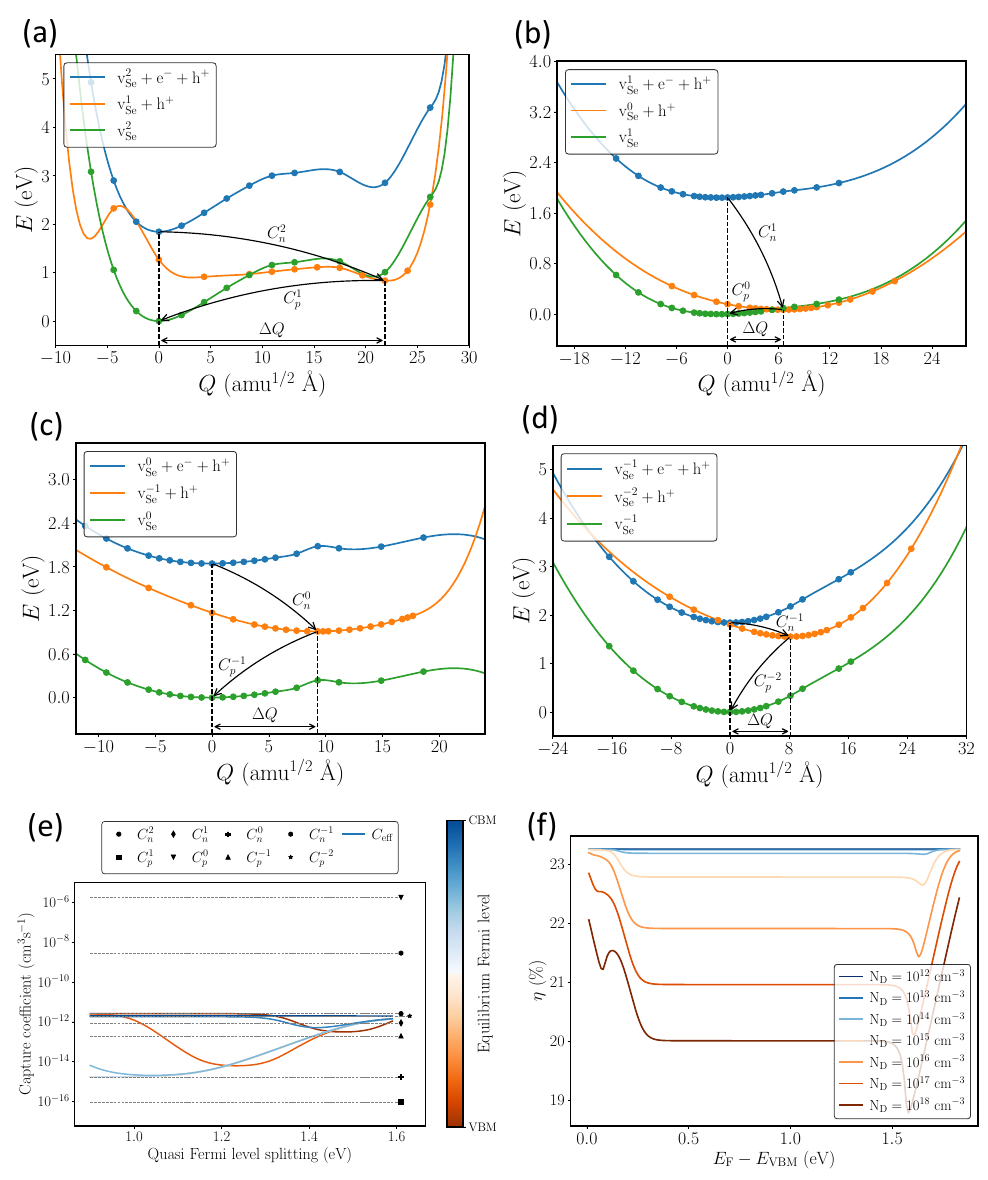}
	\caption{(a)-(d) One-dimensional configuration coordinate diagrams for the $2/1$, $1/0$, $0/-1$, $-1/-2$ charge state transitions of the v$_{\mathrm{Se}}$ defect in t-Se, respectively. Solid circles denote the data points calculated using HSE06 functional and solid lines are obtained by fitting them with quadratic spline functions accounting for anharmonicity. (e) Carrier capture coefficients  (including both radiative and nonradiative capture processes) at different charge states and associated effective capture coefficient ($C_{\mathrm{eff}}$) with respect to quasi-Fermi level splitting at different majority carrier types (equilibrium Fermi levels denoted by the color scale) at 300 K. (f) Photovoltaic device efficiency ($\mathrm{\eta}$) with respect to the equilibrium Fermi level position at different defect concentrations at 300 K and 500 nm device thickness. }
	\label{Fig3}
\end{figure*}

\begin{figure*}[t!]
	\centering
	\includegraphics[scale=1.0]{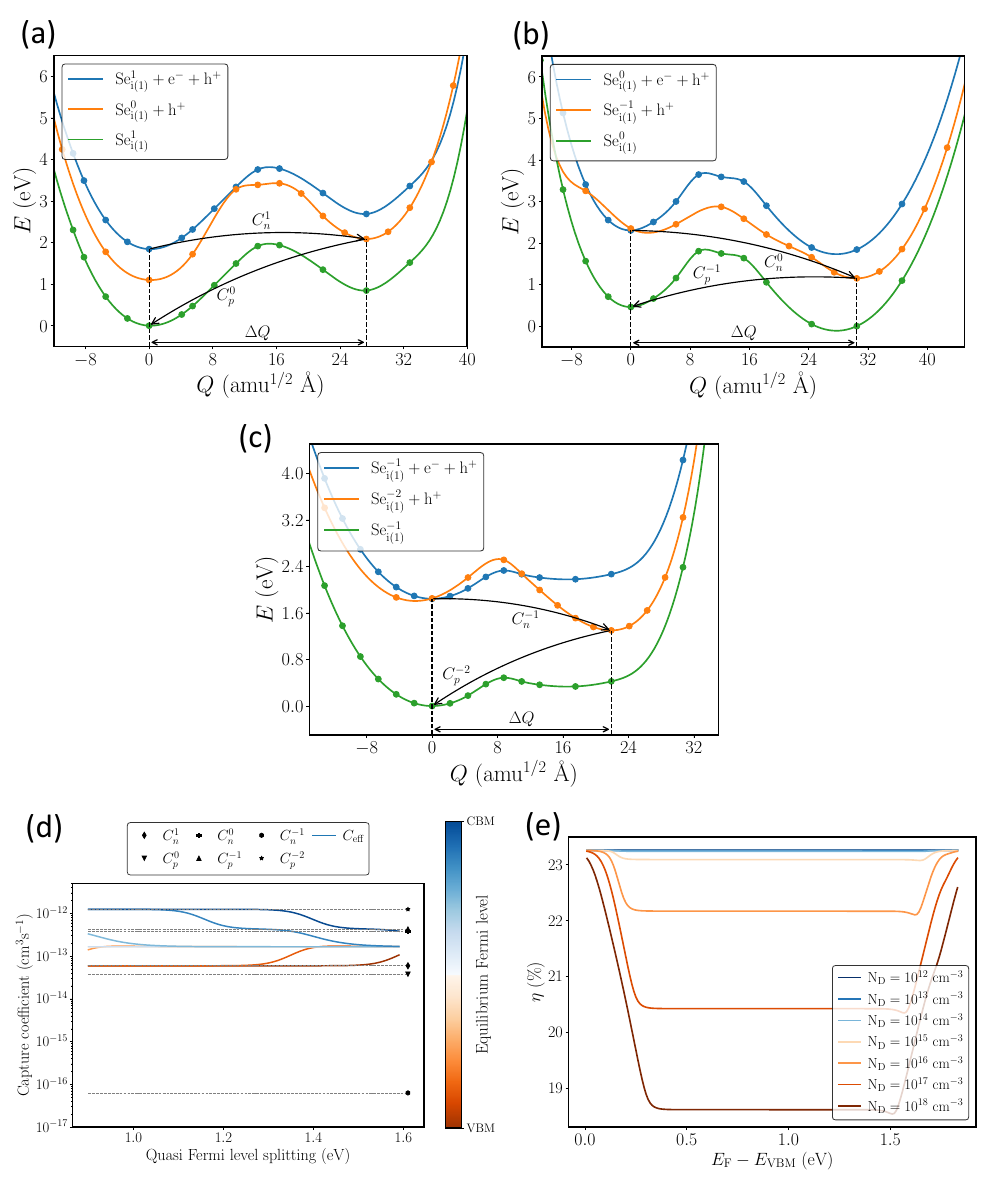}
	\caption{(a)-(c) One-dimensional configuration coordinate diagrams for the $1/0$, $0/-1$, $-1/-2$ charge state transitions of the Se$_{\mathrm{i(1)}}$ defect in t-Se, respectively. Solid circles denote the data points calculated using HSE06 functional and solid lines are obtained by fitting them with quadratic spline functions accounting for anharmonicity. (d) Carrier capture coefficients  (including both radiative and nonradiative capture processes) at different charge states and associated effective capture coefficient ($C_{\mathrm{eff}}$) with respect to quasi-Fermi level splitting at different majority carrier types (equilibrium Fermi levels denoted by the color scale) at 300 K. (e) Photovoltaic device efficiency ($\mathrm{\eta}$) with respect to the equilibrium Fermi level position at different defect concentrations at 300 K and 500 nm device thickness.}
	\label{FigS3}
\end{figure*}

\begin{figure*}[t!]
	\centering
	\includegraphics[scale=1.0]{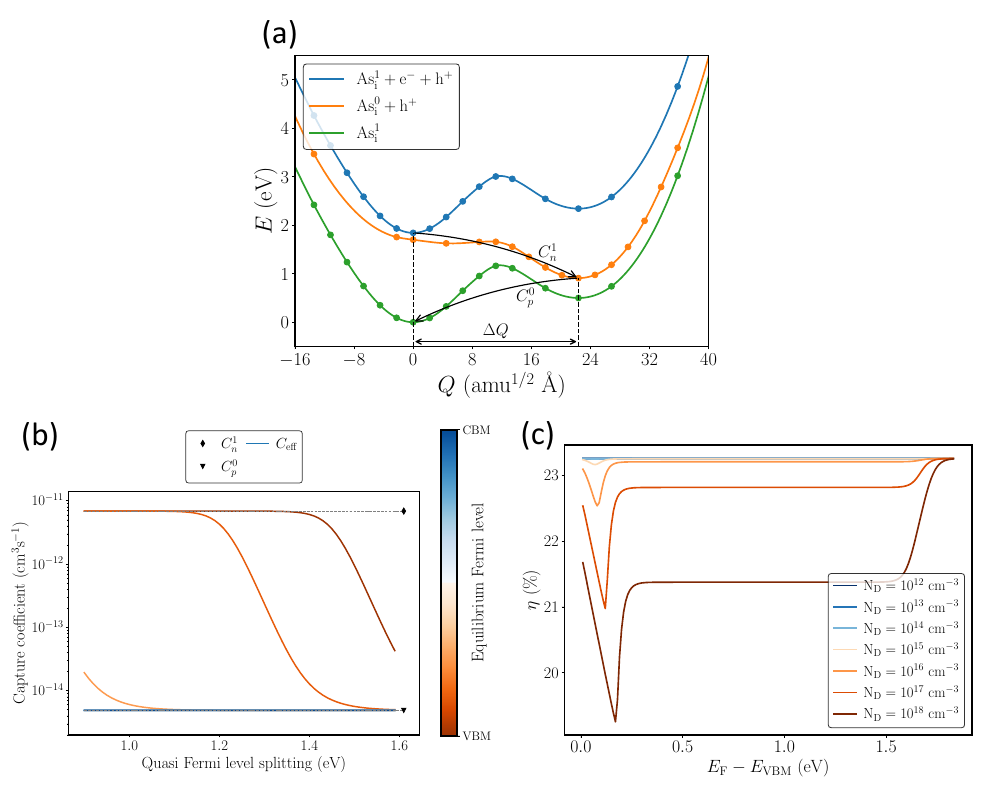}
	\caption{(a) One-dimensional configuration coordinate diagrams for the $1/0$ charge state transitions of the As$_{\mathrm{i}}$ defect in t-Se. Solid circles denote the data points calculated using HSE06 functional and solid lines are obtained by fitting them with quadratic spline functions accounting for anharmonicity. (b) Carrier capture coefficients  (including both radiative and nonradiative capture processes) at different charge states and associated effective capture coefficient ($C_{\mathrm{eff}}$) with respect to quasi-Fermi level splitting at different majority carrier types (equilibrium Fermi levels denoted by the color scale) at 300 K. (c) Photovoltaic device efficiency ($\mathrm{\eta}$) with respect to the equilibrium Fermi level position at different defect concentrations at 300 K and 500 nm device thickness. }
	\label{Fig4}
\end{figure*}

\begin{figure*}[t!]
	\centering
	\includegraphics[scale=1.0]{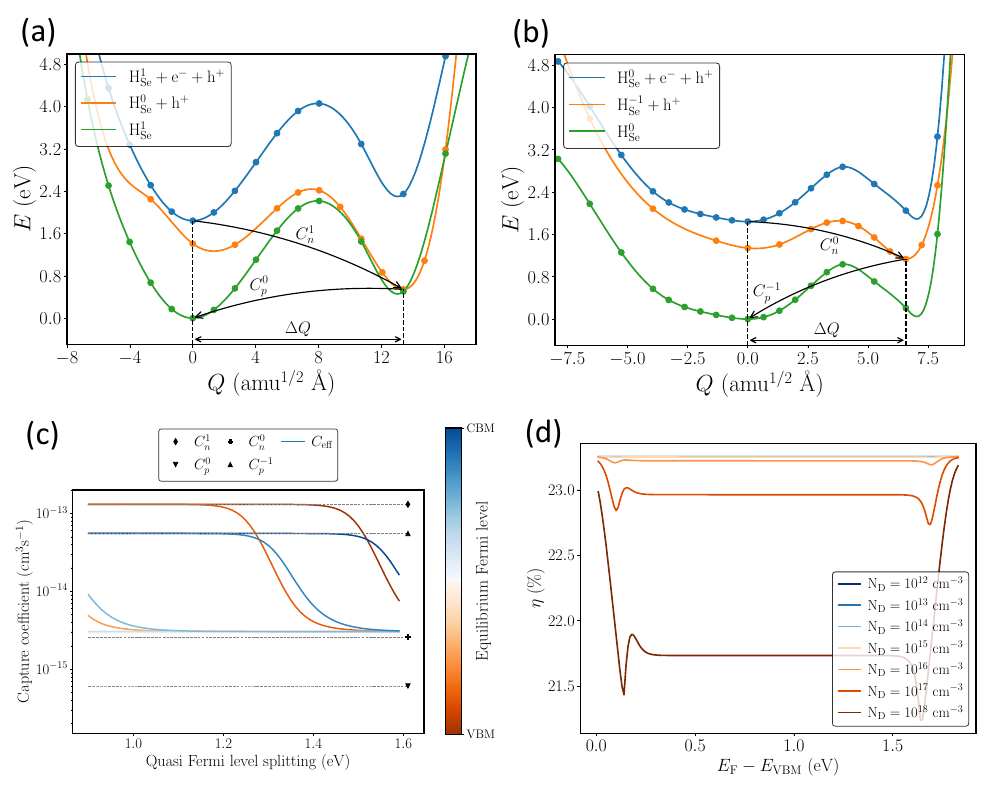}
	\caption{(a)-(b) One-dimensional configuration coordinate diagrams for the $1/0$, $0/-1$ charge state transitions of the H$_{\mathrm{Se}}$ defect in t-Se, respectively. Solid circles denote the data points calculated using HSE06 functional and solid lines are obtained by fitting them with quadratic spline functions accounting for anharmonicity. (c) Carrier capture coefficients  (including both radiative and nonradiative capture processes) at different charge states and associated effective capture coefficient ($C_{\mathrm{eff}}$) with respect to quasi-Fermi level splitting at different majority carrier types (equilibrium Fermi levels denoted by the color scale) at 300 K. (d) Photovoltaic device efficiency ($\mathrm{\eta}$) with respect to the equilibrium Fermi level position at different defect concentrations at 300 K and 500 nm device thickness.}
	\label{FigS4}
\end{figure*}

\begin{figure*}[t!]
	\centering
	\includegraphics[scale=1.0]{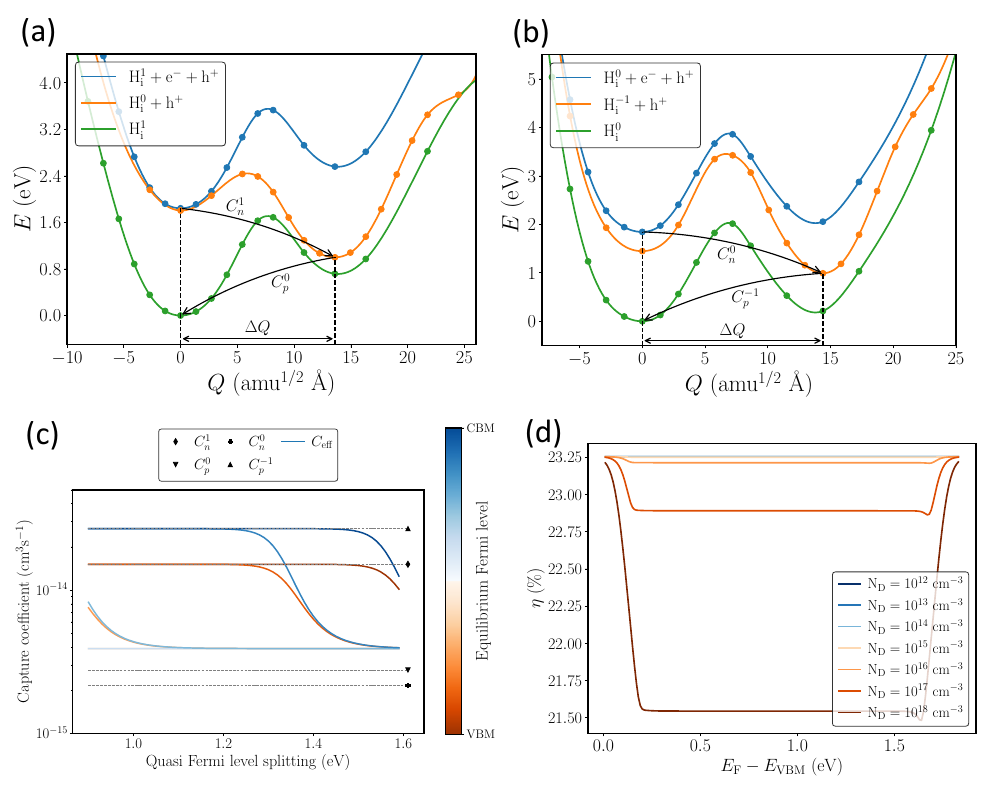}
	\caption{(a)-(b) One-dimensional configuration coordinate diagrams for the $1/0$, $0/-1$ charge state transitions of the H$_{\mathrm{i}}$ defect in t-Se, respectively. Solid circles denote the data points calculated using HSE06 functional and solid lines are obtained by fitting them with quadratic spline functions accounting for anharmonicity. (c) Carrier capture coefficients  (including both radiative and nonradiative capture processes) at different charge states and associated effective capture coefficient ($C_{\mathrm{eff}}$) with respect to quasi-Fermi level splitting at different majority carrier types (equilibrium Fermi levels denoted by the color scale) at 300 K. (d) Photovoltaic device efficiency ($\mathrm{\eta}$) with respect to the equilibrium Fermi level position at different defect concentrations at 300 K and 500 nm device thickness.}
	\label{FigS5}
\end{figure*}

\begin{figure*}[t!]
	\centering
	\includegraphics[scale=1.0]{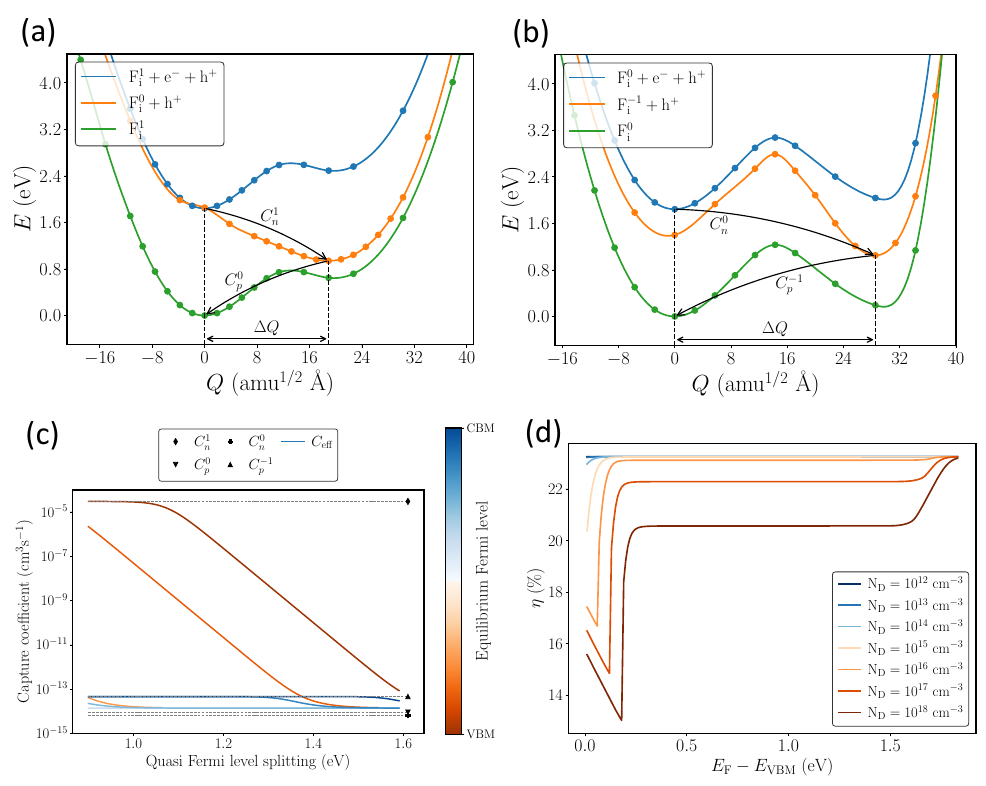}
	\caption{(a)-(b) One-dimensional configuration coordinate diagrams for the $1/0$, $0/-1$ charge state transitions of the F$_{\mathrm{i}}$ defect in t-Se, respectively. Solid circles denote the data points calculated using HSE06 functional and solid lines are obtained by fitting them with quadratic spline functions accounting for anharmonicity. (c) Carrier capture coefficients  (including both radiative and nonradiative capture processes) at different charge states and associated effective capture coefficient ($C_{\mathrm{eff}}$) with respect to quasi-Fermi level splitting at different majority carrier types (equilibrium Fermi levels denoted by the color scale) at 300 K. (d) Photovoltaic device efficiency ($\mathrm{\eta}$) with respect to the equilibrium Fermi level position at different defect concentrations at 300 K and 500 nm device thickness.}
	\label{FigS6}
\end{figure*}

\begin{figure*}[t!]
	\centering
	\includegraphics[scale=1.0]{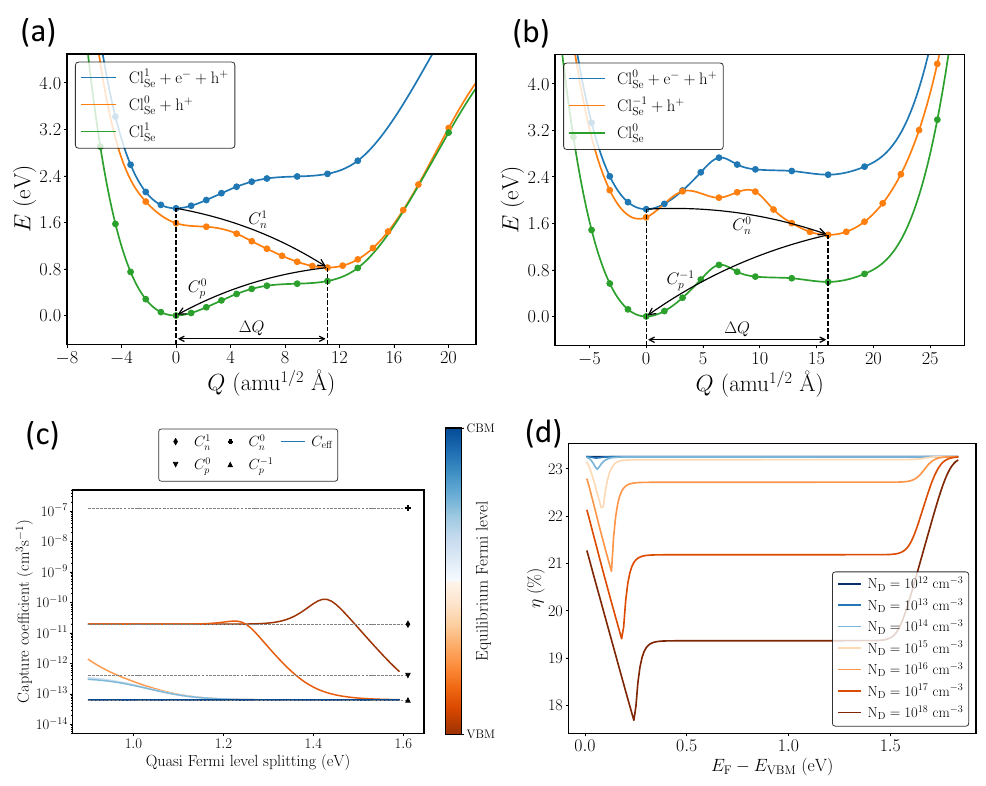}
	\caption{(a)-(b) One-dimensional configuration coordinate diagrams for the $1/0$, $0/-1$ charge state transitions of the Cl$_{\mathrm{Se}}$ defect in t-Se, respectively. Solid circles denote the data points calculated using HSE06 functional and solid lines are obtained by fitting them with quadratic spline functions accounting for anharmonicity. (c) Carrier capture coefficients  (including both radiative and nonradiative capture processes) at different charge states and associated effective capture coefficient ($C_{\mathrm{eff}}$) with respect to quasi-Fermi level splitting at different majority carrier types (equilibrium Fermi levels denoted by the color scale) at 300 K. (d) Photovoltaic device efficiency ($\mathrm{\eta}$) with respect to the equilibrium Fermi level position at different defect concentrations at 300 K and 500 nm device thickness.}
	\label{FigS7}
\end{figure*}

\begin{figure*}[t!]
	\centering
	\includegraphics[scale=1.0]{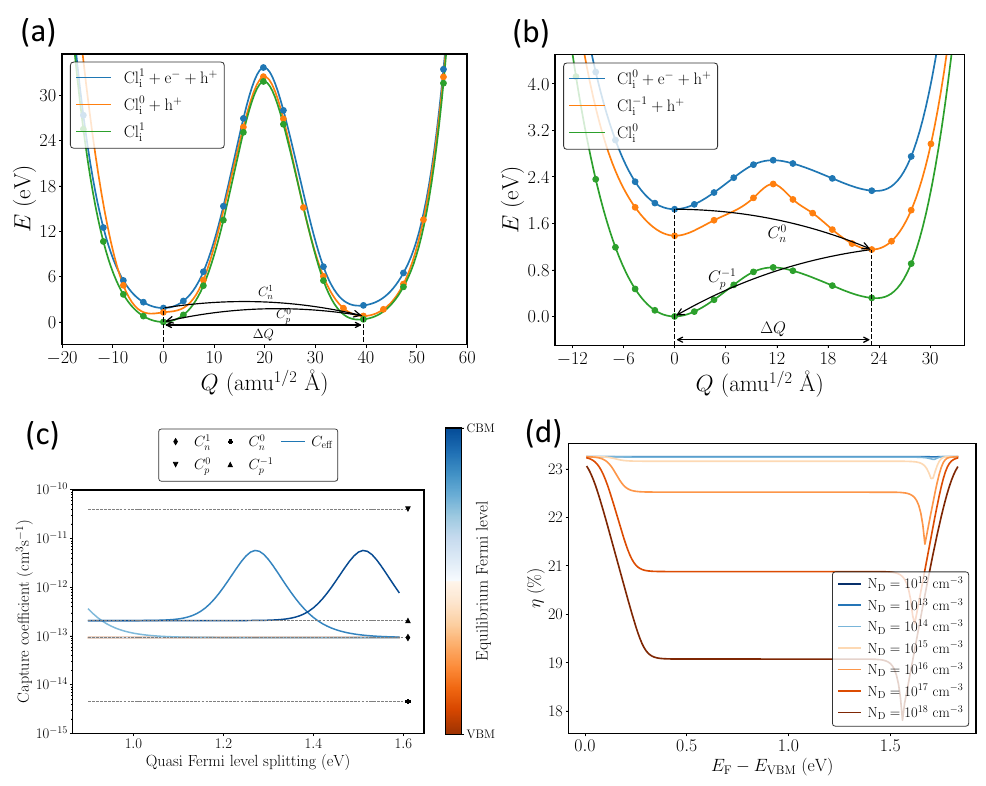}
	\caption{(a)-(b) One-dimensional configuration coordinate diagrams for the $1/0$, $0/-1$ charge state transitions of the Cl$_{\mathrm{i}}$ defect in t-Se, respectively. Solid circles denote the data points calculated using HSE06 functional and solid lines are obtained by fitting them with quadratic spline functions accounting for anharmonicity. (c) Carrier capture coefficients  (including both radiative and nonradiative capture processes) at different charge states and associated effective capture coefficient ($C_{\mathrm{eff}}$) with respect to quasi-Fermi level splitting at different majority carrier types (equilibrium Fermi levels denoted by the color scale) at 300 K. (d) Photovoltaic device efficiency ($\mathrm{\eta}$) with respect to the equilibrium Fermi level position at different defect concentrations at 300 K and 500 nm device thickness.}
	\label{FigS8}
\end{figure*}

\begin{figure*}[t!]
	\centering
	\includegraphics[scale=1.0]{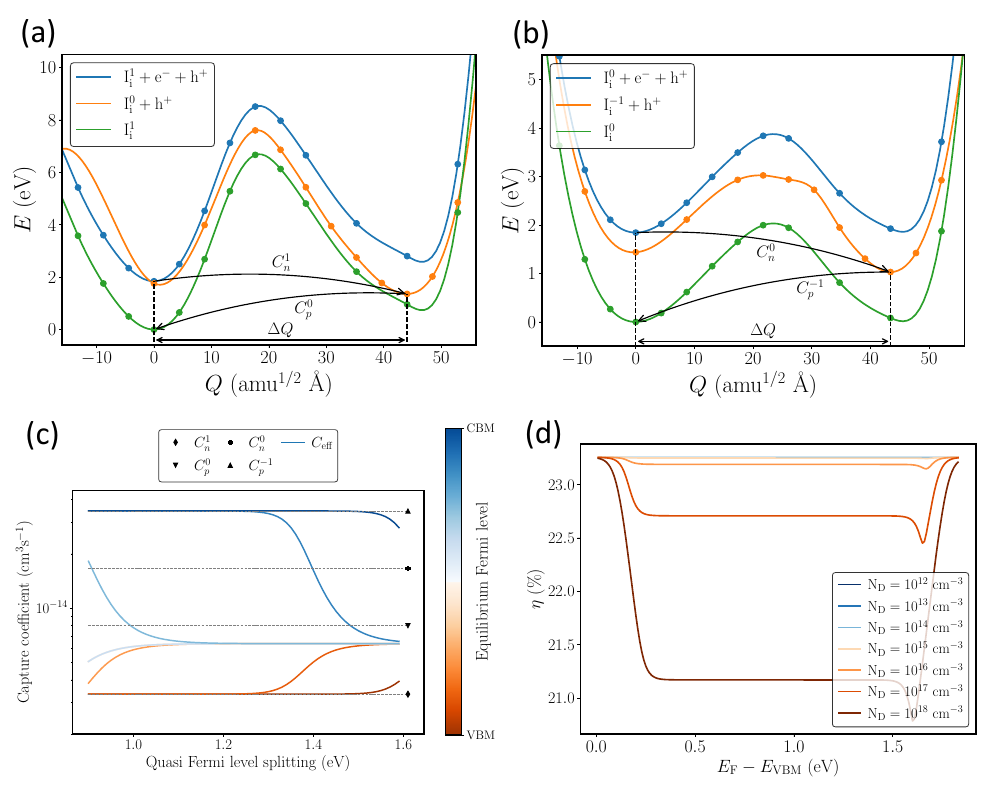}
	\caption{(a)-(b) One-dimensional configuration coordinate diagrams for the $1/0$, $0/-1$ charge state transitions of the I$_{\mathrm{i}}$ defect in t-Se, respectively. Solid circles denote the data points calculated using HSE06 functional and solid lines are obtained by fitting them with quadratic spline functions accounting for anharmonicity. (c) Carrier capture coefficients  (including both radiative and nonradiative capture processes) at different charge states and associated effective capture coefficient ($C_{\mathrm{eff}}$) with respect to quasi-Fermi level splitting at different majority carrier types (equilibrium Fermi levels denoted by the color scale) at 300 K. (d) Photovoltaic device efficiency ($\mathrm{\eta}$) with respect to the equilibrium Fermi level position at different defect concentrations at 300 K and 500 nm device thickness.}
	\label{FigS9}
\end{figure*}

\begin{figure*}[t!]
	\centering
	\includegraphics[scale=1.0]{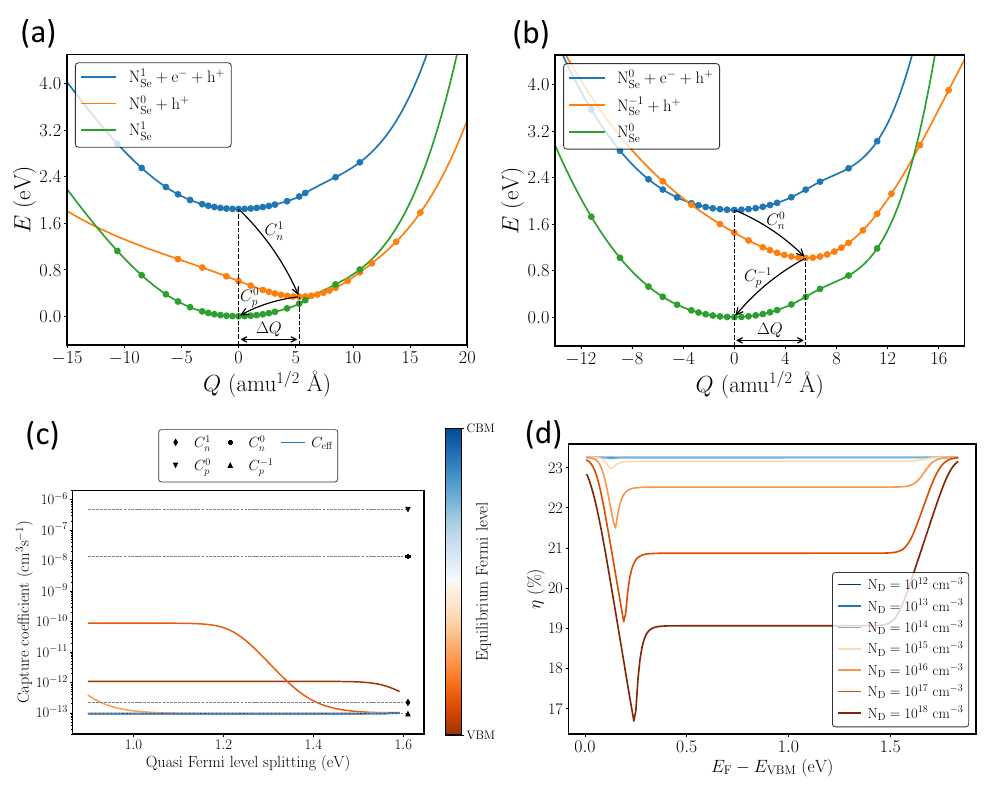}
	\caption{(a)-(b) One-dimensional configuration coordinate diagrams for the $1/0$, $0/-1$ charge state transitions of the N$_{\mathrm{Se}}$ defect in t-Se, respectively. Solid circles denote the data points calculated using HSE06 functional and solid lines are obtained by fitting them with quadratic spline functions accounting for anharmonicity. (c) Carrier capture coefficients  (including both radiative and nonradiative capture processes) at different charge states and associated effective capture coefficient ($C_{\mathrm{eff}}$) with respect to quasi-Fermi level splitting at different majority carrier types (equilibrium Fermi levels denoted by the color scale) at 300 K. (d) Photovoltaic device efficiency ($\mathrm{\eta}$) with respect to the equilibrium Fermi level position at different defect concentrations at 300 K and 500 nm device thickness.}
	\label{FigS10}
\end{figure*}

\begin{figure*}[t!]
	\centering
	\includegraphics[scale=1.0]{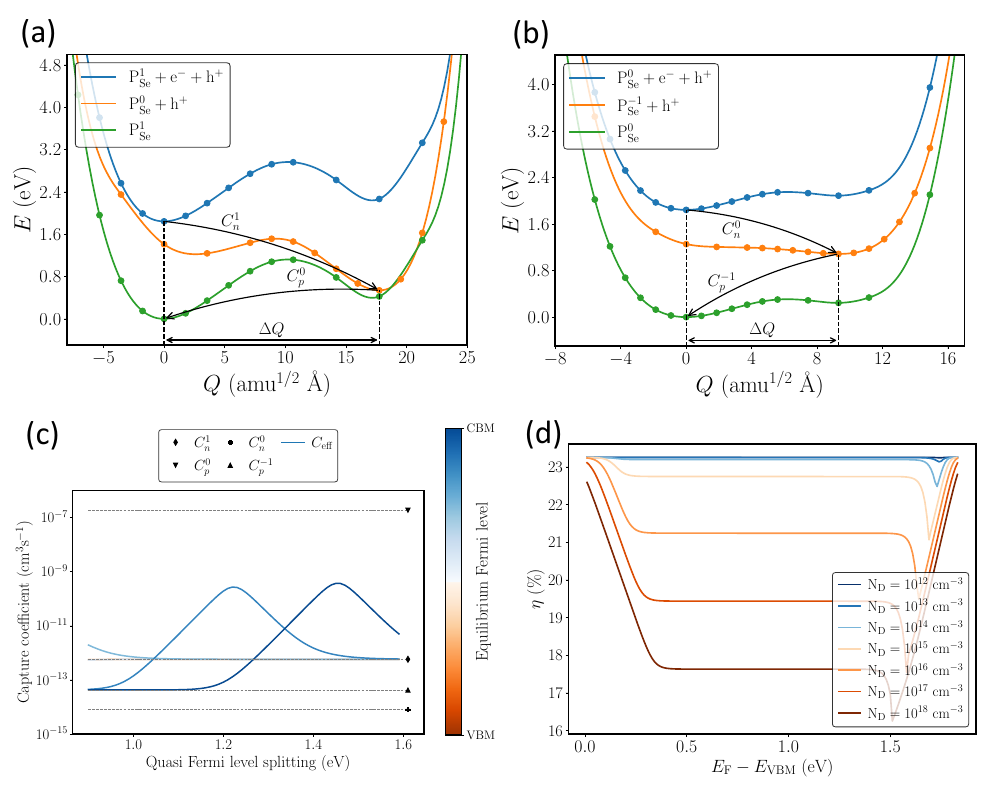}
	\caption{(a)-(b) One-dimensional configuration coordinate diagrams for the $1/0$, $0/-1$ charge state transitions of the P$_{\mathrm{Se}}$ defect in t-Se, respectively. Solid circles denote the data points calculated using HSE06 functional and solid lines are obtained by fitting them with quadratic spline functions accounting for anharmonicity. (c) Carrier capture coefficients  (including both radiative and nonradiative capture processes) at different charge states and associated effective capture coefficient ($C_{\mathrm{eff}}$) with respect to quasi-Fermi level splitting at different majority carrier types (equilibrium Fermi levels denoted by the color scale) at 300 K. (d) Photovoltaic device efficiency ($\mathrm{\eta}$) with respect to the equilibrium Fermi level position at different defect concentrations at 300 K and 500 nm device thickness.}
	\label{FigS11}
\end{figure*}

\begin{figure*}[t!]
	\centering
	\includegraphics[scale=1.0]{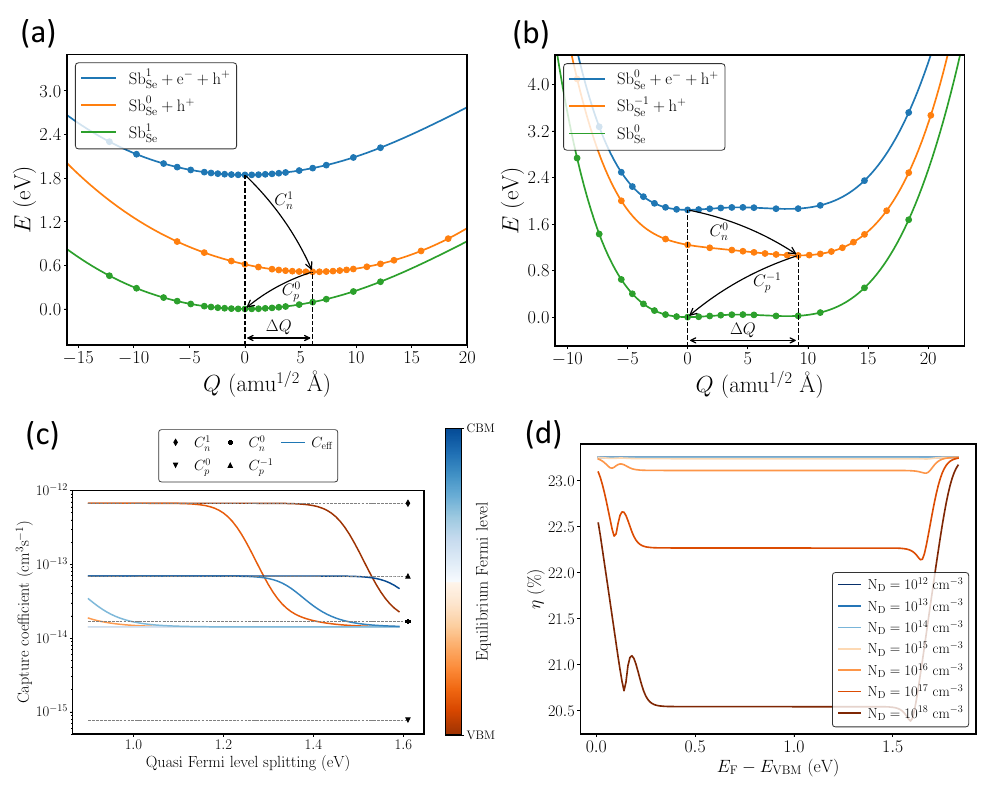}
	\caption{(a)-(b) One-dimensional configuration coordinate diagrams for the $1/0$, $0/-1$ charge state transitions of the Sb$_{\mathrm{Se}}$ defect in t-Se, respectively. Solid circles denote the data points calculated using HSE06 functional and solid lines are obtained by fitting them with quadratic spline functions accounting for anharmonicity. (c) Carrier capture coefficients  (including both radiative and nonradiative capture processes) at different charge states and associated effective capture coefficient ($C_{\mathrm{eff}}$) with respect to quasi-Fermi level splitting at different majority carrier types (equilibrium Fermi levels denoted by the color scale) at 300 K. (d) Photovoltaic device efficiency ($\mathrm{\eta}$) with respect to the equilibrium Fermi level position at different defect concentrations at 300 K and 500 nm device thickness.}
	\label{FigS12}
\end{figure*}

\begin{figure*}[t!]
	\centering
	\includegraphics[scale=1.0]{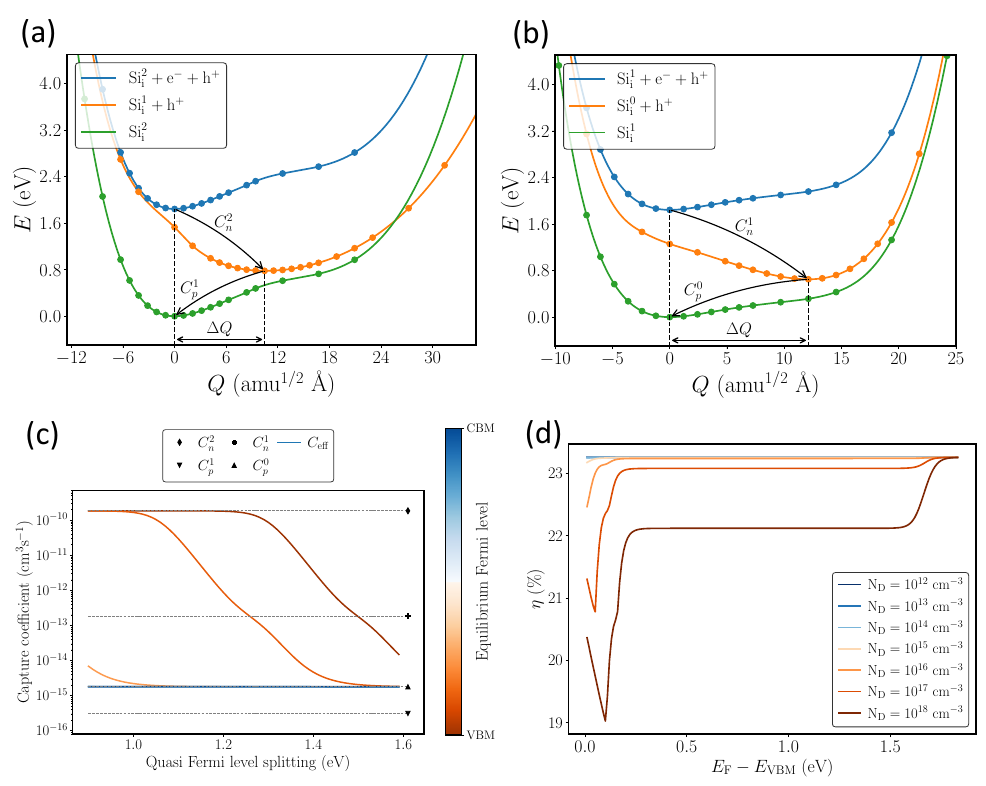}
	\caption{(a)-(b) One-dimensional configuration coordinate diagrams for the $2/1$, $1/0$ charge state transitions of the Si$_{\mathrm{i}}$ defect in t-Se, respectively. Solid circles denote the data points calculated using HSE06 functional and solid lines are obtained by fitting them with quadratic spline functions accounting for anharmonicity. (c) Carrier capture coefficients  (including both radiative and nonradiative capture processes) at different charge states and associated effective capture coefficient ($C_{\mathrm{eff}}$) with respect to quasi-Fermi level splitting at different majority carrier types (equilibrium Fermi levels denoted by the color scale) at 300 K. (d) Photovoltaic device efficiency ($\mathrm{\eta}$) with respect to the equilibrium Fermi level position at different defect concentrations at 300 K and 500 nm device thickness.}
	\label{FigS13}
\end{figure*}

\begin{figure*}[t!]
	\centering
	\includegraphics[scale=1.0]{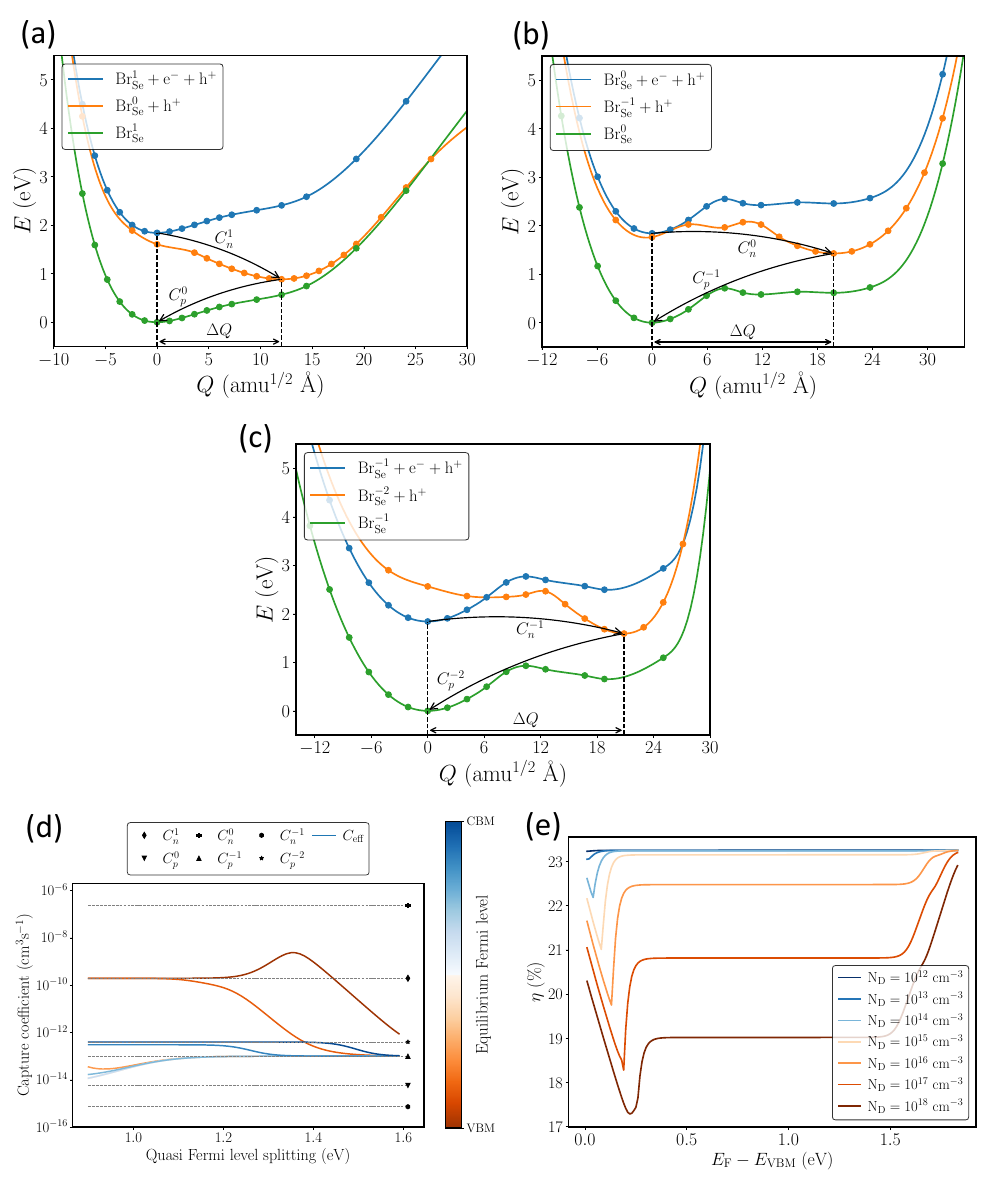}
	\caption{(a)-(c) One-dimensional configuration coordinate diagrams for the $1/0$, $0/-1$, $-1/-2$ charge state transitions of the Br$_{\mathrm{Se}}$ defect in t-Se, respectively. Solid circles denote the data points calculated using HSE06 functional and solid lines are obtained by fitting them with quadratic spline functions accounting for anharmonicity. (d) Carrier capture coefficients  (including both radiative and nonradiative capture processes) at different charge states and associated effective capture coefficient ($C_{\mathrm{eff}}$) with respect to quasi-Fermi level splitting at different majority carrier types (equilibrium Fermi levels denoted by the color scale) at 300 K. (e) Photovoltaic device efficiency ($\mathrm{\eta}$) with respect to the equilibrium Fermi level position at different defect concentrations at 300 K and 500 nm device thickness.}
	\label{Fig6}
\end{figure*}

\begin{figure*}[t!]
	\centering
	\includegraphics[scale=1.0]{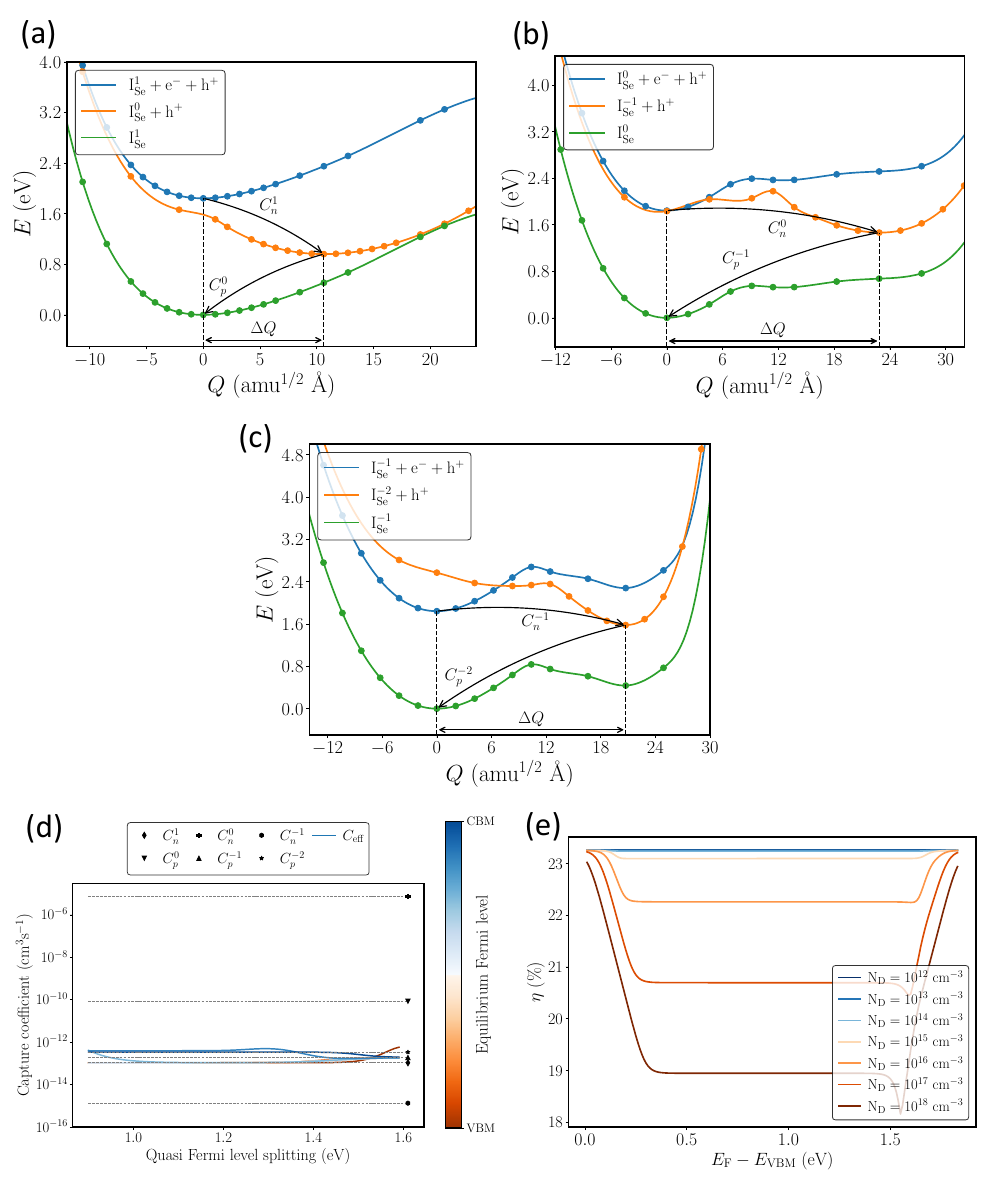}
	\caption{(a)-(c) One-dimensional configuration coordinate diagrams for the $1/0$, $0/-1$, $-1/-2$ charge state transitions of the I$_{\mathrm{Se}}$ defect in t-Se, respectively. Solid circles denote the data points calculated using HSE06 functional and solid lines are obtained by fitting them with quadratic spline functions accounting for anharmonicity. (d) Carrier capture coefficients  (including both radiative and nonradiative capture processes) at different charge states and associated effective capture coefficient ($C_{\mathrm{eff}}$) with respect to quasi-Fermi level splitting at different majority carrier types (equilibrium Fermi levels denoted by the color scale) at 300 K. (e) Photovoltaic device efficiency ($\mathrm{\eta}$) with respect to the equilibrium Fermi level position at different defect concentrations at 300 K and 500 nm device thickness.}
	\label{FigS14}
\end{figure*}

\begin{figure*}[t!]
	\centering
	\includegraphics[scale=1.0]{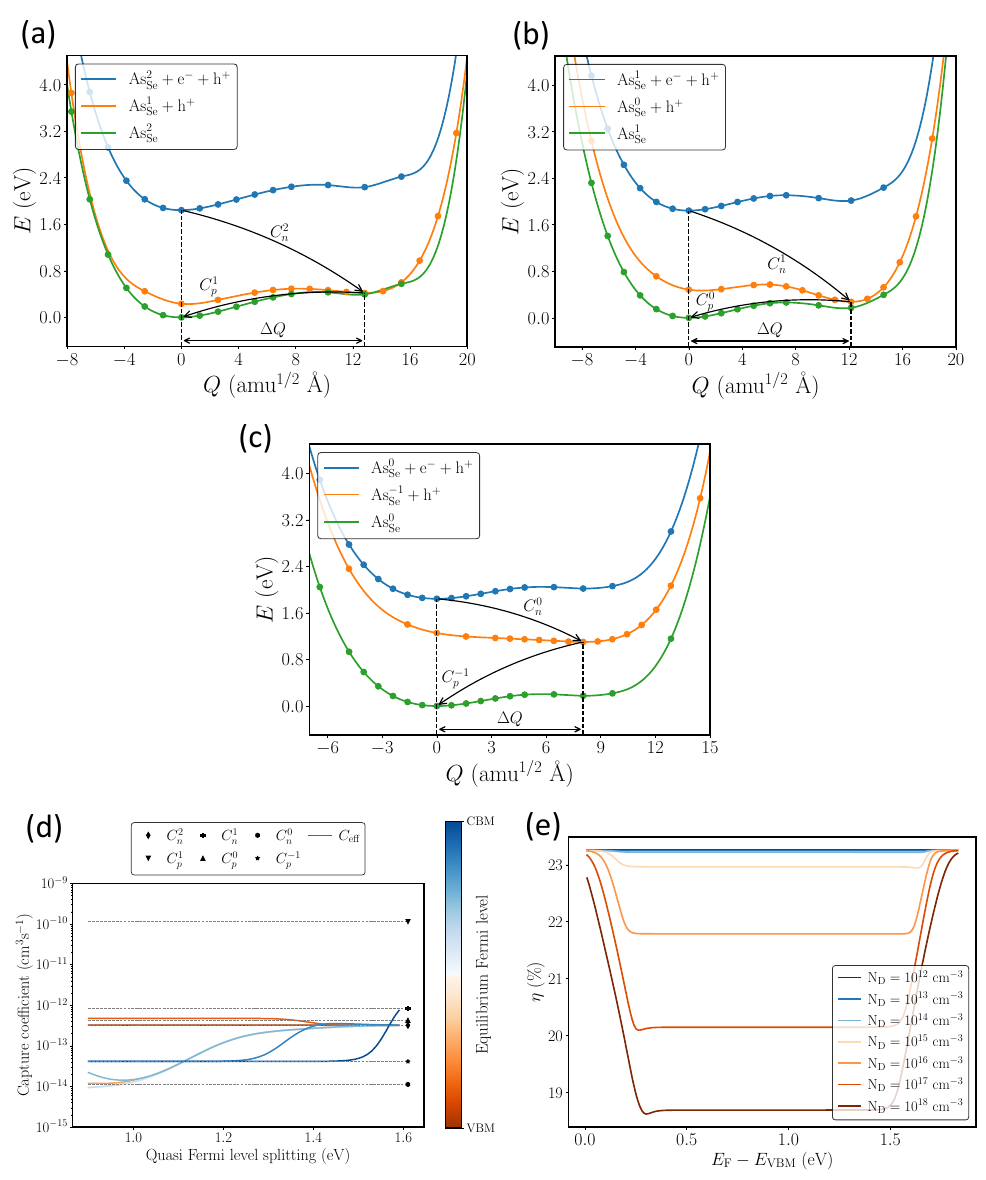}
	\caption{(a)-(c) One-dimensional configuration coordinate diagrams for the $2/1$, $1/0$, $0/-1$ charge state transitions of the As$_{\mathrm{Se}}$ defect in t-Se, respectively. Solid circles denote the data points calculated using HSE06 functional and solid lines are obtained by fitting them with quadratic spline functions accounting for anharmonicity. (d) Carrier capture coefficients  (including both radiative and nonradiative capture processes) at different charge states and associated effective capture coefficient ($C_{\mathrm{eff}}$) with respect to quasi-Fermi level splitting at different majority carrier types (equilibrium Fermi levels denoted by the color scale) at 300 K. (e) Photovoltaic device efficiency ($\mathrm{\eta}$) with respect to the equilibrium Fermi level position at different defect concentrations at 300 K and 500 nm device thickness.}
	\label{FigS15}
\end{figure*}

\begin{figure*}[t!]
	\centering
	\includegraphics[scale=1.0]{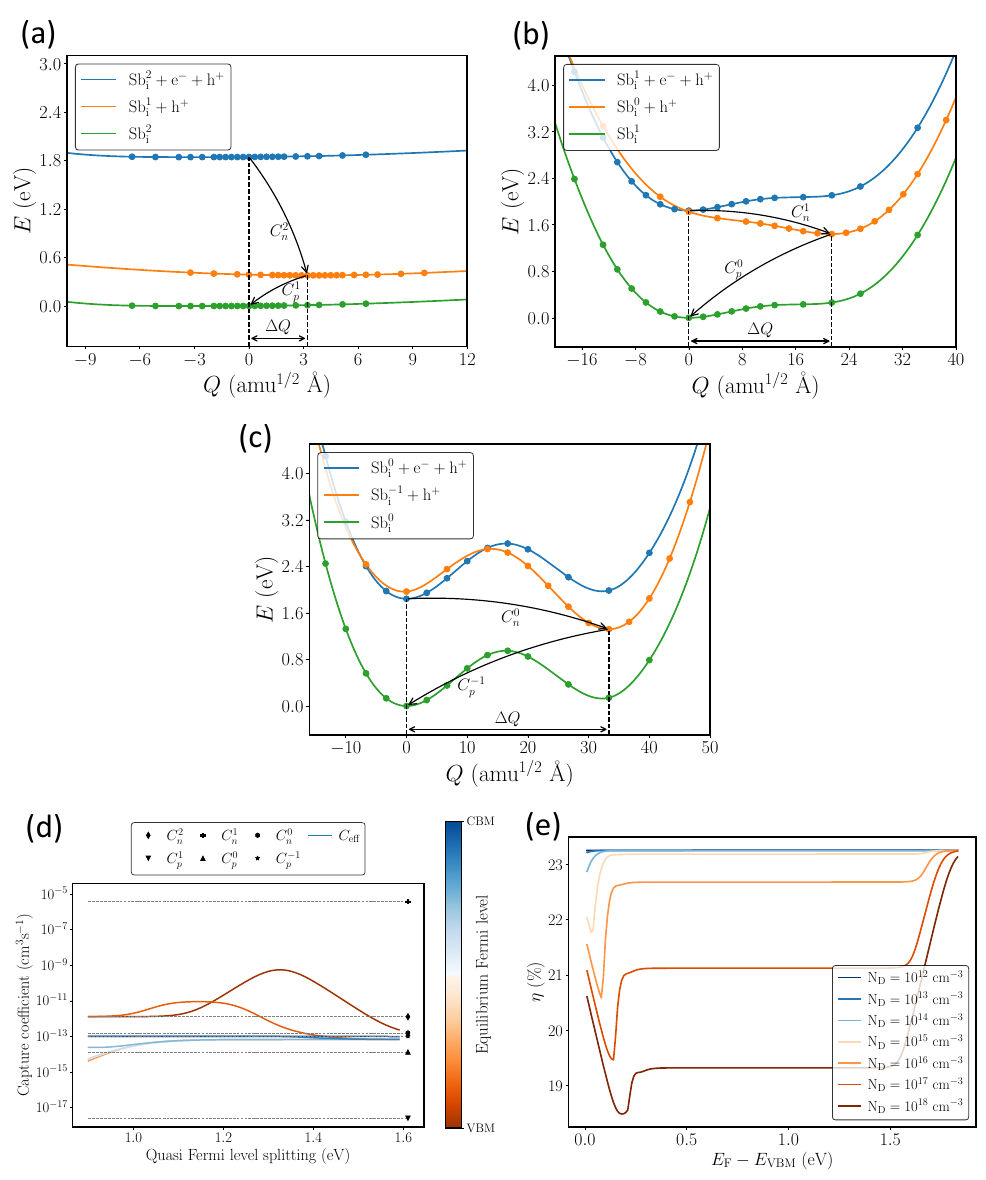}
	\caption{(a)-(c) One-dimensional configuration coordinate diagrams for the $2/1$, $1/0$, $0/-1$ charge state transitions of the Sb$_{\mathrm{i}}$ defect in t-Se, respectively. Solid circles denote the data points calculated using HSE06 functional and solid lines are obtained by fitting them with quadratic spline functions accounting for anharmonicity. (d) Carrier capture coefficients  (including both radiative and nonradiative capture processes) at different charge states and associated effective capture coefficient ($C_{\mathrm{eff}}$) with respect to quasi-Fermi level splitting at different majority carrier types (equilibrium Fermi levels denoted by the color scale) at 300 K. (e) Photovoltaic device efficiency ($\mathrm{\eta}$) with respect to the equilibrium Fermi level position at different defect concentrations at 300 K and 500 nm device thickness.}
	\label{FigS16}
\end{figure*}

\begin{figure*}[t!]
	\centering
	\includegraphics[scale=1.0]{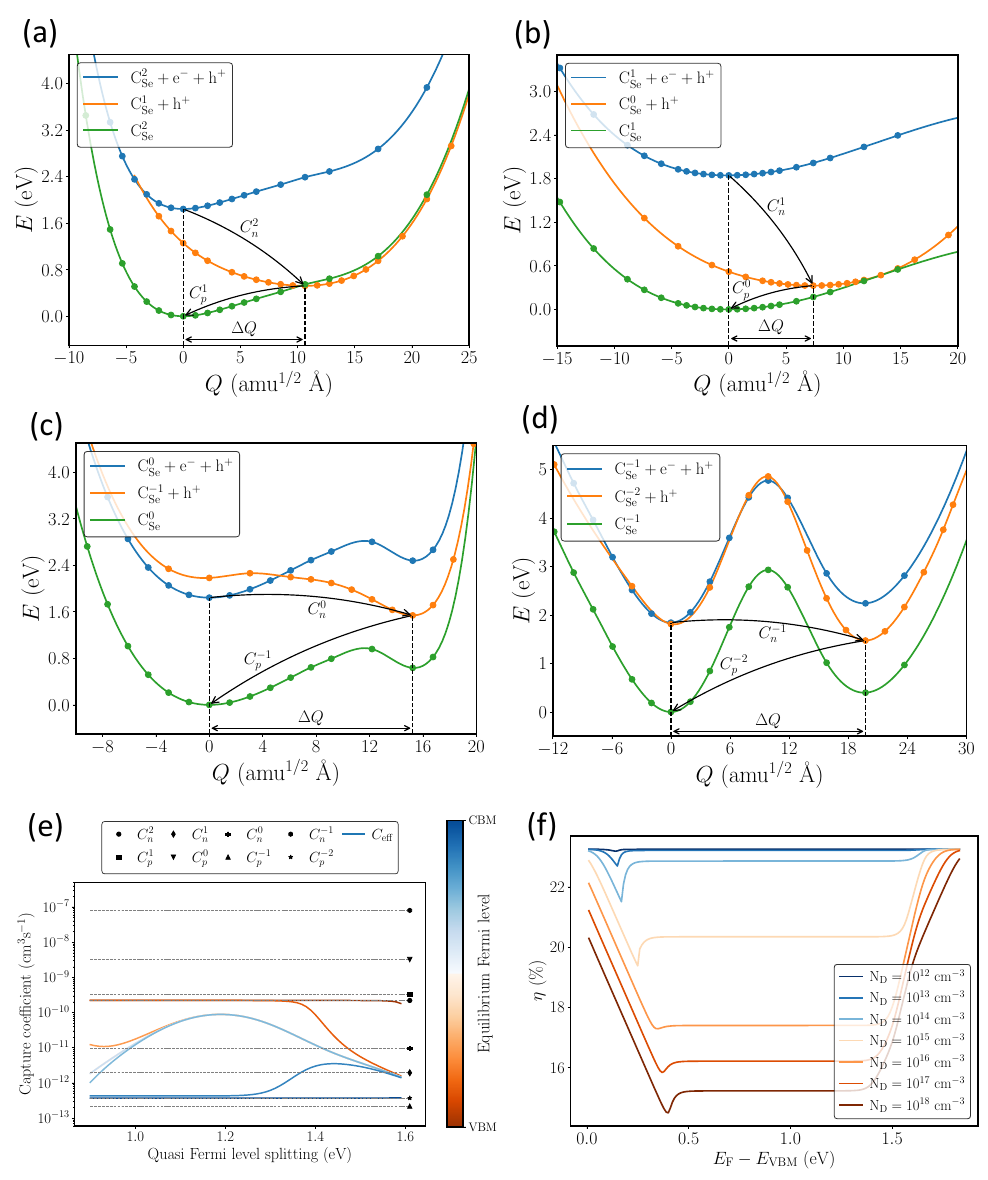}
	\caption{(a)-(d) One-dimensional configuration coordinate diagrams for the $2/1$, $1/0$, $0/-1$, $-1/-2$ charge state transitions of the C$_{\mathrm{Se}}$ defect in t-Se, respectively. Solid circles denote the data points calculated using HSE06 functional and solid lines are obtained by fitting them with quadratic spline functions accounting for anharmonicity. (e) Carrier capture coefficients  (including both radiative and nonradiative capture processes) at different charge states and associated effective capture coefficient ($C_{\mathrm{eff}}$) with respect to quasi-Fermi level splitting at different majority carrier types (equilibrium Fermi levels denoted by the color scale) at 300 K. (f) Photovoltaic device efficiency ($\mathrm{\eta}$) with respect to the equilibrium Fermi level position at different defect concentrations at 300 K and 500 nm device thickness.}
	\label{FigS17}
\end{figure*}

\begin{figure*}[t!]
	\centering
	\includegraphics[scale=1.0]{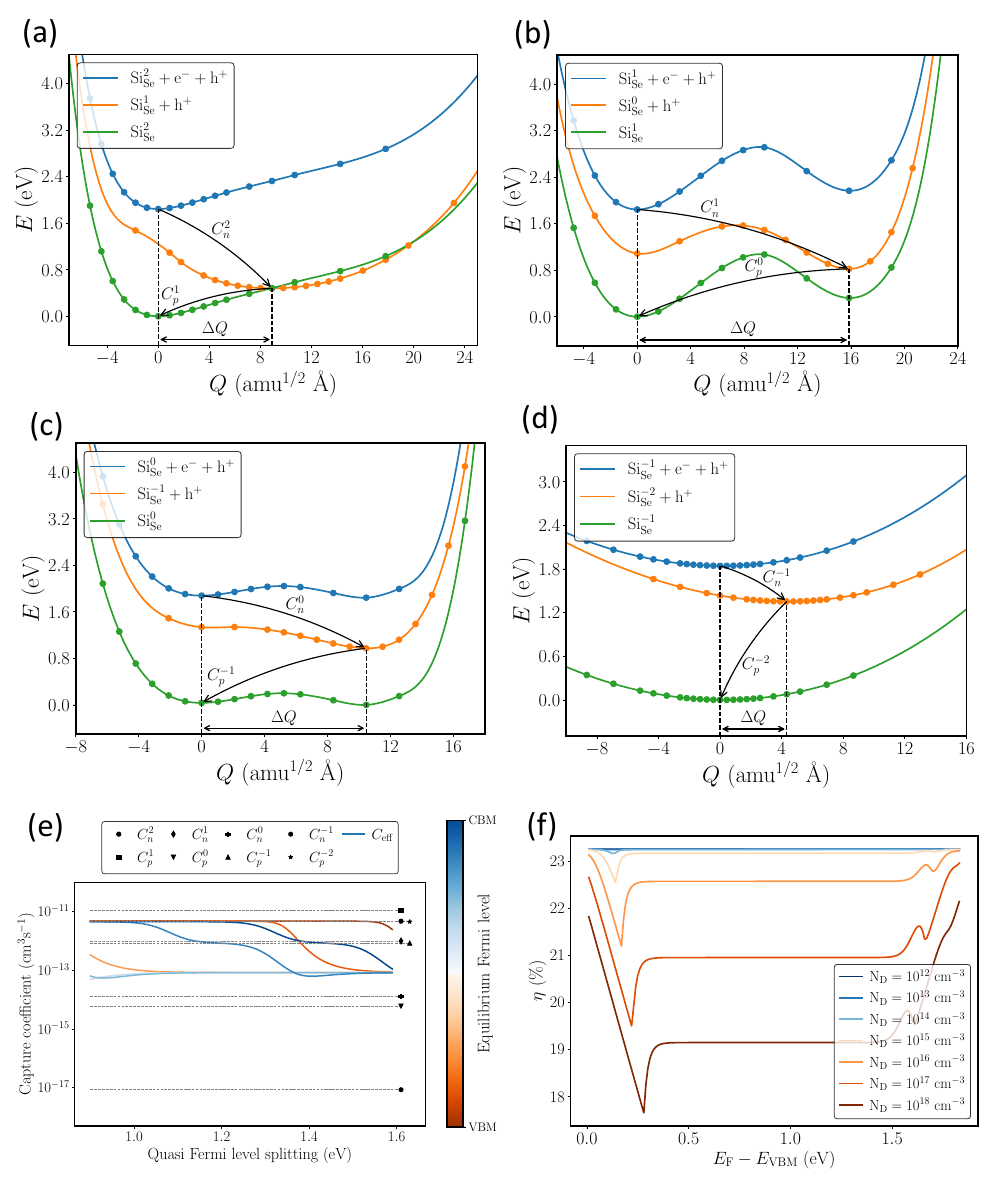}
	\caption{(a)-(d) One-dimensional configuration coordinate diagrams for the $2/1$, $1/0$, $0/-1$, $-1/-2$ charge state transitions of the Si$_{\mathrm{Se}}$ defect in t-Se, respectively. Solid circles denote the data points calculated using HSE06 functional and solid lines are obtained by fitting them with quadratic spline functions accounting for anharmonicity. (e) Carrier capture coefficients  (including both radiative and nonradiative capture processes) at different charge states and associated effective capture coefficient ($C_{\mathrm{eff}}$) with respect to quasi-Fermi level splitting at different majority carrier types (equilibrium Fermi levels denoted by the color scale) at 300 K. (f) Photovoltaic device efficiency ($\mathrm{\eta}$) with respect to the equilibrium Fermi level position at different defect concentrations at 300 K and 500 nm device thickness.}
	\label{Fig7}
\end{figure*}

\begin{figure*}[t!]
	\centering
	\includegraphics[scale=1.0]{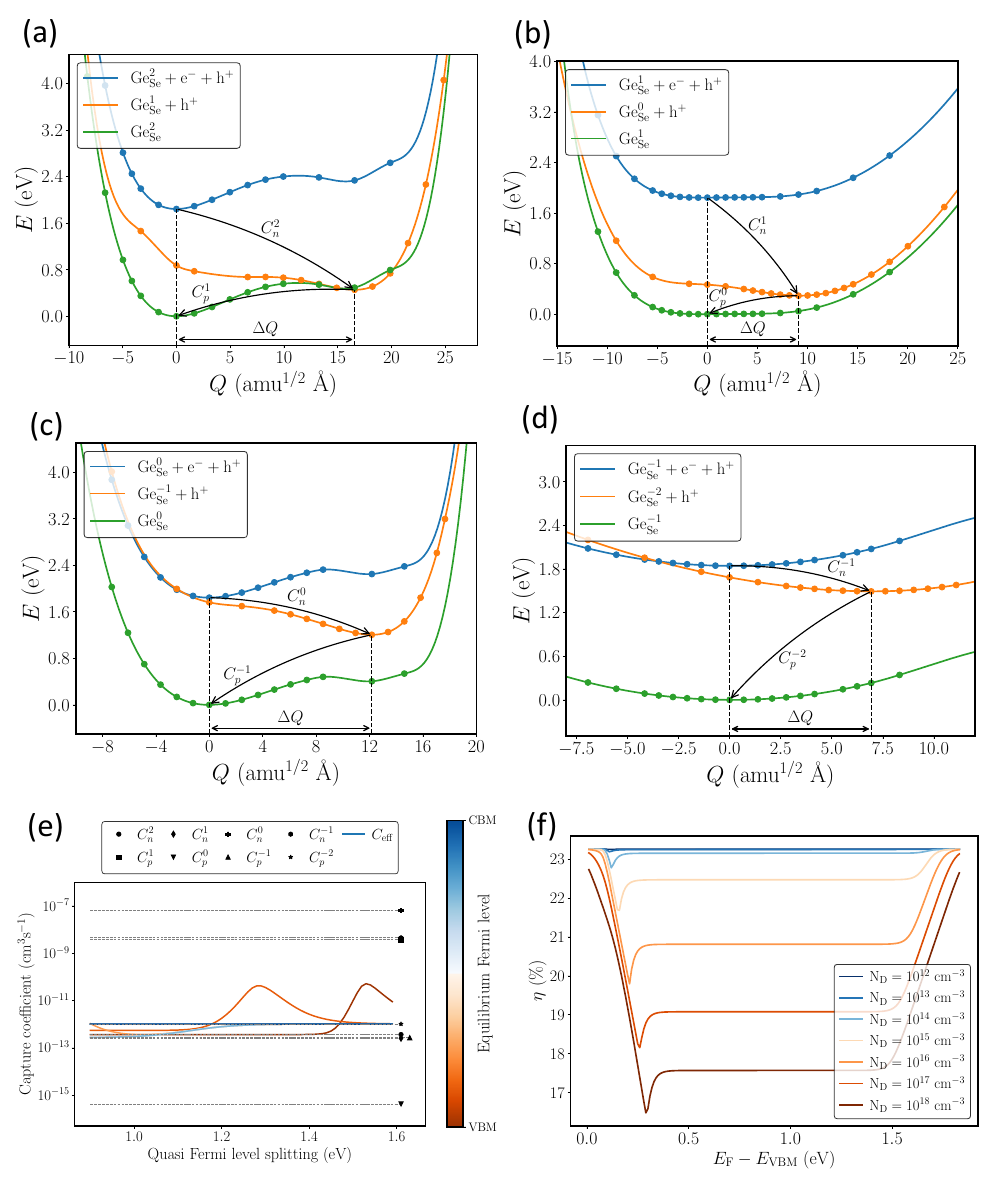}
	\caption{(a)-(d) One-dimensional configuration coordinate diagrams for the $2/1$, $1/0$, $0/-1$, $-1/-2$ charge state transitions of the Ge$_{\mathrm{Se}}$ defect in t-Se, respectively. Solid circles denote the data points calculated using HSE06 functional and solid lines are obtained by fitting them with quadratic spline functions accounting for anharmonicity. (e) Carrier capture coefficients  (including both radiative and nonradiative capture processes) at different charge states and associated effective capture coefficient ($C_{\mathrm{eff}}$) with respect to quasi-Fermi level splitting at different majority carrier types (equilibrium Fermi levels denoted by the color scale) at 300 K. (f) Photovoltaic device efficiency ($\mathrm{\eta}$) with respect to the equilibrium Fermi level position at different defect concentrations at 300 K and 500 nm device thickness.}
	\label{FigS18}
\end{figure*}

\begin{figure*}[t!]
	\centering
	\includegraphics[scale=1.1]{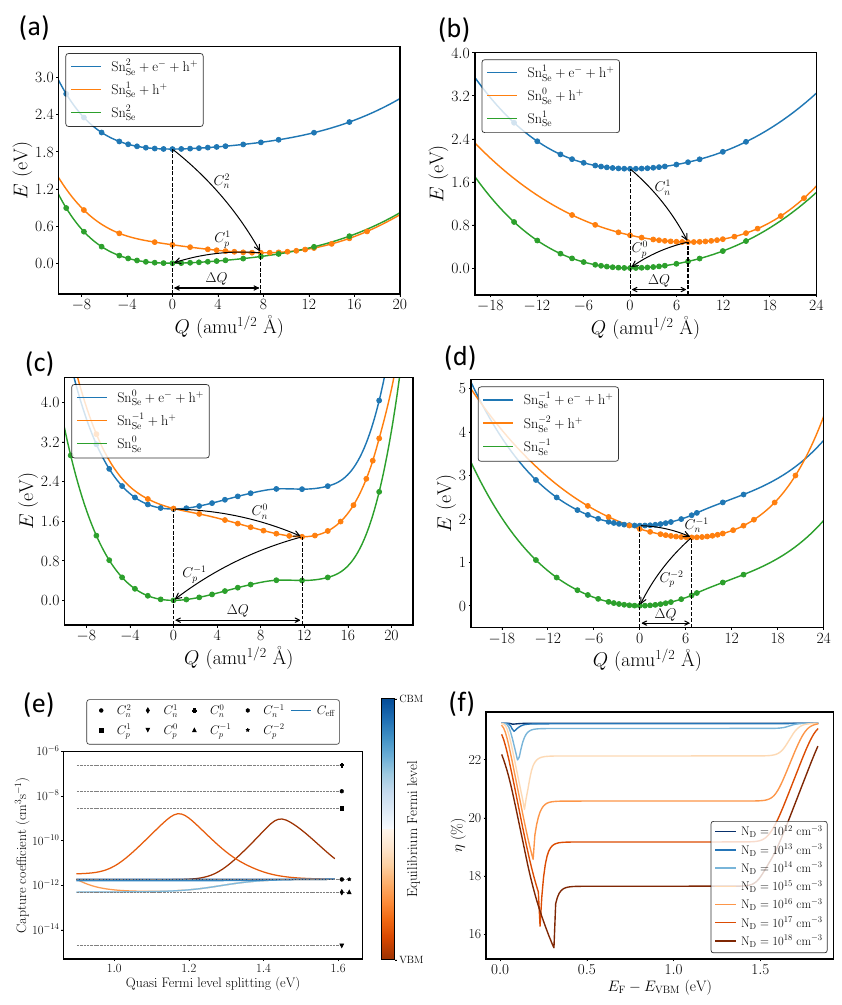}
	\caption{(a)-(d) One-dimensional configuration coordinate diagrams for the $2/1$, $1/0$, $0/-1$, $-1/-2$ charge state transitions of the Sn$_{\mathrm{Se}}$ defect in t-Se, respectively. Solid circles denote the data points calculated using HSE06 functional and solid lines are obtained by fitting them with quadratic spline functions accounting for anharmonicity. (e) Carrier capture coefficients  (including both radiative and nonradiative capture processes) at different charge states and associated effective capture coefficient ($C_{\mathrm{eff}}$) with respect to quasi-Fermi level splitting at different majority carrier types (equilibrium Fermi levels denoted by the color scale) at 300 K. (f) Photovoltaic device efficiency ($\mathrm{\eta}$) with respect to the equilibrium Fermi level position at different defect concentrations at 300 K and 500 nm device thickness.}
	\label{FigS19}
\end{figure*}

\clearpage

\subsection*{C. Photovoltaic device parameters}\label{pvdev}

\begin{figure*}[h!]
	\centering
	\includegraphics[scale=0.89]{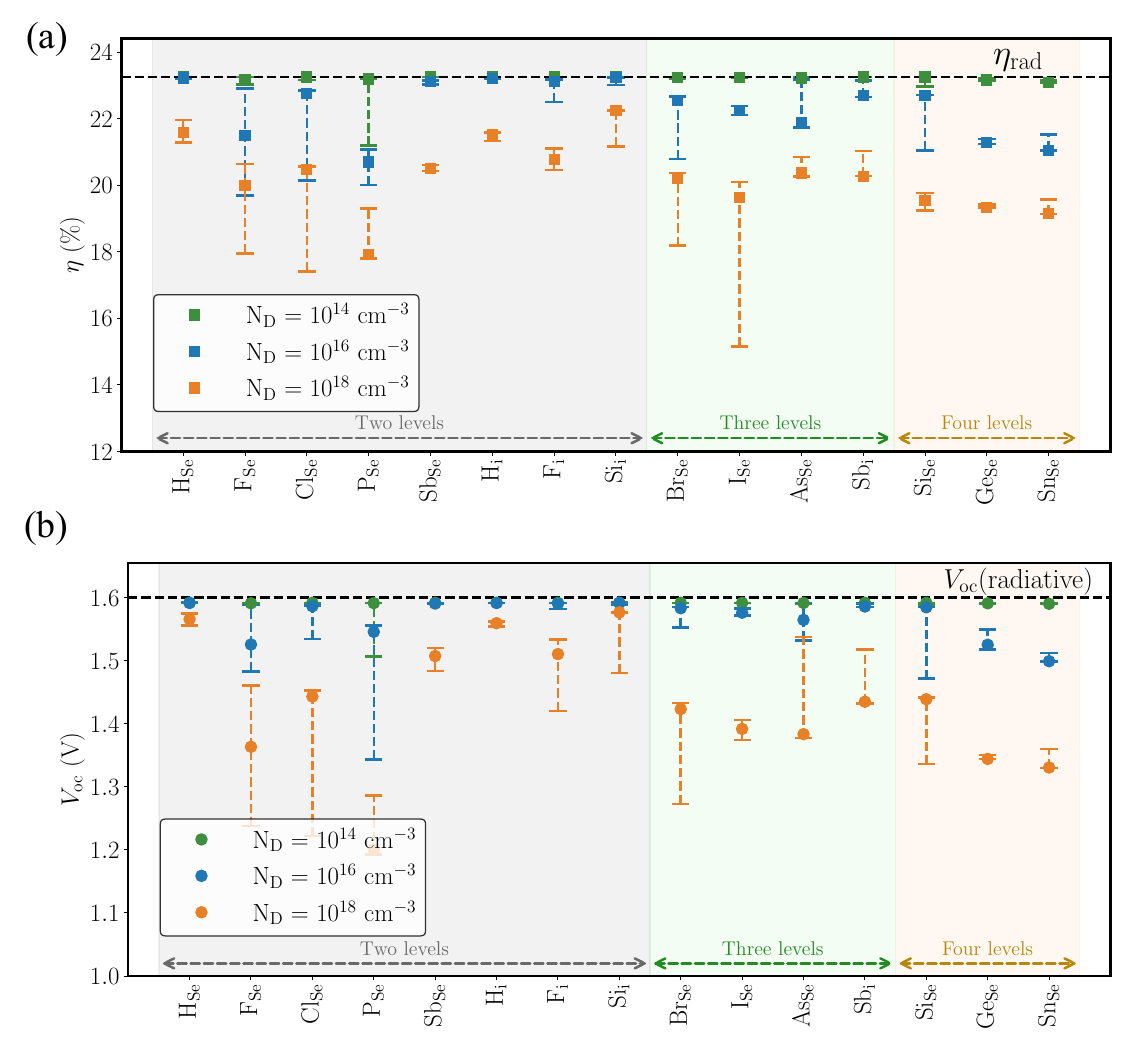}
	\caption{(a) Photovoltaic device efficiency $\eta$ for Se, accounting for the SRH recombination at considered defects, at different defect concentrations (10$^{14}$, 10$^{16}$, and 10$^{18}$ cm$^{-3}$) and at n-type majority carrier concentration (1.35x10$^{16}$ cm$^{-3}$) (initial equilibrium Fermi level at 0.23 eV below CBM). (b) Photovoltaic open-circuit voltage ($V_{\mathrm{oc}}$) for the same as above. Defects which can have defect concentration 10$^{14}$-10$^{18}$ cm$^{-3}$, based on defect formation energies and equilibrium Fermi level, at possible different chemical growth environments, are considered and categorized based on number of charge transition levels inside the band gap. PV device parameters are plotted at 300 K and 500 nm film thickness. Here we have considered a tolerance of 0.2 eV in the defect charge transition levels. The corresponding carrier capture coefficients are calculated by rigid shifting the CC curves at the considered levels. Associated device parameters are shown as error bars.}
	\label{FigS20}
\end{figure*}

\begin{figure*}[t!]
	\centering
	\includegraphics[scale=0.89]{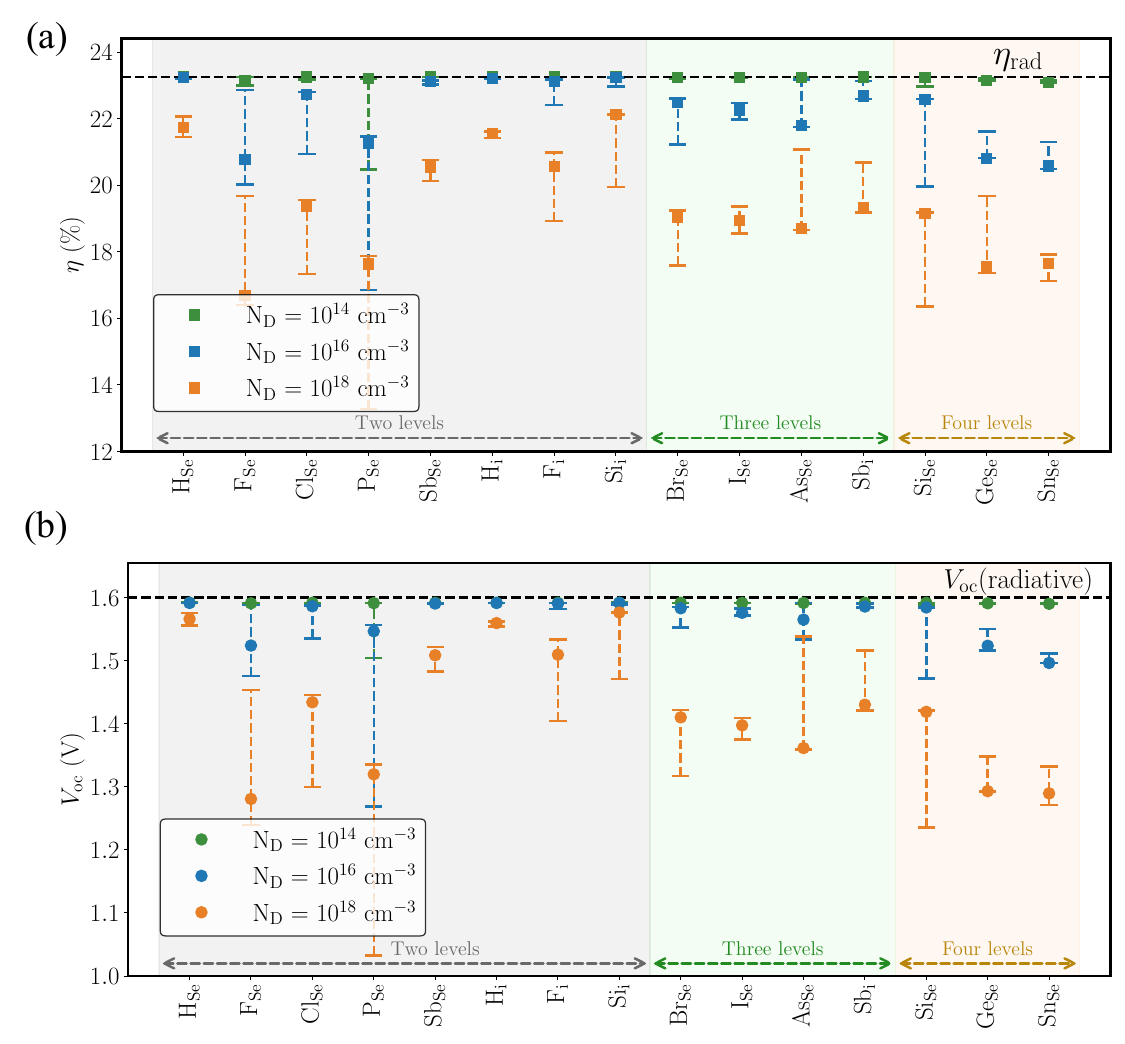}
	\caption{(a) Photovoltaic device efficiency $\eta$ for Se, accounting for the SRH recombination at considered defects, at different defect concentrations (10$^{14}$, 10$^{16}$, and 10$^{18}$ cm$^{-3}$) and at midgap (intrinsic) carrier concentration (initial equilibrium Fermi level at 0.91 eV above VBM). (b) Photovoltaic open-circuit voltage ($V_{\mathrm{oc}}$) for the same as above. Defects which can have defect concentration 10$^{14}$-10$^{18}$ cm$^{-3}$, based on defect formation energies and equilibrium Fermi level, at possible different chemical growth environments, are considered and categorized based on number of charge transition levels inside the band gap. PV device parameters are plotted at 300 K and 500 nm film thickness. Here we have considered a tolerance of 0.2 eV in the defect charge transition levels. The corresponding carrier capture coefficients are calculated by rigid shifting the CC curves at the considered levels. Associated device parameters are shown as error bars.}
	\label{FigS21}
\end{figure*}

\clearpage

\subsection*{D. Trends in nonradiative capture coefficients for different defect types}

\begin{figure*}[h!]
	\centering
	\includegraphics[scale=0.74]{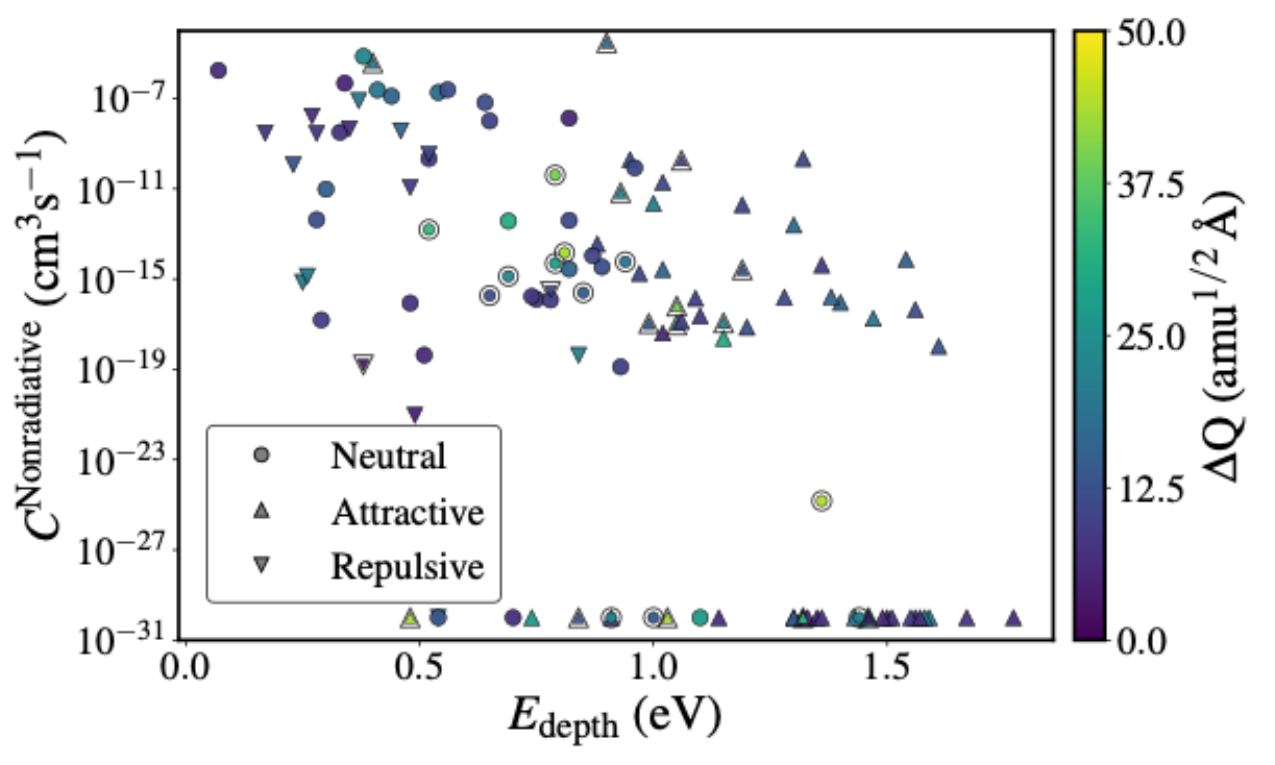}
	\caption{Nonradiative capture coefficients for all the defects are plotted with respect to distance from the band edges (for electron (e$^-$) capture it is from CBM, and for hole (h$^+$) capture, it is from VBM). Data is distributed as as neutral, attractive and repulsive centres, e.g. D$^{-1}$ will be repulsive for e$^-$, but attractive for h$^+$. All the value below 10$^{-30}$ is assumed as negligible and plotted as 10$^{-30}$. Color bar represents the configurational coordinate distance from the initial and final geometry of the defect during a capture. Interstitial defects are distinguished by an outer boundary (double-circled and triangular symbols).}
	\label{FigS22}
\end{figure*}

\clearpage

\subsection*{E. Chemical potential limits, defect capture parameters, and defect-limited photovoltaic performance}

\begin{table*}[h]
\caption{\label{tab:tables1}
Allowed chemical potential region for extrinsic defects and associated secondary phases considered. Here $\mu_{D}$= $E_{D}^{el}$$+$$\Delta \mu_{D}$, where, $\mu_{D}$ is the total chemical potential of element D, and $E_{D}^{el}$ is the total energy of D at its elemental phases. [All data are calculated using HSE06 functional]}
\begin{ruledtabular}
\begin{tabular}{c|c|c|c}
\textbf{Extrinsic Dopant (D)} & $\boldsymbol{\Delta \mu}_{\bf D}^{\bf rich}$ & $\boldsymbol{\Delta \mu}_{\bf D}^{\bf poor}$ & \textbf{Secondary phases} \\[5.2pt]
& (eV) & (eV)  &  \\[5.2pt]
\hline
H & 0.0 & -5.04 & H$_2$Se  \\[5.2pt]
F & 0.0 & -2.34 & SeF$_4$  \\[5.2pt]
Cl & 0.0 & -0.43 & SeCl, SeCl$_4$  \\[5.2pt]
Br & 0.0 & -0.23 & Se$_3$Br, SeBr, SeBr$_4$  \\[5.2pt]
I & 0.0 & 0.0 & SeI$_2$  \\[5.2pt]
O & 0.0 & -1.47 & SeO$_2$, Se$_2$O$_5$, SeO$_3$, SeO$_4$\\[5.2pt]
S & 0.0 & 0.0 & SeS, SeS$_7$  \\[5.2pt]
Te & 0.0 & 0.0 & TeSe, Te$_3$Se  \\[5.2pt]
N & 0.0 & 0.0 & SeN  \\[5.2pt]
P & 0.0 & -0.22 & PSe, P$_2$Se$_5$, P$_4$Se$_5$, P$_4$Se$_3$  \\[5.2pt]
As & 0.0 & -0.34 & AsSe, As$_2$Se$_3$, As$_4$Se$_3$  \\[5.2pt]
Sb & 0.0 & -0.64 & Sb$_2$Se$_3$  \\[5.2pt]
C & 0.0 & 0.0 & CSe$_2$  \\[5.2pt]
Si & 0.0 & -1.66 & SiSe$_2$  \\[5.2pt]
Ge & 0.0 & -1.09 & Ge$_4$Se$_9$, GeSe$_2$, GeSe  \\[5.2pt]
Sn & 0.0 & -1.08 & SnSe$_2$, SnSe  \\[5.2pt]
\end{tabular}
\end{ruledtabular}
\end{table*}

\clearpage

\begin{longtable*}{c@{\hspace{0.5cm}}c@{\hspace{0.7cm}}c@{\hspace{0.7cm}}c@{\hspace{0.7cm}}c@{\hspace{0.7cm}}c@{\hspace{0.7cm}}c@{\hspace{0.7cm}}c@{\hspace{0.7cm}}c}
\caption{Detailed data about SRH capture\label{tab:tables2}
Key parameters describing defects explored in t-Se. These include: The charge transition levels (CTLs), $\Delta Q$, Electron-phonon coupling ($W_{if}$) The carrier capture coefficients for electron and hole capture in different charge states ($C_{n/p}^q$), The semiclassical capture energy barrier ($\Delta E^\ddagger$) related to defect assisted carrier capture and carrier capture coefficients for associated defects. The Total capture coefficient were calculated by adding the radiative and non-radiative contributions.} \\
\hline
\textbf{Defect (D)} & \textbf{CTL (\boldsymbol{$q_1/q_2$})} & $\boldsymbol{\Delta Q}$ & \textbf{Capture} & $\boldsymbol{E_b}$  & $\boldsymbol{W_{if}}$ & $\boldsymbol{C^\mathrm{Nonrad}}$ & $\boldsymbol{C^\mathrm{Rad}}$ & $\boldsymbol{C^\mathrm{tot}}$ \\
 & (eV) & (amu$^{1/2}$\AA) & & (eV) &  & (cm$^3$s$^{-1}$) &(cm$^3$s$^{-1}$) & (cm$^3$s$^{-1}$) \\
\hline
\endfirsthead

\multicolumn{9}{c}{{\bfseries Table \thetable\ continued from previous page}} \\
\hline
\textbf{Defect (D)} & \textbf{CTL ($q_1/q_2$)} & $\boldsymbol{\Delta Q}$ & \textbf{Capture} & $\boldsymbol{E_b}$ & $\boldsymbol{W_{if}}$ & $\boldsymbol{C^\mathrm{Nonrad}}$ & $\boldsymbol{C^\mathrm{Rad}}$ & $\boldsymbol{C^\mathrm{tot}}$ \\
 & (eV) & (amu$^{1/2}$\AA) & & (eV) & & (cm$^3$s$^{-1}$) & (cm$^3$s$^{-1}$) & (cm$^3$s$^{-1}$) \\
\hline
\endhead

\hline \multicolumn{9}{r}{{Continued on next page}} \\ \hline
\endfoot

\hline \hline
\endlastfoot

v$_\mathrm{Se}$ & (2/1)=0.84 & 21.86 & $C_n^2$ & 0.19 & 0.04 & 2.26e-12 & 3.25e-13 & 2.58e-12  \\[5.2pt]
v$_\mathrm{Se}$ &  & & $C_p^1$ & 0.14 & 0.02 & 4.18e-19 & 9.56e-17 & 9.60e-17  \\[5.2pt]
v$_\mathrm{Se}$ & (1/0)=0.07 & 6.54 & $C_n^1$ & $>$5 & 0.005 & 8.13e-127 & 8.86e-13 & 8.86e-13  \\[5.2pt]
v$_\mathrm{Se}$ &  & & $C_p^0$ & 0.002 & 0.10 & 1.75e-06 & 5.62e-18 & 1.75e-06  \\[5.2pt]
v$_\mathrm{Se}$ & (0/-1)=0.91 & 9.30 & $C_n^0$ & $>$5 & 0.03 & 1.29e-19 & 1.72e-15 & 1.72e-15  \\[5.2pt]
v$_\mathrm{Se}$ &  & & $C_p^{-1}$ & $>$5 & 0.13 & 2.41e-33 & 1.98e-13 & 1.98e-13  \\[5.2pt]
v$_\mathrm{Se}$ & (-1/-2)=1.56 & 8.18 & $C_n^{-1}$ & 0.003 & 0.04 & 2.81e-09 & 1.12e-18 & 2.81e-09  \\[5.2pt]
v$_\mathrm{Se}$ &  & & $C_p^{-2}$ & $>$5 & 0.12 & 2.77e-276 & 1.98e-12 & 1.98e-12  \\[5.2pt]
Se$_\mathrm{i(1)}$ &(1/0)=1.10 & 27.31 & $C_n^1$ & 2.19 & 0.14 & 8.13e-127 & 5.93e-14 & 5.93e-14  \\[5.2pt]
Se$_\mathrm{i(1)}$ & & & $C_p^0$ & $>$5 & 0.19 & 1.00e-127 & 3.78e-14 & 3.78e-14  \\[5.2pt]
Se$_\mathrm{i(1)}$ &(0/-1)=1.15 & 30.49 & $C_n^0$ & 0.47 & 0.14 & 3.73e-13 & 5.92e-15 & 3.79e-13   \\[5.2pt]
Se$_\mathrm{i(1)}$ & & & $C_p^{-1}$ & 2.75 & 0.19 & 2.41e-18 & 4.30e-13 & 4.30e-13  \\[5.2pt]
Se$_\mathrm{i(1)}$ &(-1/-2)=1.30 & 21.91 & $C_n^{-1}$ & 0.0003 & 0.03 & 2.21e-129 & 6.31e-17 & 6.31e-17   \\[5.2pt]
Se$_\mathrm{i(1)}$ & & & $C_p^{-2}$ & $>$5 & 0.18 & 1.99e-126 & 1.24e-12 & 1.24e-12  \\[5.2pt]
H$_\mathrm{Se}$ & (1/0)=0.54 & 13.40 & $C_n^1$ & 4.64 & 0.08 & 8.13e-127 & 1.30e-13 & 1.30e-13   \\[5.2pt]
H$_\mathrm{Se}$ & & & $C_p^0$ & 0.007 & 0.10 & 1.00e-127 & 6.09e-16 & 6.09e-16  \\[5.2pt]
H$_\mathrm{Se}$ & (0/-1)=1.14 & 6.57 & $C_n^0$ & 2.77 & 0.02 & 1.00e-127 & 2.59e-15 & 2.59e-15  \\[5.2pt]
H$_\mathrm{Se}$ & & & $C_p^{-1}$ & 2.72 & 0.06 & 9.95e-127 & 5.50e-14 & 5.50e-14  \\[5.2pt]
H$_\mathrm{i}$ &(1/0)=1.00 & 13.60 & $C_n^1$ & 4.26 & 0.09 & 8.13e-127 & 1.51e-14 & 1.51e-14   \\[5.2pt]
H$_\mathrm{i}$ & & & $C_p^0$ & 2.85 & 0.04 & 1.00e-127 & 2.77e-15 & 2.77e-15  \\[5.2pt]
H$_\mathrm{i}$ &(0/-1)=0.99 & 14.40 & $C_n^0$ & 3.62 & 0.04 & 2.49e-16 & 1.92e-15 & 2.17e-15  \\[5.2pt]
H$_\mathrm{i}$ & & & $C_p^{-1}$ & $>$5 & 0.03 & 1.08e-17 & 2.68e-14 & 2.68e-14  \\[5.2pt]
F$_\mathrm{Se}$ &(1/0)=0.65 & 10.23 & $C_n^1$ & 1.58 & 0.18 & 1.98e-12 & 4.93e-14 & 2.03e-12   \\[5.2pt]
F$_\mathrm{Se}$ & & & $C_p^0$ & $>$5 & 0.12 & 1.02e-08 & 5.06e-16 & 1.02e-08   \\[5.2pt]
F$_\mathrm{Se}$ &(0/-1)=1.32 & 8.47 & $C_n^0$ & 0.26 & 0.02 & 2.18e-10 & 5.27e-16 & 2.18e-10   \\[5.2pt]
F$_\mathrm{Se}$ & & & $C_p^{-1}$ & $>$5 & 0.03 & 5.43e-38 & 4.12e-14 & 4.12e-14  \\[5.2pt]
F$_\mathrm{i}$ &(1/0)=0.94 & 18.92 & $C_n^1$ & 0.0007 & 0.20 & 2.98e-05 & 2.04e-14 & 2.98e-05  \\[5.2pt]
F$_\mathrm{i}$ & & & $C_p^0$ & $>$5 & 0.11 & 5.80e-15 & 3.35e-15 & 9.15e-15   \\[5.2pt]
F$_\mathrm{i}$ &(0/-1)=1.05 & 28.54 & $C_n^0$ & $>$5 & 0.33 & 4.99e-15 & 1.70e-15 & 6.69e-15   \\[5.2pt]
F$_\mathrm{i}$ & & & $C_p^{-1}$ & 3.81 & 0.13 & 9.38e-18 & 4.64e-14 & 4.64e-14  \\[5.2pt]
Cl$_\mathrm{Se}$ &(1/0)=0.82 & 11.12 & $C_n^1$ & $>$5 & 0.22 & 1.93e-11 & 3.44e-14 & 1.93e-11   \\[5.2pt]
Cl$_\mathrm{Se}$ & & & $C_p^0$ & 1.05 & 0.13 & 3.98e-13 & 1.28e-15 & 3.99e-13  \\[5.2pt]
Cl$_\mathrm{Se}$ &(0/-1)=1.40 & 16.01 & $C_n^0$ & 0.10 & 0.11 & 1.27e-07 & 3.43e-16 & 1.27e-07  \\[5.2pt]
Cl$_\mathrm{Se}$ & & & $C_p^{-1}$ & $>$5 & 0.12 & 9.34e-17 & 6.25e-14 & 6.26e-14  \\[5.2pt]
Cl$_\mathrm{i}$ &(1/0)=0.79 & 39.49 & $C_n^1$ & 4.63 & 0.12 & 6.40e-17 & 9.27e-14 & 9.27e-14   \\[5.2pt]
Cl$_\mathrm{i}$ & & & $C_p^0$ & $>$5 & 0.24 & 4.04e-11 & 6.75e-15 & 4.04e-11  \\[5.2pt]
Cl$_\mathrm{i}$ &(0/-1)=1.15 & 23.14 & $C_n^0$ & $>$5 & 0.24 & 1.33e-15 & 3.24e-15 & 4.57e-15  \\[5.2pt]
Cl$_\mathrm{i}$ & & & $C_p^{-1}$ & $>$5 & 0.19 & 1.16e-17 & 2.07e-13 & 2.07e-13  \\[5.2pt]
Br$_\mathrm{Se}$ &(1/0)=0.89 & 12.04 & $C_n^1$ & 4.29 & 0.27 & 1.96e-10 & 2.93e-14 & 1.96e-10   \\[5.2pt]
Br$_\mathrm{Se}$ & & & $C_p^0$ & 2.25 & 0.14 & 3.42e-15 & 2.39e-15 & 5.81e-15   \\[5.2pt]
Br$_\mathrm{Se}$ &(0/-1)=1.43 & 19.77 & $C_n^0$ & $>$5 & 0.12 & 2.39e-07 & 2.94e-16 & 2.39e-07   \\[5.2pt]
Br$_\mathrm{Se}$ & & & $C_p^{-1}$ & $>$5 & 0.14 & 4.78e-71 & 9.95e-14 & 9.95e-14   \\[5.2pt]
Br$_\mathrm{Se}$ &(-1/-2)=1.59 & 20.86 & $C_n^{-1}$ & 0.50 & 0.05 & 7.15e-16 & 1.03e-17 & 7.25e-16  \\[5.2pt]
Br$_\mathrm{Se}$ & & & $C_p^{-2}$ & $>$5 & 0.20 & 6.61e-163 & 3.98e-13 & 3.98e-13   \\[5.2pt]
I$_\mathrm{Se}$ &(1/0)=0.96 & 10.62 & $C_n^1$ & $>$5 & 0.04 & 3.62e-14 & 7.12e-14 & 1.07e-13   \\[5.2pt]
I$_\mathrm{Se}$ & & & $C_p^0$ & $>$5 & 0.16 & 8.47e-11 & 5.40e-15 & 8.47e-11  \\[5.2pt]
I$_\mathrm{Se}$ &(0/-1)=1.46 & 22.84 & $C_n^0$ & 0.004 & 0.20 & 7.36e-06 & 6.90e-16 & 7.36e-06   \\[5.2pt]
I$_\mathrm{Se}$ & & & $C_p^{-1}$ & $>$5 & 0.21 & 9.95e-127 & 1.90e-13 & 1.90e-13  \\[5.2pt]
I$_\mathrm{Se}$ &(-1/-2)=1.58 & 20.74 & $C_n^{-1}$ & 0.48 & 0.04 & 1.33e-15 & 1.53e-17 & 1.34e-15  \\[5.2pt]
I$_\mathrm{Se}$ & & & $C_p^{-2}$ & $>$5 & 0.16 & 1.99e-126 & 3.42e-13 & 3.42e-13  \\[5.2pt]
I$_\mathrm{i}$ &(1/0)=1.36 & 44.06 & $C_n^1$ & 0.04 & 0.09 & 3.70e-111 & 3.33e-15 & 3.33e-15   \\[5.2pt]
I$_\mathrm{i}$ & & & $C_p^0$ & 4.28 & 0.10 & 1.48e-25 & 8.00e-15 & 8.00e-15   \\[5.2pt]
I$_\mathrm{i}$ &(0/-1)=1.03 & 43.42 & $C_n^0$ & $>$5 & 0.15 & 1.47e-14 & 1.96e-15 & 1.67e-14   \\[5.2pt]
I$_\mathrm{i}$ & & & $C_p^{-1}$ & 4.28 & 0.02 & 1.60e-35 & 3.46e-14 & 3.46e-14   \\[5.2pt]
N$_\mathrm{Se}$ &(1/0)=0.34 & 5.30 & $C_n^1$ & $>$5 & 0.03 & 2.12e-46 & 2.19e-13 & 2.19e-13   \\[5.2pt]
N$_\mathrm{Se}$ & & & $C_p^0$ & 0.03 & 0.07 & 4.77e-07 & 3.38e-16 & 4.77e-07   \\[5.2pt]
N$_\mathrm{Se}$ &(0/-1)=1.02 & 5.59 & $C_n^0$ & 0.14 & 0.05 & 1.35e-08 & 4.45e-15 & 1.35e-08   \\[5.2pt]
N$_\mathrm{Se}$ & & & $C_p^{-1}$ & 1.73 & 0.01 & 4.11e-18 & 9.38e-14 & 9.38e-14   \\[5.2pt]
P$_\mathrm{Se}$ &(1/0)=0.54 & 17.74 & $C_n^1$ & 3.10 & 0.17 & 2.52e-13 & 3.28e-13 & 5.80e-13  \\[5.2pt]
P$_\mathrm{Se}$ & & & $C_p^0$ & 0.19 & 0.16 & 1.86e-07 & 5.23e-16 & 1.86e-07   \\[5.2pt]
P$_\mathrm{Se}$ &(0/-1)=1.09 & 9.31 & $C_n^0$ & 3.94 & 0.14 & 1.28e-16 & 7.91e-15 & 8.04e-15   \\[5.2pt]
P$_\mathrm{Se}$ & & & $C_p^{-1}$ & $>$5 & 0.06 & 1.43e-16 & 4.26e-14 & 4.27e-14  \\[5.2pt]
As$_\mathrm{Se}$ &(2/1)=0.23 & 12.82 & $C_n^2$ & $>$5 & 0.04 & 1.06e-18 & 3.22e-13 & 3.22e-13   \\[5.2pt]
As$_\mathrm{Se}$ & & & $C_p^1$ & 0.20 & 0.02 & 1.15e-10 & 1.64e-17 & 1.15e-10   \\[5.2pt]
As$_\mathrm{Se}$ &(1/0)=0.28 & 12.15 & $C_n^1$ & $>$5 & 0.21 & 4.47e-17 & 8.45e-13 & 8.45e-13  \\[5.2pt]
As$_\mathrm{Se}$ & & & $C_p^0$ & $>$5 & 0.003 & 4.28e-13 & 6.46e-17 & 4.28e-13   \\[5.2pt]
As$_\mathrm{Se}$ &(0/-1)=1.10 & 8.03 & $C_n^0$ & $>$5 & 0.21 & 1.70e-16 & 1.11e-14 & 1.13e-14   \\[5.2pt]
As$_\mathrm{Se}$ & & & $C_p^{-1}$ & $>$5 & 0.04 & 2.34e-17 & 4.13e-14 & 4.13e-14   \\[5.2pt]
As$_\mathrm{i}$ &(1/0)=0.91 & 22.38 & $C_n^1$ & $>$5 & 0.09 & 6.82e-12 & 5.86e-14 & 6.88e-12   \\[5.2pt]
As$_\mathrm{i}$ & & & $C_p^0$ & $>$5 & 0.01 & 6.69e-31 & 4.88e-15 & 4.88e-15  \\[5.2pt]
Sb$_\mathrm{Se}$ &(1/0)=0.51 & 6.09 & $C_n^1$ & $>$5 & 0.01 & 2.89e-289 & 6.71e-13 & 6.71e-13   \\[5.2pt]
Sb$_\mathrm{Se}$ & & & $C_p^0$ & $>$5 & 0.01 & 4.30e-19 & 7.77e-16 & 7.77e-16   \\[5.2pt]
Sb$_\mathrm{Se}$ &(0/-1)=1.06 & 9.19 & $C_n^0$ & $>$5 & 0.12 & 1.19e-16 & 1.67e-14 & 1.68e-14   \\[5.2pt]
Sb$_\mathrm{Se}$ & & & $C_p^{-1}$ & $>$5 & 0.09 & 1.36e-17 & 6.94e-14 & 6.94e-14   \\[5.2pt]
Sb$_\mathrm{i}$ &(2/1)=0.38 & 3.21 & $C_n^2$ & $>$5 & 0.01 & 1.63e-126 & 1.28e-12 & 1.28e-12  \\[5.2pt]
Sb$_\mathrm{i}$ & & & $C_p^1$ & $>$5 & 0.01 & 1.57e-19 & 2.40e-18 & 2.55e-18   \\[5.2pt]
Sb$_\mathrm{i}$ &(1/0)=1.44 & 21.41 & $C_n^1$ & 0.002 & 0.12 & 3.81e-06 & 9.10e-15 & 3.81e-06   \\[5.2pt]
Sb$_\mathrm{i}$ & & & $C_p^0$ & $>$5 & 0.05 & 1.00e-127 & 1.33e-14 & 1.33e-14   \\[5.2pt]
Sb$_\mathrm{i}$ &(0/-1)=1.32 & 33.34 & $C_n^0$ & 0.75 & 0.13 & 1.56e-13 & 2.45e-15 & 1.58e-13   \\[5.2pt]
Sb$_\mathrm{i}$ & & & $C_p^{-1}$ & $>$5 & 0.06 & 9.95e-127 & 1.03e-13 & 1.03e-13   \\[5.2pt]
C$_\mathrm{Se}$ &(2/1)=0.52 & 10.65 & $C_n^2$ & 0.40 & 0.04 & 2.18e-10 & 2.57e-12 & 2.20e-10   \\[5.2pt]
C$_\mathrm{Se}$ & & & $C_p^1$ & 0.001 & 0.04 & 3.34e-10 & 1.89e-17 & 3.34e-10   \\[5.2pt]
C$_\mathrm{Se}$ &(1/0)=0.33 & 7.38 & $C_n^1$ & 1.35 & 0.07 & 2.49e-196 & 1.94e-12 & 1.94e-12   \\[5.2pt]
C$_\mathrm{Se}$ & & & $C_p^0$ & 0.14 & 0.005 & 3.20e-09 & 4.44e-16 & 3.20e-09   \\[5.2pt]
C$_\mathrm{Se}$ &(0/-1)=1.54 & 15.23 & $C_n^0$ & 0.38 & 0.10 & 9.61e-12 & 3.90e-16 & 9.61e-12   \\[5.2pt]
C$_\mathrm{Se}$ & & & $C_p^{-1}$ & 4.97 & 0.26 & 7.51e-15 & 2.12e-13 & 2.20e-13   \\[5.2pt]
C$_\mathrm{Se}$ &(-1/-2)=1.47 & 19.75 & $C_n^{-1}$ & 0.04 & 0.22 & 7.97e-08 & 1.53e-17 & 7.97e-08   \\[5.2pt]
C$_\mathrm{Se}$ & & & $C_p^{-2}$ & $>$5 & 0.12 & 1.82e-17 & 3.72e-13 & 3.72e-13   \\[5.2pt]
Si$_\mathrm{Se}$ &(2/1)=0.48 & 8.90 & $C_n^2$ & $>$5 & 0.10 & 4.21e-15 & 4.75e-12 & 4.75e-12   \\[5.2pt]
Si$_\mathrm{Se}$ & & & $C_p^1$ & 0.00001 & 0.02 & 1.13e-11 & 6.83e-18 & 1.13e-11   \\[5.2pt]
Si$_\mathrm{Se}$ &(1/0)=0.82 & 15.86 & $C_n^1$ & $>$5 & 0.28 & 2.69e-15 & 1.00e-12 & 1.00e-12   \\[5.2pt]
Si$_\mathrm{Se}$ & & & $C_p^0$ & $>$5 & 0.18 & 2.77e-15 & 3.28e-15 & 6.05e-15  \\[5.2pt]
Si$_\mathrm{Se}$ &(0/-1)=0.97 & 10.46 & $C_n^0$ & 4.81 & 0.14 & 1.09e-14 & 2.17e-15 & 1.31e-14   \\[5.2pt]
Si$_\mathrm{Se}$ & & & $C_p^{-1}$ & $>$5 & 0.24 & 1.80e-15 & 8.48e-13 & 8.50e-13   \\[5.2pt]
Si$_\mathrm{Se}$ &(-1/-2)=1.35 & 4.33 & $C_n^{-1}$ & 1.62 & 0.02 & 9.14e-22 & 8.59e-18 & 8.59e-18   \\[5.2pt]
Si$_\mathrm{Se}$ & & & $C_p^{-2}$ & $>$5 & 0.02 & 5.46e-59 & 4.56e-12 & 4.56e-12   \\[5.2pt]
Si$_\mathrm{i}$ &(2/1)=0.78 & 10.46 & $C_n^2$ & $>$5 & 0.06 & 1.84e-10 & 2.57e-13 & 1.84e-10   \\[5.2pt]
Si$_\mathrm{i}$ & & & $C_p^1$ & 0.86 & 0.10 & 2.80e-16 & 2.92e-17 & 3.10e-16   \\[5.2pt]
Si$_\mathrm{i}$ &(1/0)=0.65 & 12.11 & $C_n^1$ & $>$5 & 0.17 & 2.60e-15 & 1.82e-13 & 1.85e-13   \\[5.2pt]
Si$_\mathrm{i}$ & & & $C_p^0$ & $>$5 & 0.05 & 1.86e-16 & 1.61e-15 & 1.80e-15   \\[5.2pt]
Ge$_\mathrm{Se}$ &(2/1)=0.46 & 16.57 & $C_n^2$ & $>$5 & 0.17 & 1.63e-16 & 3.66e-13 & 3.66e-13   \\[5.2pt]
Ge$_\mathrm{Se}$ & & & $C_p^1$ & 0.07 & 0.03 & 3.71e-09 & 1.62e-17 & 3.71e-09   \\[5.2pt]
Ge$_\mathrm{Se}$ &(1/0)=0.29 & 9.10 & $C_n^1$ & $>$5 & 0.02 & 8.13e-127 & 2.55e-13 & 2.55e-13   \\[5.2pt]
Ge$_\mathrm{Se}$ & & & $C_p^0$ & 1.85 & 0.01 & 1.55e-17 & 4.10e-16 & 4.26e-16  \\[5.2pt]
Ge$_\mathrm{Se}$ &(0/-1)=1.20 & 12.17 & $C_n^0$ & 0.08 & 0.03 & 6.62e-08 & 7.16e-15 & 6.62e-08   \\[5.2pt]
Ge$_\mathrm{Se}$ & & & $C_p^{-1}$ & $>$5 & 0.01 & 7.39e-18 & 2.69e-13 & 2.69e-13   \\[5.2pt]
Ge$_\mathrm{Se}$ &(-1/-2)=1.49 & 6.93 & $C_n^{-1}$ & 0.06 & 0.04 & 4.49e-09 & 2.63e-17 & 4.49e-09   \\[5.2pt]
Ge$_\mathrm{Se}$ & & & $C_p^{-2}$ & $>$5 & 0.03 & 5.24e-246 & 1.02e-12 & 1.02e-12   \\[5.2pt]
Sn$_\mathrm{Se}$ &(2/1)=0.17 & 7.78 & $C_n^2$ & $>$5 & 0.02 & 1.63e-126 & 1.84e-12 & 1.84e-12   \\[5.2pt]
Sn$_\mathrm{Se}$ & & & $C_p^1$ & 0.03 & 0.02 & 2.91e-09 & 8.23e-19 & 2.91e-09   \\[5.2pt]
Sn$_\mathrm{Se}$ &(1/0)=0.48 & 7.47 & $C_n^1$ & $>$5 & 0.02 & 8.13e-127 & 4.92e-13 & 4.92e-13   \\[5.2pt]
Sn$_\mathrm{Se}$ & & & $C_p^0$ & $>$5 & 0.05 & 8.61e-17 & 1.86e-15 & 1.95e-15   \\[5.2pt]
Sn$_\mathrm{Se}$ &(0/-1)=1.28 & 11.80 & $C_n^0$ & 0.0004 & 0.18 & 2.43e-07 & 5.01e-15 & 2.43e-07   \\[5.2pt]
Sn$_\mathrm{Se}$ & & & $C_p^{-1}$ & $>$5 & 0.12 & 1.58e-16 & 5.06e-13 & 5.06e-13   \\[5.2pt]
Sn$_\mathrm{Se}$ &(-1/-2)=1.57 & 6.77 & $C_n^{-1}$ & 0.01 & 0.07 & 1.65e-08 & 1.26e-17 & 1.65e-08   \\[5.2pt]
Sn$_\mathrm{Se}$ & & & $C_p^{-2}$ & $>$5 & 0.02 & 1.99e-126 & 1.86e-12 & 1.86e-12   \\[5.2pt]
\end{longtable*}

\begin{table*}[h]
\caption{\label{tab:tables3}
Theoretical range (at $T$ = 300 K - 500 K) of extrinsic defect concentrations at different growth conditions at experimentally observed p-type majority carrier concentration (1.37x10$^{16}$ cm$^{-3}$) (initial equilibrium Fermi level at 0.21 eV above VBM). Device Efficiency (Open-circuit voltage) at low (10$^{14}$ cm$^{-3}$) and high (10$^{18}$ cm$^{-3}$) defect concentrations are represented as $\eta^\mathrm{1}$, $\eta^\mathrm{2}$ ($V_\mathrm{oc}^\mathrm{1}$, $V_\mathrm{oc}^\mathrm{2}$) respectively at 300 K and 500 nm film thickness.}
\begin{ruledtabular}
\begin{tabular}{cccccccc}
\textbf{Defect (D)} & $\boldsymbol{{n}^\mathrm{el-poor/Se-rich}}$ & $\boldsymbol{{n}^\mathrm{el-rich/Se-poor}}$ & $\boldsymbol{{n}^\mathrm{el-rich/Se-rich}}$ & $\boldsymbol{\eta^\mathrm{1}}$ & $\boldsymbol{\eta^\mathrm{2}}$ & $\boldsymbol{V_\mathrm{oc}^\mathrm{1}}$ & $\boldsymbol{V_\mathrm{oc}^\mathrm{2}}$ \\[5.2pt]
& (cm$^3$s$^{-1}$) & (cm$^{-3}$) & (cm$^{-3}$) & (\%) & (\%) & (V) & (V) \\[5.2pt]
\hline
v$_{\rm Se}$ & 9.67x10$^{5}$-7.13x10$^{11}$ & 9.67x10$^{5}$-7.13x10$^{11}$ & 9.67x$10^{5}$-7.13x10$^{11}$ & 23.19 & 20.90 & 1.59 & 1.40 \\[5pt]
Se$_{\rm i(1)}$ & 2.92x10$^{3}$-2.19x10$^{10}$ & 2.92x10$^{3}$-2.19x10$^{10}$ & 2.92x10$^{3}$-2.19x10$^{10}$ & 23.24 & 20.40 & 1.59 & 1.40 \\[5pt]
H$_{\rm Se}$ & 2.82x10$^{14}$-8.55x10$^{16}$ & $>$10$^{18}$ & $>$10$^{18}$ & 23.25 & 21.83 & 1.59 & 1.57 \\[5pt]
H$_{\rm i}$ & 2.27x10$^{17}$-$>$10$^{18}$ & $>$10$^{18}$ & $>$10$^{18}$ & 23.25 & 21.55 & 1.59 & 1.56 \\[5pt]
F$_{\rm Se}$ & 1.96x10$^{9}$-6.88x10$^{13}$  & $>$10$^{18}$ & $>$10$^{18}$ & 23.14 & 19.06 & 1.59 & 1.32 \\[5pt]
F$_{\rm i}$ & 7.14x10$^{5}$-5.96x10$^{11}$ & $>$10$^{18}$ & $>$10$^{18}$ & 23.25 & 19.75 & 1.59 & 1.51 \\[5pt]
Cl$_{\rm Se}$ & 9.23x10$^{14}$-1.74x10$^{17}$    & $>$10$^{18}$ & $>$10$^{18}$ & 23.25 & 18.14 & 1.59 & 1.42 \\[5pt]
Cl$_{\rm i}$ & 4.20x10$^{2}$-6.85x10$^{9}$ & 7.02x10$^{9}$-1.48x10$^{14}$ & 7.02x10$^{9}$-1.48x10$^{14}$ & 23.25 & 20.31 & 1.59 & 1.43 \\[5pt]
Br$_{\rm Se}$ & 1.07x10$^{15}$1.91x10$^{17}$ & $>$10$^{18}$ & $>$10$^{18}$ & 23.24 & 17.33 & 1.59 & 1.38 \\[5pt]
I$_{\rm Se}$ & 3.88x10$^{13}$-2.60x10$^{16}$  & 3.88x10$^{13}$-2.60x10$^{16}$ & 3.88x10$^{13}$-2.60x10$^{16}$ & 23.24 & 20.25 & 1.59 & 1.41 \\[5pt]
I$_{\rm i}$ & 2.58x10$^{1}$-1.29x10$^{9}$ & 2.58x10$^{1}$-1.29x10$^{9}$ & 2.58x10$^{1}$-1.29x10$^{09}$  & 23.25 & 21.65 & 1.59 & 1.54 \\[5pt]
N$_{\rm Se}$ & 5.46x10$^{-26}$-1.34x10$^{-7}$ & 5.46x10$^{-26}$-1.34x10$^{-7}$ & 5.46x10$^{-26}$-1.34x10$^{-7}$ & 23.24 & 17.57 & 1.59 & 1.39 \\[5pt]
P$_{\rm Se}$ & 3.40x10$^{8}$-2.42x10$^{13}$   & 8.39x10$^{15}$-$>$10$^{18}$ & 1.69x10$^{12}$-3.98x10$^{15}$ & 23.20 & 19.57 & 1.59 & 1.36 \\[5pt]
As$_{\rm Se}$ & 3.52x10$^{6}$-2.12x10$^{12}$ & 1.33x10$^{16}$-$>$10$^{18}$ & 1.81x10$^{12}$-5.68x10$^{15}$ & 23.22 & 19.65 & 1.59 & 1.37 \\[5pt]
As$_{\rm i}$ & 2.61x10$^{0}$-3.25x10$^{8}$  & 1.34x10$^{6}$-8.69x10$^{11}$ & 1.34x10$^{6}$-8.69x10$^{11}$ & 23.25 & 20.92 & 1.59 & 1.55 \\[5pt]
Sb$_{\rm Se}$ & 7.34x10$^{5}$-6.05x10$^{11}$  & $>$10$^{18}$ & 4.14x10$^{16}$-$>$10$^{18}$ & 23.25 & 20.93 & 1.59 & 1.51 \\[5pt]
Sb$_{\rm i}$ & 7.19x10$^{8}$-3.83x10$^{13}$ & $>$10$^{18}$ & $>$10$^{18}$ & 23.25 & 18.56 & 1.59 & 1.43 \\[5pt]
C$_{\rm Se}$ & 2.97x10$^{-34}$-1.45x10$^{-12}$ & 2.97x10$^{-34}$-1.45x10$^{-12}$  & 2.97x10$^{-34}$-1.45x10$^{-12}$ & 22.68 & 17.19 & 1.59 & 1.20 \\[5pt]
Si$_{\rm Se}$ & 2.94x10$^{-12}$-2.36x10$^{1}$ & $>$10$^{18}$ & 2.27x10$^{16}$-$>$10$^{18}$ & 23.24 & 18.72 & 1.59 & 1.30 \\[5pt]
Si$_{\rm i}$ & 1.46x10$^{0}$-2.30x10$^{8}$  & $>$10$^{18}$ & $>$10$^{18}$ & 23.25 & 21.96 & 1.59 & 1.58 \\[5pt]
Ge$_{\rm Se}$ & 3.40x10$^{-5}$-4.47x10$^{5}$ & $>$10$^{18}$ & 6.97x10$^{13}$-4.33x10$^{16}$ & 23.14 & 18.91 & 1.59 & 1.26 \\[5pt]
Sn$_{\rm Se}$ & 7.54x10$^{5}$-6.25x10$^{11}$ & $>$10$^{18}$ & $>$10$^{18}$ & 23.06 & 17.17 & 1.59 & 1.15 \\[5pt]

\end{tabular}
\end{ruledtabular}
\end{table*}

\clearpage

\makeatletter
\c@enumiv=0
\makeatother

%